\documentclass[pra,superscriptaddress,showpacs,preprint,aps,longbibliography]{revtex4-1}
\usepackage{epsfig}             
\usepackage{epstopdf}           
\usepackage{hyperref}           
\usepackage{color}
\usepackage{colortbl}
\usepackage{graphicx}
\usepackage{amssymb,amsmath}
\usepackage{subfigure}
\usepackage{caption}
\usepackage{enumerate}
\usepackage{gensymb}
\usepackage{url}

\begin{document}

\title[HHG]{High-order harmonic generation in atomic and molecular systems}

\author{Noslen Su\'arez}
\email[]{noslen.suarez@icfo.es}
\affiliation{ICFO - Institut de Ci\`encies Fot\`oniques, The Barcelona Institute of Science and Technology, Av. Carl Friedrich Gauss 3, 08860 Castelldefels (Barcelona), Spain}

\author{Alexis Chac\'on}
\affiliation{ICFO - Institut de Ci\`encies Fot\`oniques, The Barcelona Institute of Science and Technology, Av. Carl Friedrich Gauss 3, 08860 Castelldefels (Barcelona), Spain}

\author{Jose A. P\'erez-Hern\'andez}
\affiliation{Centro de L\'aseres Pulsados (CLPU), Parque Cient\'ifico, 37185 Villamayor, Salamanca, Spain}

\author{Jens Biegert}
\affiliation{ICFO - Institut de Ci\`encies Fot\`oniques, The Barcelona Institute of Science and Technology, Av. Carl Friedrich Gauss 3, 08860 Castelldefels (Barcelona), Spain}
\affiliation{ICREA - Pg. Llu\'is Companys 23, 08010 Barcelona, Spain}

\author{Maciej Lewenstein}
\affiliation{ICFO - Institut de Ci\`encies Fot\`oniques, The Barcelona Institute of Science and Technology, Av. Carl Friedrich Gauss 3, 08860 Castelldefels (Barcelona), Spain}
\affiliation{ICREA - Pg. Llu\'is Companys 23, 08010 Barcelona, Spain}

\author{Marcelo F. Ciappina}
\affiliation{Institute of Physics of the ASCR, ELI-Beamlines project, Na Slovance 2, 182 21 Prague, Czech Republic}

\date{\today}
\pacs{32.80.Rm,33.20.Xx,42.50.Hz}

\begin{abstract}
High-order harmonic generation (HHG) results from strong field laser matter interaction and it is 
one of the main processes that are used to extract electron structural and dynamical information 
about the atomic or molecular targets with sub-femtosecond temporal resolution. Moreover, it is 
the workhorse for the generation of attosecond pulses. Here we develop an analytical 
description of HHG, which extends the well established theoretical strong field approximation 
(SFA). Our approach involves two innovative  aspects: i) First, using a model non-local, but separable potential, we calculate  the bound-free dipole and the rescattering transition matrix elements  analytically for both atomic and molecular multicenter systems. In comparison with the standard approaches to the HHG process, these analytic
derivations of the different matrix elements allows us to study directly how the HHG spectra 
depend on the atomic target and laser pulse features. We can turn on and off contributions 
having distinct physical origins or corresponding to different physical mechanisms. This allow us 
to quantify their weights in the various regions of the HHG spectra; ii) Second, as Ref.~[Phys. Rev. A {\bf 94}, 043423 (2016)] reports, in our theory the multicenter dipole matrix elements are free from non-physical gauge and coordinate system dependent terms -- this is
achieved by adapting the coordinate system, in which the SFA is formulated, to the centre from which
the corresponding part of the time dependent wave function originates. 
Our SFA results are compared, when possible, with the direct numerical integration of the time-dependent Schr\"odinger equation 
(TDSE) in reduced and full dimensionality. Excellent agreement is found for single and multielectronic atomic systems, modeled under the single active electron approximation, and for simple diatomic molecular systems. Our model 
captures also the interference features, ubiquitously present in every strong field phenomenon involving a multicenter target. 
\end{abstract}

\maketitle

\section{Introduction}

High-order harmonic generation (HHG) is a conversion process resulting from the extremely 
high nonlinear interaction of a short and intense laser pulse with gas atoms or molecules or, 
recently, solid targets and nanostructures~\cite{ALHuillier1991, ALHuillier1996,Krausz2009, Ghimire2011,Vampa2014, Marcelo2016}. 
Nowadays, the HHG process is the conventional route for the production of 
spatially and temporally coherent extreme-ultraviolet (XUV) light, as well as light pulses in the 
sub-femtosecond and attosecond regimes~\cite{PBCorkum2007}. Coherent light sources in the 
ultraviolet (UV) to XUV spectral range are ubiquitously employed in a broad range of subjects, 
including basic research, material science, biology, and lithography~\cite{Krausz2009}. 
Furthermore, the molecular HHG process encodes electronic orbital structure information and 
presents, as a consequence, a reliable method to retrieve molecular intrinsic parameters with 
attosecond and sub-{\AA}ngstr\"om temporal and spatial resolution, 
respectively~\cite{MLein2007,PRAMLein2002, Kanai2005, Torres2007, Morishita2008}. Taking this objective in mind, several theoretical and experimental work have been conducted in order to 
optimize, improve and understand the molecular HHG process.  Furthermore, HHG in atoms is 
one of the most studied topics of strong field physics and several theoretical models, besides of 
the solution of the time-dependent Schr\"odinger equation (TDSE), have been developed to 
describe it. Amongst them the most widely used and successful is the strong field approximation 
(SFA)~\cite{Lewenstein1995, Lewenstein1994}.

The underlying physics of the HHG process is usually understood invoking the so-called ``three step model'': (i) tunnel ionization; (ii)  propagation in the laser field ``continuum'', and (iii) recombination with the parent ion~\cite{PBCorkum1993, Lewenstein1994}. According to this approach, when a strong laser pulse interacts with an atomic or molecular target a bounded electron is liberated through tunnel ionization (this happens when the laser electric field is close to its peak during an optical cycle). This ``free'' electron is then driven away from the ionic core and accelerated by the laser electric field, developing an oscillating trajectory. During this journey, the electron accumulates kinetic energy, that is released during the recombination process in the form of a high energy photon. As this three-step process usually occurs every half-cycle of the laser field, the spectrum of the generated coherent radiation consists of peaks at odd integer multiples of the driven laser frequency.

On the other hand, for multicenter molecules much less experimental~\cite{ ALHuillier1996,Schmidt1994,Normand1994, Lappas2000,Velotta2001,Hay2002, Lein_Hay2002} and theoretical~\cite{Ivanov1993, Zuo1993,  Yu1995,  Moreno1997, Kopold1998,Lappas2000, Alon1998,Alon2001, Yu1999, Kreibich2001} work have been done. The direct numerical solution of the TDSE for more complex systems with more than two centers is a quite challenging and formidable task from the numerical and computational viewpoints. Even for the simplest diatomic molecule, one has to solve a three-dimensional TDSE, that typically requires the utilization of a multicore CPUs and large amount of memory. In addition, the interpretation of the results extracted from TDSE is a not trivial task, in particular if one wants to disentangle the underlying physical mechanisms contributing to the total HHG spectrum.

As was mentioned above, the initial interest in the molecular HHG was due to the fact that it offers additional degrees of freedom and promising possibilities, such as the alignment of the molecular axis with respect to the laser field polarization axis. Specifically for a diatomic system, the existence of a distinctive quantum-interference minima pattern in the spectra and its dependence with the molecular orientation have been theoretically predicted by Lein et al. \cite{PRAMLein2002,PRAMLein2003,marcelo2007}. It is demonstrated that this pattern is due to a destructive interference from the high-harmonic emission at spatially separated centers and the internuclear distance can be accurately obtained scrutinizing the HHG spectra.  In addition, the chance of controlling the phases and improve the phase-matching condition opens a route to the investigation in this area. More importantly, research on this field revealed how the distinctive features of the molecular HHG spectra can be used to retrieve structural information in simple molecules~\cite{NatItatani2004}. Furthermore, the high harmonic generation spectroscopy has shown the possibility to extract structural and dynamical information from the molecular HHG spectra in more complex targets (for a couple of examples see e.g.~\cite{marceloc60,marceloc180,breathing}). Finally, studies in small molecules shows that the temporal evolution of the electronic wavefunction can be directly recovered~\cite{Smirnova2009, Haessler2010, Kraus2015}.  

In this paper, we use the SFA within the framework of the Lewenstein's model to study the HHG from atomic, diatomic and three atomic molecular systems in the few-cycle IR laser pulse regime. The derivation for the two and three centers molecular systems is constructed as a consecutive extension of the atomic model. For simplicity, our analytical model is based on a non-local potential which is approximately a short-range (SR) potential. We compute HHG spectra for those three systems and, for the atomic case, compare the results with the numerical solution of the 3D-TDSE in the single active electron (SAE) approximation.

Our approach involves two innovative  aspects: i) First, using a model non-local, but separable potential, we calculate  the bound-free dipole and the rescattering transition matrix elements  analytically for both atomic and molecular multicenter systems. In comparison with the standard approaches to the HHG process, these analytic
derivation of the different matrix elements allows us to study directly how the HHG spectra 
depend on the atomic target and laser-pulse features; we can turn on and off contributions 
having distinct physical origins or corresponding to different physical mechanisms. This allow us 
to quantify their weights in the various regions of the HHG spectra; ii) Second, as in Ref.~\cite{Molecule2016}, 
in our theory  the dipole matrix elements are free from non-physical gauge and coordinate system dependent terms -- this is
achieved by adapting the coordinate system, in which SFA is performed, to the centre from which
the corresponding part of the time dependent wave function originates. We compare, when possible,
our SFA results with the numerical solutions of the time-dependent Schr\"odinger equation in full dimensionality. Excellent agreement is found for atomic and molecular systems, including 
multielectronic systems modeled under the SAE. Our model 
captures also the interference features, ubiquitously present in any multicenter target. 

This article is organized as follows. In Section II, we address the main theory that describes the HHG process and, in particular, the derivation of the time-dependent dipole matrix element within the SFA. Here, we make use of the results previously presented in \cite{PRANoslen2015, Molecule2016} to obtain the analytical expressions needed to compute each of the individual contributions. Particularly for the three-center molecular system, we develop a new set of equations to compute the time-dependent dipole matrix element, making use of the non-local short-range potential bound states. In section III, HHG spectra for the atomic and molecular cases are numerically calculated. Results in hydrogen and argon atoms are presented, comparing them with those obtained from the 3D-TDSE. For diatomics, we analyze two systems: H$_2^+$ and H$_2$. The basic analysis of the interference minima of the harmonic spectra with respect to the alignment for H$_2^+$  is discussed. In addition, the contribution of the different processes to the total spectra is assessed. A time analysis of the HHG spectra using a Gabor transformation is performed and the influence of the short and long trajectories is investigated. In addition, CO$_2$ and H$_2$O define our three-center molecular systems. For both cases, we investigate the dependence of the HHG spectra with respect to the molecular orientation and extract information about the different mechanism contributing to the total HHG spectra. Finally, in section IV, we summarize the main ideas and present our conclusions. Atomic units will be uses throughout the manuscript otherwise stated.

\section{Theory of HHG within the SFA}

In this section we develop a quantum mechanical approach of HHG using the generalized SFA model described in Ref.~\cite{Lewenstein1995, Lewenstein1994}.  The source of the additional frequencies that are generated during the interaction of a strong laser pulse with an atomic or molecular target, is the nonlinear dipole oscillation of the medium. Therefore, the aim is to calculate this dipole response by mean of the solution of the time dependent Schr\"odinger equation. The time-dependent radiation dipole moment reads:
\begin{equation}
\vec{\mu} (t)= - \langle \Psi(t) |\textbf{r}|\Psi(t) \rangle,
\label{Eq:d(t)_A}
\end{equation}
where $| \Psi(t) \rangle$, is the state describing the time-evolution of the atomic or molecular system under study.
In general, within the SFA statement, we can write the wavefunction of the whole system as a superposition of the ground, $|0\rangle$, and continuum states, $|\textbf{v}\rangle$, as
$| \Psi (t)\rangle= e^{\textit{i}I_p\textit{t}}(a(t) |0 \rangle + \: \int{\textit{d}^3 \textbf{v} \:  \textit{b}( \textbf{v},t) |\textbf{v}\rangle} )$,
where the transition amplitude of the continuum states is denoted by $\textit{b}(\textbf{v},t)$.
After some algebra with the above equations and only considering transitions from the bound to the continuum states the time-dependent dipole radiation moment reads
\begin{eqnarray}
\vec{\mu}(t)&=&\int{\textit{d}^3 \textbf{v}\: \textit{b}( \textbf{v},t) \: \textbf{d}^*( \textbf{v})} + \textit{c}.\textit{c}.,
\label{Eq:Dipole}
\end{eqnarray}
where the bound-continuum transition dipole matrix element is defined as $\textbf{d}( \textbf{v})=-\langle \textbf{v} |\textbf{r}|0 \rangle$. The radiation emitted by a single atom is proportional to the time-dependent dipole moment $\vec{\mu}(t)$. In this way the harmonic spectrum, $I(\omega)$, is calculated as the modulus squared of the Fourier transformed dipole acceleration -$\textbf{a}(t)$- related to the defined time-dependent dipole matrix element, Eq.~(\ref{Eq:Dipole}), by the Ehrenfest theorem as $|\tilde{\textbf{a}}(\omega)|=|\omega^2\: \tilde{\mu}(\omega)|$. In this way we can compute the harmonic spectra as:
\begin{equation}
I_{x{\rm N}}(\omega)\propto\: \Bigg|\int_{-\infty}^{\infty}{\textit{d}t \: e^{i \omega t}\: \vec{\mu}_{x{\rm N}}(t)}\Bigg|^2.
\label{Eq:HHG}
\end{equation}

In here the subscript $x$ represents the total numbers of atoms in the system to study and it will count as $x=1,2,\dots,n$, where $n$ is the total number of atoms of the molecule. For case of a diatomic molecule, i.e.~constituted by two atoms, the subscript reads as: $x=2$ meaning $\vec{\mu}_{\rm 2N}(t)$ and $I_{\rm 2N}(\omega)$. Notice that both the atomic and molecular harmonic spectra depend directly on the time-dependent dipole moment which in turn depends on the form of the bound-continuum matrix element and the continuum states transition amplitude, that are different for each of atomic, diatomic or multiatomic system under study.

\subsection{Calculation of the time-dependent dipole moment for atomic systems: $\vec{\mu}_{\rm 1N}(t)$}

In order to have all the ingredients to compute the harmonic spectrum for an atomic system, $I_{1N}(\omega)$, using the Eqs.~(\ref{Eq:Dipole}) and (\ref{Eq:HHG}) we need to know the exact dependency of the bound-continuum matrix element and the continuum states transition amplitude. The method to find the transition amplitude of the continuum states and bound-continuum matrix element for an atom under the influence of an intense laser pulse has been described in our previous work~\cite{PRANoslen2015}. We therefore take advantage of those results and only explain here the new derivations needed to tackle the HHG problem. 

The transition amplitude for the continuum states of the atomic system reads as:
\begin{equation}
b( \textbf{p},t) = i \int_0^t{\textit{d} \textit{t}^{\prime}\:\textbf{E}(t^{\prime})}\:\cdot \textbf{d}\left[ \textbf{p}+\textbf{A}(t^{\prime})\right]  e^{-\textit{i} \: {S}({\bf p},t,t^{\prime})},
 \label{Eq:IntDirecTerm}
\end{equation}
where the exponent phase factor is ``the semiclassical action'' ${S}({\bf p},t,t^{\prime})=\int_{t^{\prime}}^{t}{\:d{\tilde t}\left\{[{\bf p}+\textbf{A}({\tilde t})]^2/2 +I_p \right \}}$ defining all the possible electron trajectories from the birth time $t'$ until the ``recombination'' one $t$. 

The explicit expression for the bound-continuum transition dipole matrix element obtained in~\cite{PRANoslen2015} is
\begin{equation}
\textbf{d}_{\rm1N}( \textbf{p}_0) = \textit{i}\:\textbf{p}_0 \: \frac{(p_0^2 + \Gamma^2) + (\frac{p_0^2}{2} + I_p)}{(p_0^2 + \Gamma^2)^{\frac{3}{2}}(\frac{p_0^2}{2} + I_p)^2} \Bigg [\frac{ \Gamma +  \sqrt{2I_p}}{2\pi \: (2I_p)^{-1/4}} \Bigg].
 \label{Eq:d1NFn}
\end{equation}
Inserting Eqs.~(\ref{Eq:IntDirecTerm}) and~(\ref{Eq:d1NFn}) in the 
time-dependent dipole moment, Eq.~(\ref{Eq:Dipole}), and changing variables to the canonical 
momentum defined by $\textbf{p}=\textbf{v}-\textbf{A}(t)$ we get,
\begin{eqnarray}
\vec{ \mu}_{\rm 1N}(t)= i \int_0^t{\textit{d} \textit{t}^{\prime} }\:\int{\textit{d}^3 \textbf{p}}&&\:\textbf{E}(t^{\prime})\:\cdot \textbf{d}_{\rm 1N}\left[ \textbf{p}+\textbf{A}(t^{\prime})\right] e^{-\textit{i} \: {S}({\bf p},t,t^{\prime})} \:\textbf{d}_{\rm 1N}^*[ \textbf{p}+\textbf{A}(t)]+ \textit{c}.\textit{c}..
 \label{Eq:Dipole2}
 \end{eqnarray}
Equation (\ref{Eq:Dipole2}) has to be understood as follows: the electron is ionized at 
time $t'$ with a certain probability defined by, 
$\textbf{E}(t^{'})\cdot \textbf{d}_{\rm 1N}\left[ \textbf{p}+\textbf{A}(t^{'})\right]$. During its excursion in the continuum the electronic wavefunction is then propagated until the time $t$ acquiring a classical phase $ \textit{S}(\textbf{p},t,t') $ to finally recombine with the ion core at time $t$ with a rate given by $\textbf{d}_{\rm 1N}^*[ \textbf{p}+\textbf{A}(t)]$.
 All possible combinations of birth time and momenta must be considered and therefore a multidimensional integration is required, where their contributions are added up coherently. Note that Eq.~(\ref{Eq:Dipole2}) configures a highly oscillatory integral, both in the momentum $\textbf{p}$ and $t'$ variables. As a consequence it is convenient to rewrite the integral over ${\bf p}$ using the stationary-phase approximation or saddle point method. In order to do that is necessary  to find the extremal points over the exponential phase. The extrema $\mathbf{p} = {\bf p}_s$ are found from the solutions of  $\nabla_{{\bf p}} \textit{S}(\textbf{p})|_{{\bf p}_s}={\bf 0} $. These saddle point momenta ${\bf p}_s$ thus can be written as $\textbf{p}_s = -\frac{1}{\tau}\int_{t^{{\prime}}}^{t}  { \textbf{A}(\tilde{t}) \textbf{d}\tilde{t}}$.
Here, $\tau=t-t^{\prime}$ is the excursion time of
the electron in the continuum. Expanding, the function $ \textit{S}(\textbf{p},t,t') $ in a Taylor series around the roots $\textbf{p}_s$ and then applying the standard saddle point method to the momentum integral over ${\bf p}$, the time-dependent dipole matrix element for the atomic system reads as:
\begin{eqnarray}
\vec{\mu}_{\rm1N}(t)&=& i  \int_0^t{\textit{d} \textit{t}^{\prime}\: \left( \frac{\pi}{\varepsilon +{\frac{\textit{i}(t-t')}{2}}}  \right)^{\frac{3}{2}} \:\textbf{E}(t^{\prime})}\:\cdot \textbf{d}_{\rm1N}\left[ \textbf{p}_s+\textbf{A}(t^{\prime})\right] \nonumber\\
 &&\times \: e^{-\textit{i} \: {S}(\textbf{p}_s,t,t^{\prime})} \: \: \textbf{d}_{\rm1N}^*[ \textbf{p}_s+\textbf{A}(t)]+ \textit{c}.\textit{c}.,
 \label{Eq:m_HH1N}
\end{eqnarray}
where we have introduced an infinitesimal  parameter, $\varepsilon$, to avoid the divergence 
at  $t=t'$ (for a detailed discussion see~\cite{PRANoslen2015, Molecule2016}). The harmonic 
spectrum, $I_{1N}(\omega)$, is then numerically computed inserting Eq.~(\ref{Eq:m_HH1N}) 
in Eq.~(\ref{Eq:HHG}).

\subsection{Calculation of the time-dependent dipole moment for diatomic molecular systems: $\vec{\mu}_{\rm 2N}(t)$}
In order to calculate the harmonic spectrum generated by a diatomic molecule we use the results obtained in Ref.~\cite{Molecule2016}. As we can extract from that reference the general wavefunction describing the state for a diatomic molecule can be written as:
\begin{eqnarray}
|\Psi(t) \rangle &=e^{\textit{i}I_p\textit{t}}\bigg(a(t) |0\rangle  +\:\sum_{j=1}^2 \int{\textit{d}^3 \textbf{v} \textit{b}_j( \textbf{v},t) |\textbf{v}\rangle} \bigg),
\label{Eq:PWfull}
\end{eqnarray}
from which the molecular time-dependent dipole moment $\vec{\mu}_{\rm 2N}(t)$ is easily obtained and have the following form:
\begin{eqnarray}
\vec{\mu}_{\rm 2N}(t)&=&  \sum_{j=1}^2\int{\textit{d}^3 \textbf{v}\: \textbf{d}^*_{\rm 2N}( \textbf{v}) \textit{b}_{j}( \textbf{v},t)} + \textit{c.c.}.
 \label{Eq:dipMole}
\end{eqnarray}
In the above equation we require to insert the explicit expression for the continuum states transition amplitude $b(\textbf{p},t)$: 
\begin{eqnarray}
b( \textbf{p},t) &= & b_{0,1}( \textbf{p},t)+b_{0,2}( \textbf{p},t),\nonumber\\
&=& i \sum_{j=1}^2\:\int_0^t{\textit{d} \textit{t}^{\prime}\:\textbf{E}(t^{\prime})}\cdot \textbf{d}_{j}\left[\textbf{p}+\textbf{A}(t^{\prime})\right]e^{-i \left\{S({\bf p},t,t') +\textbf{R}_j \cdot \left[ \textbf{A}(t) -\textbf{A}(t')\right] \right\}} 
\label{Eq:b_0R}
\end{eqnarray}
and the bound-continuum dipole matrix element ${\bf d}_j$. 

In the derivation of the length-gauge SFA model for HHG in diatomic molecules, and in particular for the computation of the bound-continuum dipole matrix element $\textbf{d}( \textbf{v}) = - \langle \textbf{v} |\textbf{r}|0 \rangle$, an unphysical term is neglected, without give a consistent reason/argument (see~\cite{CarlaPRA2007,LeinPRA2006,MilosevicPRL2008} for more details). This term, that is a linear function of the internuclear distance $\textbf{R}$, immediately introduces convergence problems as ${\textbf R}\to \infty$. Clearly, this behavior introduces conflicts between the length and velocity gauges predictions, observed in the case of above-threshold ionization (ATI) as well. The root of the problem relies in the degree of approximation to handle the continuum states, considered as a set of plane waves for the molecular system, without considering the relative position of each atomic core. This creates an unphysical treatment and therefore the appearance of such unphysical term. In our approach we solve this issue by computing ${\bf d}_j( \textbf{v})= - \langle \textbf{v} |(\hat{\textbf{r}} -\textbf{R}_j) | 0_j \rangle$, where here the bound-continuum dipole matrix element is calculated with respect to each atomic center located at $\textbf{R}_j$. Note that if no approximations are done, i.e.~if we consider the case where $\langle \textbf{v}|$ is a scattering wave of the field free Hamiltonian H$_0$, the above mentioned problem will not arise -- the scattering waves are orthonormal to the ground states $|0_j\rangle$. However, as the main core of the SFA is to handle the continuum states as Volkov states, i.e.~neglecting the influence of the residual molecular potential once the electron is in the continuum, the convergence problems would remain if we do not correct  the bound-continuum dipole matrix element. The full derivation of the bound-continuum dipole matrix element for the ATI problem in a two-center molecular system was introduced in Ref.~\cite{Molecule2016} - this bound-continuum dipole matrix element is the same used for the computation of HHG. In addition, an extended derivation for a three-center molecular system is presented in the Appendix A.

The bound-continuum dipole matrix element is defined by $\textbf{d}_{\rm 2N}( \textbf{p}_0)=\sum_j^2{\bf d}_j({\bf p}_0)$, where ${\bf d}_j$ denotes the bound-continuum dipole matrix element related to the nucleus located at the position ${\bf R}_j$ and is given by: 
\begin{eqnarray}
{\bf d}_j({\bf p}_0)=-2\textit{i}\: \mathcal{M} \frac{- \textbf{p}_0\: (3p_0^2 + 2I_p +2\Gamma^2)}{(p_0^2 + \Gamma^2)^{\frac{3}{2}}(p_0^2 + 2I_p)^2} \:  e^{-i\textbf{R}_j \cdot \textbf{p}_0}.
\end{eqnarray}
Here $\mathcal{M}$ is a normalization constant (for details see Ref.~\cite{Molecule2016}). Note that the index $j$ can take the value of 1 (or 2), referring to the nucleus located at the position ${\bf R}_1$ (or ${\bf R}_2$) on the left (and right).

The time-dependent radiation dipole moment $\vec{\mu}_{\rm2N}(t)$ thus reads: 
\begin{eqnarray}
\hspace{-0.75cm}\vec{\mu}_{\rm 2N}(t)= i \sum_{j=1}^2\sum_{j'=1}^2 && \int_0^t \textit{d} \textit{t}^{\prime} \int \textit{d}^3 \textbf{p} \, \textbf{E}(t^{\prime}) \cdot \textbf{d}_{j}\left[\textbf{p}+\textbf{A}(t^{\prime})\right] 
 \nonumber\\
 &&\times\: e^{-i\left\{S({\bf p},t,t')+ \textbf{R}_j \cdot \left[ \textbf{A}(t) -\textbf{A}(t')\right] \right\}} \: \textbf{d}^*_{j'}[ \textbf{p}+\textbf{A}(t)],  
\label{Eq:HHGd2N}
\end{eqnarray}
where subscript $j$ and $j'$ represent the ionization the recombination atom positions, respectively.

Equation (\ref{Eq:HHGd2N}) contains information about all the recombination processes occurring in the entire molecule during the HHG phenomenon and can then be written as a sum of components as:
\begin{eqnarray}
\vec{\mu}_{\rm 2N}(t) &=& \sum_{j=1}^2\sum_{j'=1}^2\vec{\mu}_{j j'}(t).
\label{Eq:HHGd2N1}
\end{eqnarray}
The four terms in the above equation encode all the individual molecular recombination processes. Our physical interpretation of those contributions is as follows:
\begin{enumerate}[(i)]

\item An electron is ionized from the atom placed at the \textit{Left} 
with respect to the coordinate origin at time $t'$ with certain probability: 
$\textbf{E}(t^{\prime}) \cdot \textbf{d}_{1} \left[\textbf{p}+\textbf{A}(t^{\prime})\right]$. During its excursion in the continuum this electron accumulates a phase which depends on the position from where it was detached, in this case $\textbf{R}_1$. Finally, because the electric field changes its sign and the electron returns to the parent ion, the probability of recombination results $\textbf{d}^*_{1}[ \textbf{p}+\textbf{A}(t)]$. In this step the excess of energy acquired from the laser electric field is converted into a high energy photon. This whole process is described by:
\begin{eqnarray}
\hspace{-.5cm}\vec{\mu}_{11}(t)&=& i \int_0^t{\textit{d} \textit{t}^{\prime} \int{  \textit{d}^3  \textbf{p}}} \, \textbf{E}(t^{\prime}) \cdot \textbf{d}_{1} \left[\textbf{p}+\textbf{A}(t^{\prime})\right]\nonumber\\
&&\times \: e^{-i\left\{S({\bf p},t,t') +\textbf{R}_1 \cdot \left[ \textbf{A}(t) -\textbf{A}(t')\right]\right\}} \, \textbf{d}^*_{1}[ \textbf{p}+\textbf{A}(t)].
\label{Eq:H2N_LL}
\end{eqnarray}
 \item The second term is understood in a similar way. In this case the ionization and recombination processes occur in the core placed at the \textit{Right}. The equation describing this process, $\vec{\mu}_{22}$, is similar to Eq.~(\ref{Eq:H2N_LL}) but considering the dipole matrix element $\textbf{d}_{2}$ and now the electron is detached from the position $\textbf{R}_2$. The two processes described before are spatially localized (involving only one core placed at a fixed position $\textbf{R}_1$ or $\textbf{R}_2$) and we then refer to them as ``Local Processes''.

 \item The last two terms, $\vec{\mu}_{21}(t)$ and $\vec{\mu}_{12}(t)$, describe events involving two atoms at two different positions $\textbf{R}_1$ or $\textbf{R}_2$. Here $\vec{\mu}_{21}$ can be understood as follows: the electron is tunnel-ionized  from the atom on the \textit{Right} with certain probability given by: $\textbf{E}(t^{\prime}) \cdot \textbf{d}_{2} \left[\textbf{p}+\textbf{A}(t^{\prime})\right]$. After this ionization event the electron starts to move under the laser electric field influence accumulating energy and acquire a phase: $e^{-i\left\{S({\bf p},t,t') + \textbf{R}_2 \cdot \left[\textbf{A}(t) -\textbf{A}(t')\right]\right\}}$. Finally the electron returns back to the other core ($\textit{Left}$) at the time $t$ to end up its journey in a recombination process that has an amplitude  proportional to: $\textbf{d}^*_{1}[ \textbf{p}+\textbf{A}(t)]$. As in previous cases the excess energy is emitted in a form of a high energy photon. Considering both centers are involved in the HHG process, we call these terms as ``Cross processes''. 
The equation describing these processes reads:
\begin{eqnarray}
\hspace{-0.75cm}\vec{\mu}_{j j'}(t)&=& i \int_0^t{\textit{d} \textit{t}^{\prime}\: \int{\textit{d}^3 \textbf{p}}}\: \textbf{E}(t^{\prime}) \cdot \textbf{d}_{ {j}} \left[\textbf{p}+\textbf{A}(t^{\prime})\right] \nonumber\\
&&\times\: e^{-i\left\{S({\bf p},t,t')+\textbf{R}_{j} \cdot \left[ \textbf{A}(t) -\textbf{A}(t')\right]\right\}} \: \textbf{d}^*_{ {j'}}[ \textbf{p}+\textbf{A}(t)],
\label{Eq:H2N_RL/LR}
\end{eqnarray}
where now $j\neq j'$ denotes the nucleus-index located at left ($j$=1) 
or right ($j$=2).
\end{enumerate} 

Note from the above description that we have to account four different possible processes corresponding to four different time-dependent transition dipole moments. Two of them are ``Localized'' and two `` Cross'' representing all the possible recombination scenarios in our diatomic molecule.

Similarly to the atomic case, in order to obtain the molecular time-dependent dipole matrix $\vec{\mu}_{\rm2N}(t)$ we apply the saddle point method in the momentum variable $\textbf{p}$. In fact, the phases of the local contributions in Eq.~(\ref{Eq:H2N_LL}), function on the relative positions $\textbf{R}_{1/2}$ of the atoms, cancel each other defining a saddle point momentum ${\bf p}_s$ equivalent to the one presented in the atomic case (see Sec.~II.A). On the other hand, the cross process presents more complex phases, that directly depend on the position variables. For instance, in the process $\vec{\mu}_{21}\:(\:\vec{\mu}_{12})$  the saddle-point momentum can be found to be: $\textbf{p}_{s+} = -\frac{1}{\tau} \left[\textbf{R} + \int_{t^\prime}^{t}  { \textbf{A}(\tilde{t}) \textbf{d}\tilde{t}} \right] \left(\textbf{p}_{s-} = -\frac{1}{\tau} \left[-\textbf{R} + \int_{t^\prime}^{t}{\textbf{A}(\tilde{t}) \textbf{d}\tilde{t}}\right] \right)$.  In all our cases we are going to work with shorter internuclear distances, where the following condition is fulfilled: $R<\mathcal{E}_0/\omega^2$, with $\mathcal{E}_0$ and $\omega_0$ being the laser electric field peak amplitude and  carrier frequency, respectively. As a consequence, it is not needed to consider this saddle-point momentum definition (for more details about the validity of this approximation see \cite{LeinPRA2006}). Thus, we proceed by applying the standard saddle point momentum to all the local and cross contributions. The total time-dependent dipole moment for our diatomic molecule then reads as: 
\begin{eqnarray}
\vec{\mu}_{\rm2N}(t)&= &i \sum_{j,j'}\int_0^t{ \textit{d} \textit{t}^{\prime}\: \left( \frac{\pi}{\varepsilon +\frac{\textit{i}(t-t')}{2}}  \right)^\frac{3}{2}    \textbf{E}(t^{\prime})}\cdot \textbf{d}_{j} \left[{\bf p}_s+\textbf{A}(t^{\prime})\right] \nonumber \\
&&  \times \: e^{-i \left\{ S({\bf p}_s,t,t') + \textbf{R}_j \cdot \left[ \textbf{A}(t) -\textbf{A}(t')\right]\right\}}\:\textbf{d}^*_{j'}[ {\bf p}_s +\textbf{A}(t)] .
\label{Eq:m_HH2N}
 \end{eqnarray} 
 
Finally the total HHG spectrum can be calculated using Eq.~(\ref{Eq:HHG}), similarly to the atomic case, but using the time-dependent dipole matrix obtained in Eq.~(\ref{Eq:m_HH2N}). As it was discussed, four terms are needed to compute each molecular harmonic spectrum. Each term represents a different process and this is equivalent to the split made in the time-dependent dipole matrix element. We label each contribution depending on the position of the atoms, e.g.~from the $\textit{Left}-\textit{Left}$ term we obtain the $I_{\rm2N, 11}(\omega)$ spectrum. Similarly we write the other three terms as $I_{\rm2N, 22}(\omega)$, $I_{\rm2N, 12}(\omega)$ and $I_{\rm2N, 21}(\omega)$, respectively.

It is convenient to identify two main contributions in the total harmonic spectrum, Eq.~(\ref{Eq:m_HH2N}), namely, (i) one generated for the Local processes and (ii) other developed by the Cross processes. In this way we can write the total harmonic spectrum as:
\begin{eqnarray}
I_{\rm 2N}(\omega)=I_{\rm 2N-Local}(\omega) +I_{\rm 2N-Cross}(\omega),
\label{Eq:I2N_1}
\end{eqnarray} 
where $I_{\rm 2N-Local}(\omega)=I_{\rm 2N, 11}(\omega) +I_{\rm 2N, 22}(\omega)$ and $I_{\rm2N-Cross}(\omega)=I_{\rm2N, 12}(\omega) +I_{\rm2N, 21}(\omega)$ denote the local and cross terms, respectively. 

\subsection{Calculation of the time-dependent dipole moment for three-center molecular systems: $\vec{\mu}_{\rm3N}(t)$}

The computation of the HHG spectrum generated by a three-center molecule using the definition in Eq.~(\ref{Eq:HHG}) involves the search of the exact bound states describing the whole system. In order to do so we use a method similar to the one presented in Ref.~\cite{PRANoslen2015, Molecule2016}. In short, we consider a three-center molecule as a set of three atoms placed at $\textbf{R}_{1} = -\frac{\textbf{R}}{2}$, $\textbf{R}_2=0$ and $\textbf{R}_3= \frac{\textbf{R}}{2}$, respectively, where $\textbf{R}$ is the so-called internuclear distance, defined as the distance between the atoms placed at $\textbf{R}_1$ and $\textbf{R}_3$, when the molecule is linear.
The state describing the time evolution of a three-center molecule can be written as a coherent superposition of the states $|\Psi(t) \rangle$  as $|\Psi(t) \rangle = e^{\textit{i}I_p\textit{t}}\left(a(t) |0\rangle + \sum_{j=1}^{3} \int{\textit{d}^3 \textbf{v} \:  \textit{b}_j ( \textbf{v},t) |\textbf{v}\rangle}\right)$, where the subscript $j=1,2,3$, refers to the contributions of the spatially localized nuclei at $\textbf{R}_1$,  $\textbf{R}_2$ and $\textbf{R}_3$, respectively. By employing the Schr\"odinger equation on that state and our basic SFA approach, the molecular time-dependent dipole moment $\vec{\mu}_{3N}(t)$ reads:
\begin{eqnarray}
\vec{\mu}_{\rm3N}(t)&=&\int{\textit{d}^3 \textbf{v}\: \textbf{d}^*_{\rm3N}( \textbf{v}) \textit{b}( \textbf{v},t)} + \textit{c.c}.
\label{Eq:miu3N}
\end{eqnarray}
$\vec{\mu}_{\rm3N}(t)$ is defined as a superposition of the bound-continuum dipole matrix of each atom on the molecule, i.e.~$\textbf{d}_{3N}( \textbf{v})= \sum_{j=1}^{3} \textbf{d}_j( \textbf{v})$. The exact dependency of the bound-continuum matrix element is presented  in the Appendix A (see Eq. (\ref{Eq:d3m1}) for more details).

Using the exact definition of the bound-continuum matrix element, the total continuum states transition amplitude, $\textit{b}( \textbf{v},t)=\sum_{j=1}^{3} b_{0,\mathit{j}}( \textbf{p},t)$, reads as: 
\begin{eqnarray}
b( \textbf{p},t) &=& i  \sum_{j=1}^{3}\int_0^t{\textit{d} \textit{t}^{\prime}\:\textbf{E}(t^{\prime})}\cdot \textbf{d}_\mathit{j}\left[\textbf{p}+\textbf{A}(t^{\prime})\right] e^{-i \left\{S({\bf p},t,t') + \textbf{R}_j \cdot \left[ \textbf{A}(t) -\textbf{A}(t')\right] \right\}}.
\label{Eq:b_0R}
\end{eqnarray}
The explicit expression for the molecular time-dependent dipole matrix element $\vec{\mu}_{3N}(t)$ is obtained inserting Eq.~(\ref{Eq:b_0R}) in Eq.~(\ref{Eq:miu3N}). As in the case of diatomics, it is also possible here to disentangle each of the recombination processes contributing to the total harmonic spectrum. In order to do so we write $\vec{\mu}_{\rm3N}(t)$ as a sum of nine terms as:
\begin{eqnarray}
\vec{\mu}_{\rm3N}(t)&=& \sum_{j=1}^{3}\sum_{j'=1}^{3} \vec{\mu}_{jj'}(t).
\label{Eq:mu3N_total}
\end{eqnarray}

The above equation contains information about all the possible recombination scenarios present in our three-center 
molecule. In order to make clearer the interpretation let us write the individual time-dependent dipole matrix element $\vec{\mu}_{jj'}(t)$ explicitly as
\begin{eqnarray}
\vec{\mu}_{jj'}(t)&=&i \int_0^t{\textit{d} }\textit{t}^{\prime} \left( \frac{\pi}{\varepsilon +\frac{\textit{i}(t-t')}{2}}  \right)^\frac{3}{2} \textbf{E}(t^{\prime}) \cdot \textbf{d}_{j} \left[\textbf{p}_s+\textbf{A}(t^{\prime})\right] \nonumber\\
&&\times \, e^{-i\left\{ S({\bf p}_s,t,t') + \textbf{R}_{j} \cdot \left[ \textbf{A}(t) -\textbf{A}(t')\right] \right\}}\,  \textbf{d}^*_{j'}[ \textbf{p}_s+\textbf{A}(t)],
\label{Eq:m_HH3N}
\end{eqnarray}
where the subscripts ``$j$" and ``$j'$" refer to the position $\textbf{R}_1$, $\textbf{R}_2$ and $\textbf{R}_3$ of each of the atoms in the three-center molecule. In Eq.~(\ref{Eq:m_HH3N}) the first subscript $j$ represents the atom from where the electron is detached and can be $j=1,2,3$. In addition, the second one, $j'$, labels the atom where the recombination process occurs, and can also take the values 1,2 or 3.

Note that, as in the case of atoms and diatomic molecules, in Eq.~(\ref{Eq:m_HH3N}) we have applied the saddle point method in the momentum $\textbf{p}$ integral. In Eq.~(\ref{Eq:m_HH3N}) we have followed the same criteria as in the diatomic system (see Sec.~II.B) and in this way the saddle point momentum $\mathbf{p}_s$ is the conventional one.  

As in the case of diatomics, the nine terms of Eq.~(\ref{Eq:mu3N_total}) represent the Local and Cross processes. These different terms should be understood as follows:

\begin{enumerate}[(i)]
 \item The first term, $\vec{\mu}_{11}$, describes the process of an electron ionized from the atom placed at $\textbf{R}_1$ at time $t'$ with probability: $\textbf{E}(t^{\prime}) \cdot \textbf{d}_1 \left[\textbf{p}_s+\textbf{A}(t^{\prime})\right]$. This electron, during its excursion in the continuum, accumulates a phase which depends on the position from where it was detached, in this case $\textbf{R}_1$. Finally, because the change in the sign of the laser electric field, the electron returns to the parent ion, with a recombination probability given by $\textbf{d}^*_1[ \textbf{p}_s+\textbf{A}(t)]$. As a result of this recombination stage the excess of energy acquired from the laser electric field is converted into a high energy photon. As an example the time dependent dipole equation describing this process, where $j=1$ and $j'=1$, reads as:
 \begin{eqnarray}
\vec{\mu}_{11}(t)&=& i \int_0^t{\textit{d} \textit{t}^{\prime} \left( \frac{\pi}{\varepsilon +\frac{\textit{i}(t-t')}{2}}  \right)^\frac{3}{2}} \textbf{E}(t^{\prime}) \cdot \textbf{d}_1 \left[\textbf{p}_s+\textbf{A}(t^{\prime})\right]\nonumber\\
&& \times \: e^{-i \left\{ S({\bf p}_s,t,t') + \textbf{R}_1 \cdot \left[ \textbf{A}(t) -\textbf{A}(t')\right]\right\}} \:  \textbf{d}^*_1[ \textbf{p}_s+\textbf{A}(t)].
 \end{eqnarray} 
 \item The second and third terms, $\vec{\mu}_{22}$ and $\vec{\mu}_{33}$, describe the same process, but for the atoms located at $\textbf{R}_2$ and $\textbf{R}_3$, respectively. These three process are spatially localized: the electron starts and ends at the same point, the same ion core. We then refer to them as ``Local processes''. 
\item From the fourth to the seventh terms we have the cross processes with the closer neighbor in one and other direction. In this case notice that in our reference frame the second atom is placed at $\textbf{R}_2=0$. These processes are understood as in the diatomic cases.    
\item The last two terms are also cross processes. For instance, the eighth term can be understood as follows: one electron tunnels ionize from the atom located at $\textbf{R}_1$ with probability: $\textbf{E}(t^{\prime}) \cdot \textbf{d}_1 \left[\textbf{p}_s+\textbf{A}(t^{\prime})\right]$. This electron starts to move with the electric field and acquires a phase $e^{-i\left\{S({\bf p}_s,t,t') + \textbf{R}_1 \cdot \left[\textbf{A}(t) -\textbf{A}(t')\right]\right\}}$. It then recombines at time $t$ with the farthest ion-core at $\textbf{R}_3$ with an amplitude $ \textbf{d}^*_3[ \textbf{p}_s+\textbf{A}(t)]$. The last term is understood in a similar way, but inverting the tunnel ionization and recombination positions. 
\end{enumerate} 

For our three-center molecular system is also possible to group the processes as Local and Cross. As in the diatomic case the sum of all these terms represents the total time-dependent dipole element, $\vec{\mu}_{3N}(t)=\vec{\mu}_{\mathrm{3N-Local}}(t)+\vec{\mu}_{\mathrm{3N-Cross}}(t)$. In the same way we can split the contributions depending on the excursion of the electron in the continuum before recombination. The shorter excursions are represented by the local processes where only one atom is involved. For the cross processes we have two possibilities: the recombination with (i) the closest neighbor or (ii) with the farthest one. Those contributions are denoted by:
\begin{eqnarray}
\vec{\mu}_{\mathrm{3N-Local}}(t)=\vec{\mu}_{11}(t) + \vec{\mu}_{22}(t) + \vec{\mu}_{33}(t),
\label{Eq:mu3N_local}
\end{eqnarray} 
and
\begin{eqnarray}
\vec{\mu}_{\mathrm{3N-Cross}}(t)=\vec{\mu}_{\mathrm{3N-Cross_1}}(t) + \vec{\mu}_{\mathrm{3N-Cross_2}}(t),
\label{Eq:mu3N_NLocal}
\end{eqnarray} 
where
\begin{eqnarray}
\vec{\mu}_{\mathrm{3N-Cross_1}}(t)=\vec{\mu}_{12}(t) + \vec{\mu}_{21}(t) + \vec{\mu}_{23}(t) + \vec{\mu}_{32}(t),
\label{Eq:mu3N_Nlocal1}
\end{eqnarray} 
and 
\begin{eqnarray}
\vec{\mu}_{\mathrm{3N-Cross_2}}(t)=\vec{\mu}_{13}(t) + \vec{\mu}_{31}(t).
\label{Eq:mu3N_Nlocal2}
\end{eqnarray} 
 In order to describe the direct processes we set $j=j'$. For instance, the Local process for the $\textit{Right}$ atom located at $\textbf{R}_1$ is described by the time-dependent dipole matrix element $\vec{\mu}_{11}(t)$. On the other hand, the Cross processes are those where $j\not=j'$.  

Finally, in order to compute the total time-dependent dipole element $\vec{\mu}_{\rm3N}(t)$ of our three-center molecule we need to evaluate each of the contributions defined by Eq. (\ref{Eq:m_HH3N}). The HHG spectrum can then be obtained by Fourier transforming $\vec{\mu}_{\rm3N}(t)$ (see Eq. (\ref{Eq:HHG})).  The separation of the time-dependent dipole matrix element allows us to compute the harmonic spectrum from each process separately as we will see in the next Section.

\section{Results and Discussion} \label{cap:Background}

In this section we calculate the harmonic spectra for different systems using the equations previously presented. In addition, we compare the harmonic spectra emitted from hydrogen and argon atoms with the exact numerical solution of the 3D-TDSE. A scan over different laser wavelength and peak intensities is performed in order to verify and validate the model. In a second stage, we apply our molecular approach to two prototypical diatomic systems: H$_2^+$ and H$_2$. We display the different contributions coming from the local and cross recombination processes which helps to distinguish which contributions are interfering constructively and destructively to the total high harmonic spectra.  Finally, we present results for more complex molecules: CO$_2$ and H$_2$O. For these cases, besides to disentangle the different contributions to the HHG spectra, we analyze the influence of the angular orientation. 

The numerical integration of Eqs.~(\ref{Eq:m_HH1N}),~(\ref{Eq:m_HH2N}) and~(\ref{Eq:m_HH3N}) has been performed by employing a rectangular rule with dedicated emphasis on the convergence of the results. The HHG process is driven by an ultrashort laser pulse with an electric field of the form: 
\begin{eqnarray}
\textbf{E}(t) = \mathcal{E}_0\:f(t) \sin(\omega_0 \:t + \phi_0)\,{\bf e}_{z}.
\label{Eq:Efield1}
\end{eqnarray}
The field has a carrier frequency $\omega_0 = \frac{2\pi c}{\lambda_0}$, where $c$ is the speed of light ($c\approx137$~a.u) and $\lambda_0$ the central laser wavelnegth and $\mathcal{E}_0$ is the field peak amplitude, linearly polarized in the $z$-axis. ${f(t)}=\sin^2(\omega_0 t/2 N_c)$ denotes the pulse envelope, with $N_c$ the total number of cycles, and the parameter $\phi_0$ is the carrier envelope phase (CEP). Under the dipole approximation, the influence of the magnetic field  component of the laser field is neglected.

\subsection{Atomic systems. Comparison between SFA and 3D-TDSE models}

\begin{figure}[htb]
            \subfigure[]{ \includegraphics [width=0.550\textwidth] {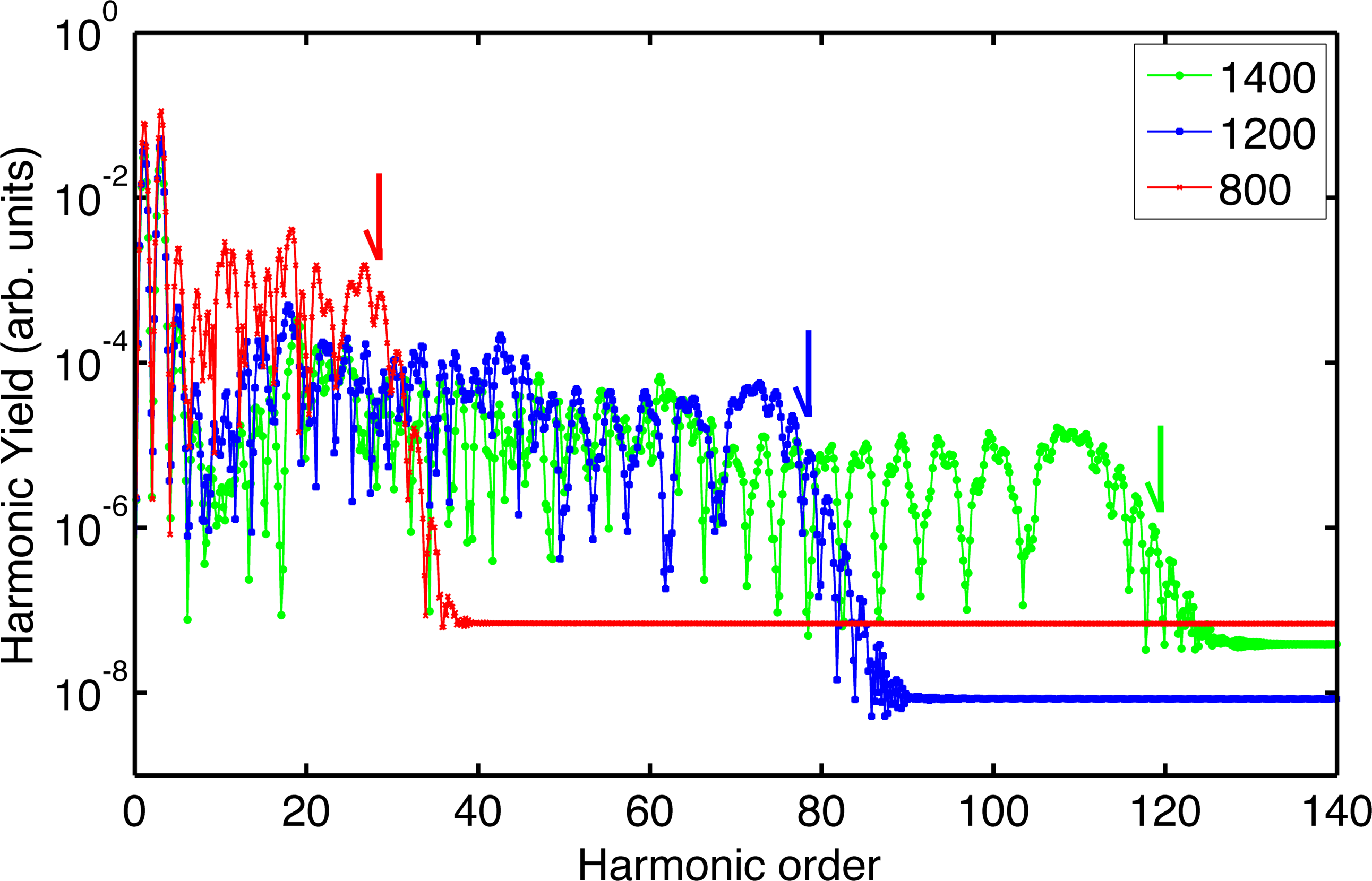}
            \label{fig:subfiga}}
            \subfigure[]{ \includegraphics [width=0.550\textwidth] {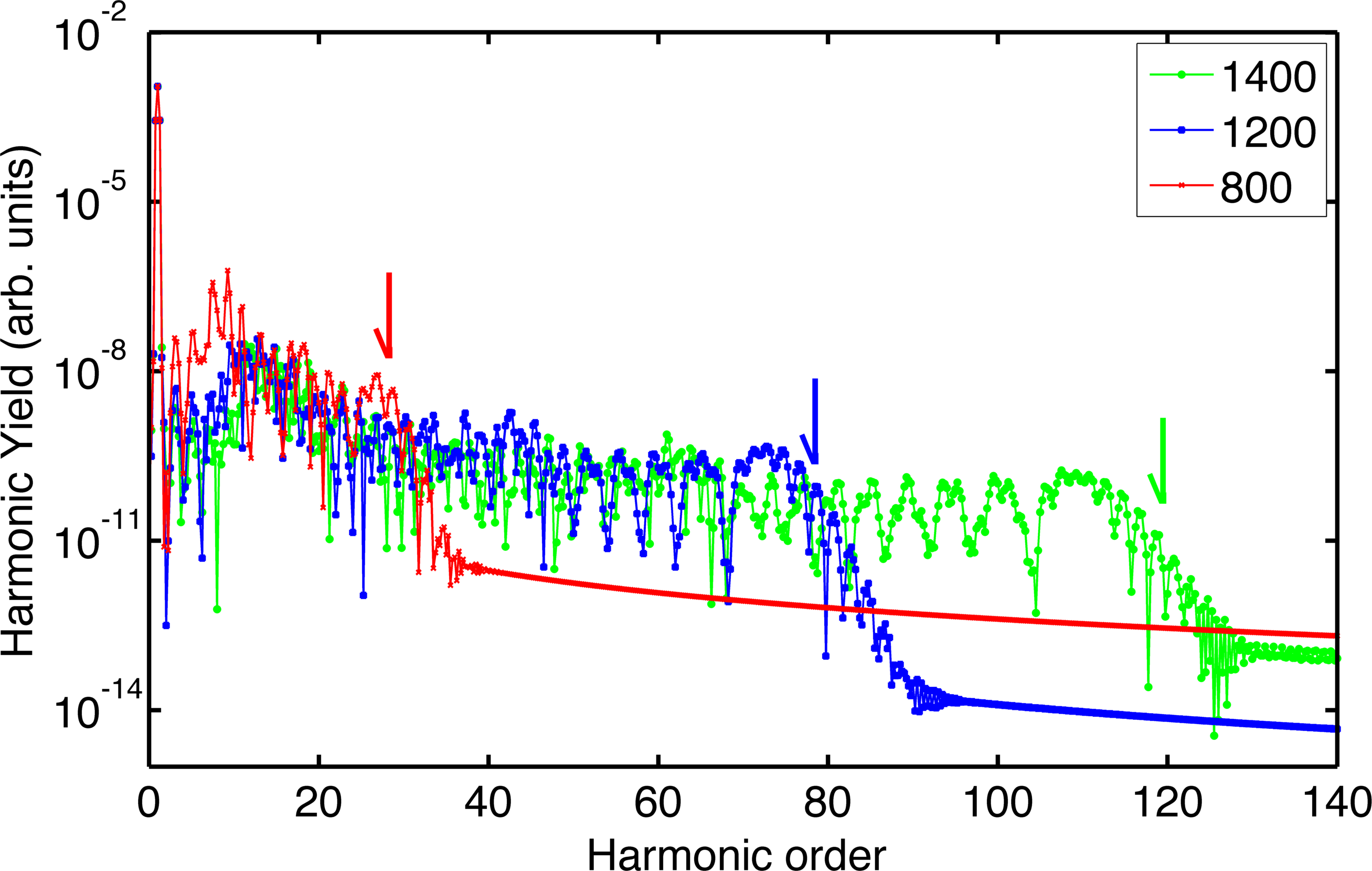}
            \label{fig:subfiga}}
		    \caption{(color online) HHG spectra $I_{\rm1N}(\omega)$ (in logarithmic scale) of hydrogen driven by a strong few-cycle pulse at different wavelengths. $\lambda_1= 800$ nm (red asterisk line), $\lambda_2= 1200$ nm (blue square line) and $\lambda_3= 1400$ nm (green circle line). (a) quasiclassical SFA model; (b) 3D-TDSE. The arrows in all the panels indicate the position of the classical HHG cutoff (see the text for details).}
		  	\label{Fig:H+lamda}
		\end{figure}

To calculate the harmonic spectra of an hydrogen atom we perform the Fourier transform of the time-dependent dipole moment presented in Eq.~(\ref{Eq:m_HH1N}).  We set $\Gamma=1$ and $\gamma=38$~a.u in our non-local potential in order to match the ionization potential $I_p=0.5$~a.u. of the hydrogen atom. We consider a pulse with $N_c=4$ total cycles and $\phi_{0}=0$~rad. A total of $131072$ points in the time window $t \in[0,t_{\rm F}]$, where $t_{\rm F}= N_cT_0$ and $T_0=2\pi/\omega_0$, are used during the numerical integration. The simulation of the harmonic spectra for H at different laser wavelengths and using our quasiclassical SFA model is shown in Fig.~\ref{Fig:H+lamda}(a). In addition, in Fig.~\ref{Fig:H+lamda}(b) we show the HHG spectra obtained by using the  numerical solution of the 3D-TDSE. 

In order to compute the HHG spectra displayed in Fig.~\ref{Fig:H+lamda} we consider the laser pulse described by Eq.~(\ref{Eq:Efield1}), with a laser peak intensity of $I_0=1.58 \times10^{14}$~W$\,\cdot$\,cm$^{-2}$ and different laser wavelengths (see the panels label for details). In order to calculate the HHG  spectra of Fig.~\ref{Fig:H+lamda}(b) we numerically solve the 3D-TDSE in the length gauge. Thus, by Fourier transform of the dipole acceleration, calculated from the time propagated electronic wave function, the HHG spectra is obtained. We have used our code, which is
based on an expansion of spherical harmonics, $Y_{lm}$ considering only the $m=0$ terms due to the cylindrical
symmetry of the problem. The numerical technique is based on a Crank-Nicolson method implemented on a
splitting of the time-evolution operator that preserves the
norm of the wave function. 

Both panels of Fig.~\ref{Fig:H+lamda} reveal the typical HHG behavior, namely (i) a rapidly decreasing of the harmonic yield for the lower harmonic orders ($<10^{\mathrm{th}}$); (ii) a plateau with almost constant yield and (iii) an abrupt end at the so-called HHG cutoff. 
The cutoff energy is one of the most important features of any HHG spectrum. It can be defined as the maximum photon energy that can be released at recollision. Classically it is possible to prove that~\cite{PBCorkum1993, Lewenstein1994}:
\begin{eqnarray}
\omega_{\textrm{cutoff}}=I_p +3.17\:U_p.
\label{Eq:HHGCutoff}
\end{eqnarray}
where $\omega_{\textrm{cutoff}}$ is the maximum photon energy and $U_p=I_0/4\omega_0^2$ is the ponderomotive energy.
As can we see from Fig.~\ref{Fig:H+lamda} both the SFA and 3D-TDSE calculations show the expected classical cutoff defined by Eq.~(\ref{Eq:HHGCutoff}), noted with a dashed line of each color at $\omega_{\textrm{cutoff}-800}=1.59$~a.u. (43.26 eV),~$\omega_{\textrm{cutoff}-1200}=2.97$~a.u. (80.8 eV) and $\omega_{\textrm{cutoff}-1400}=3.87$~a.u. (105.3 eV), respectively. 
From Eq.~(\ref{Eq:HHGCutoff}) we should note that $\omega_{\textrm{cutoff}}\propto I\lambda^2$ and this behaviour can also be observed in Fig.~\ref{Fig:H+lamda}. For instance,  the spectra at $\lambda_3=1400$~nm have a cutoff energy about 4 times higher than the one calculated using a wavelength of $\lambda_1=800$~nm. 

\begin{figure}[htb]
            \subfigure[]{ \includegraphics [width=0.55\textwidth] {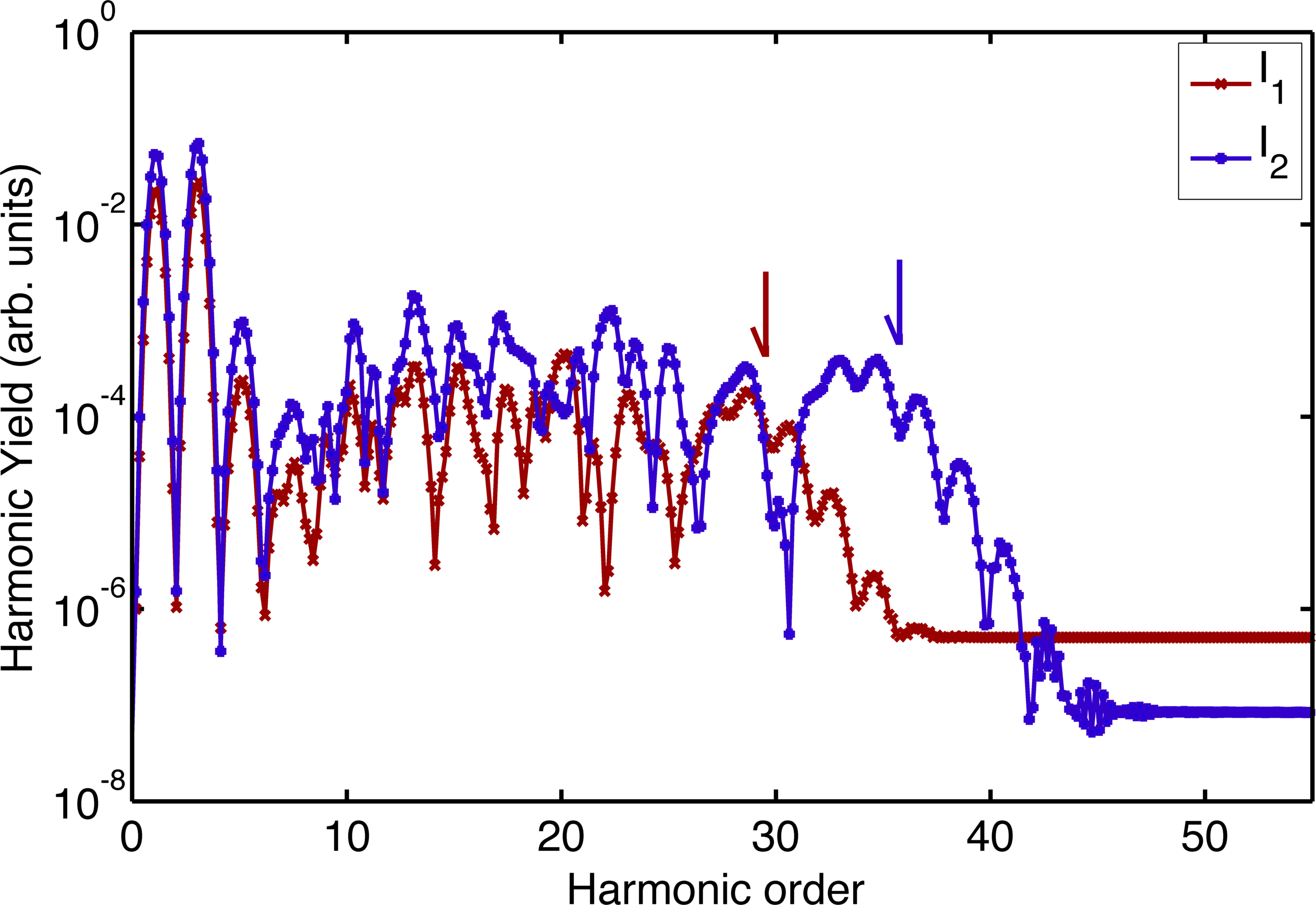}
            \label{fig:subfiga}}
                         \subfigure[]{\includegraphics[width=0.55\textwidth]{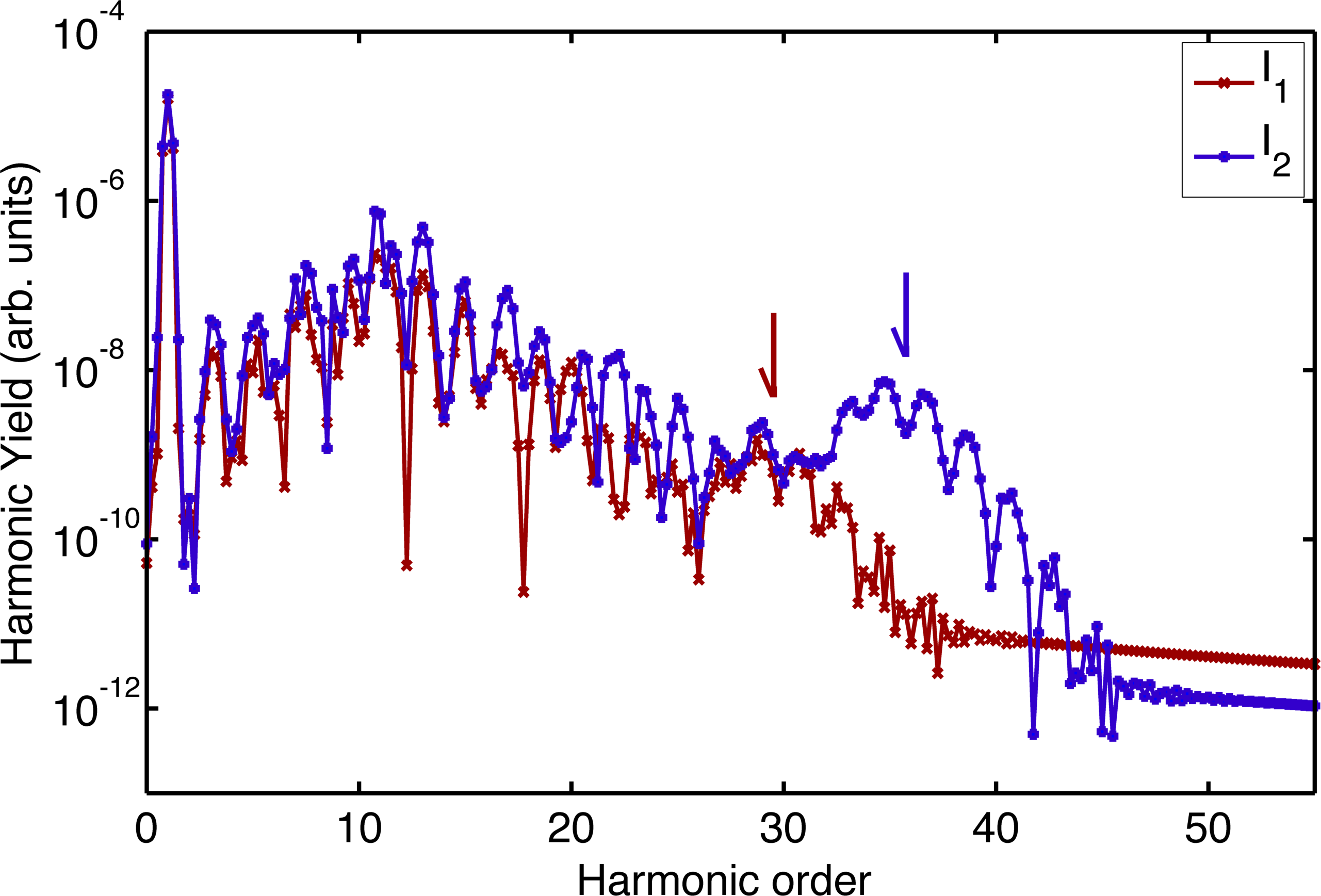}}
		    \caption{(color online) HHG spectra $I_{\mathrm{1N}}(\omega)$ (in logarithmic scale) of Ar driven by a strong few-cycle pulse with $\lambda= 800$~nm, at different laser peak intensities. (a) our quasiclassical SFA at $I_1=1.58 \times10^{14}$~W$\,\cdot$\,cm$^{-2}$  (red square line) and $I_2=2.08\times10^{14}$~W$\,\cdot$\,cm$^{-2}$ (blue cross line), (b)  same as in (a) but solving the 3D-TDSE. Note that in this case the minimum in the efficiency around the 27$^{\mathrm{th}}$ harmonic is the Cooper minimum in Ar. The arrows in all the panels indicate the position of the classical HHG cutoff (see the text for details).}
		  	\label{Fig:H+Int}
		\end{figure}

A natural next step would be to test our model with a more complex atom. In order to do so in Fig.~\ref{Fig:H+Int} we show HHG spectra for an argon atom, calculated both with by (i) our quasiclassical SFA (Fig.~\ref{Fig:H+Int}(a)) and (ii) using the numerical solution of the 3D-TDSE under the SAE approximation (Fig.~\ref{Fig:H+Int}(b)). We employ two different intensities and using a laser pulse with a central frequency of $\omega_0=0.057$ a.u., that corresponds to a wavelength of about 800 nm.  As in the previous case, we confirm that our model is capable to capture not only the dependency of the harmonic spectra with the wavelength, but also with the laser peak intensity. As we can see, and considering that $I_2>I_1$, a clear cutoff extension in the HHG spectra for $I_2$ is observed. A remarkable good agreement between both methods is clearly seen in Fig.~\ref{Fig:H+Int} and for both laser intensities. 

The HHG spectra presented both for a single electron system (H, Fig.~\ref{Fig:H+lamda}) and a complex target (Ar, Fig.~\ref{Fig:H+Int}) reveal the excellent agreement between our quasiclassical SFA model and the numerical solution of the 3D-TDSE.
\subsection{Diatomic molecular systems}

In this section we calculate HHG spectra for two prototypical diatomic molecules: H$_2^+$ and H$_2$.

\subsubsection{H$_2^+$ molecule}

\begin{figure}[htb]
            \subfigure[]{ \includegraphics [width=0.45\textwidth] {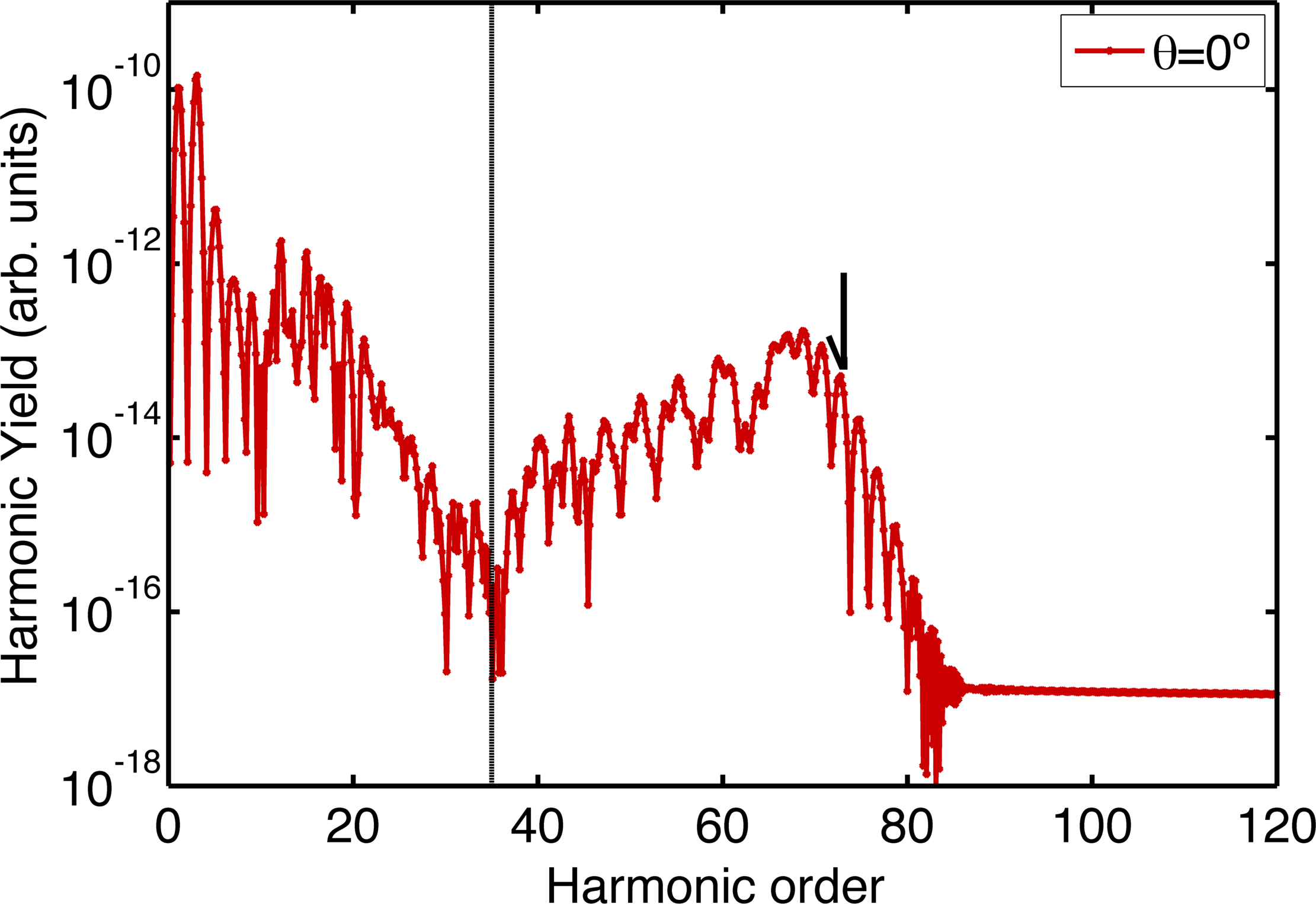}
            \label{fig:subfiga}}
             \subfigure[]{ \includegraphics [width=0.45\textwidth] {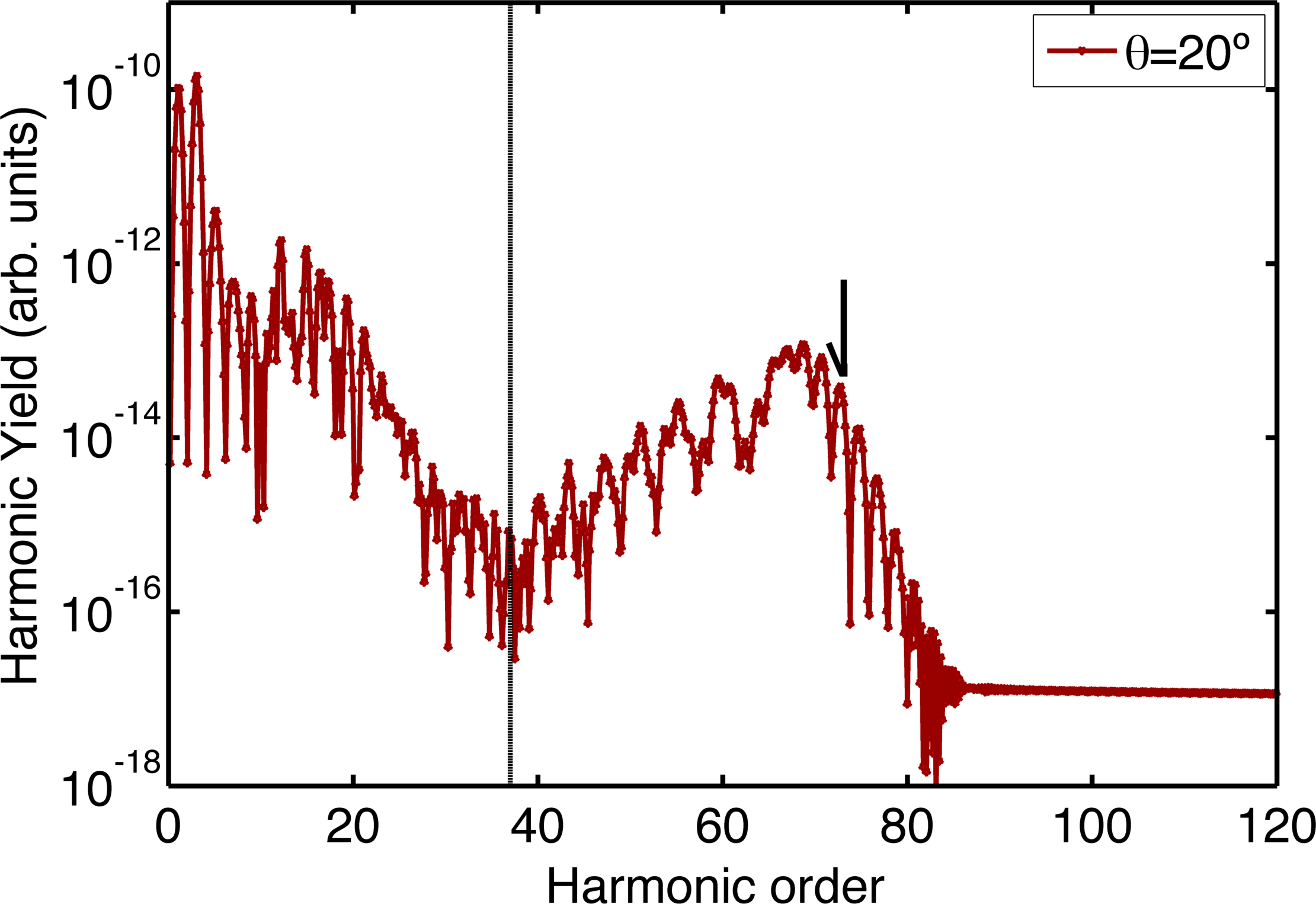}
            \label{fig:subfiga}}
             \subfigure[]{ \includegraphics [width=0.45\textwidth] {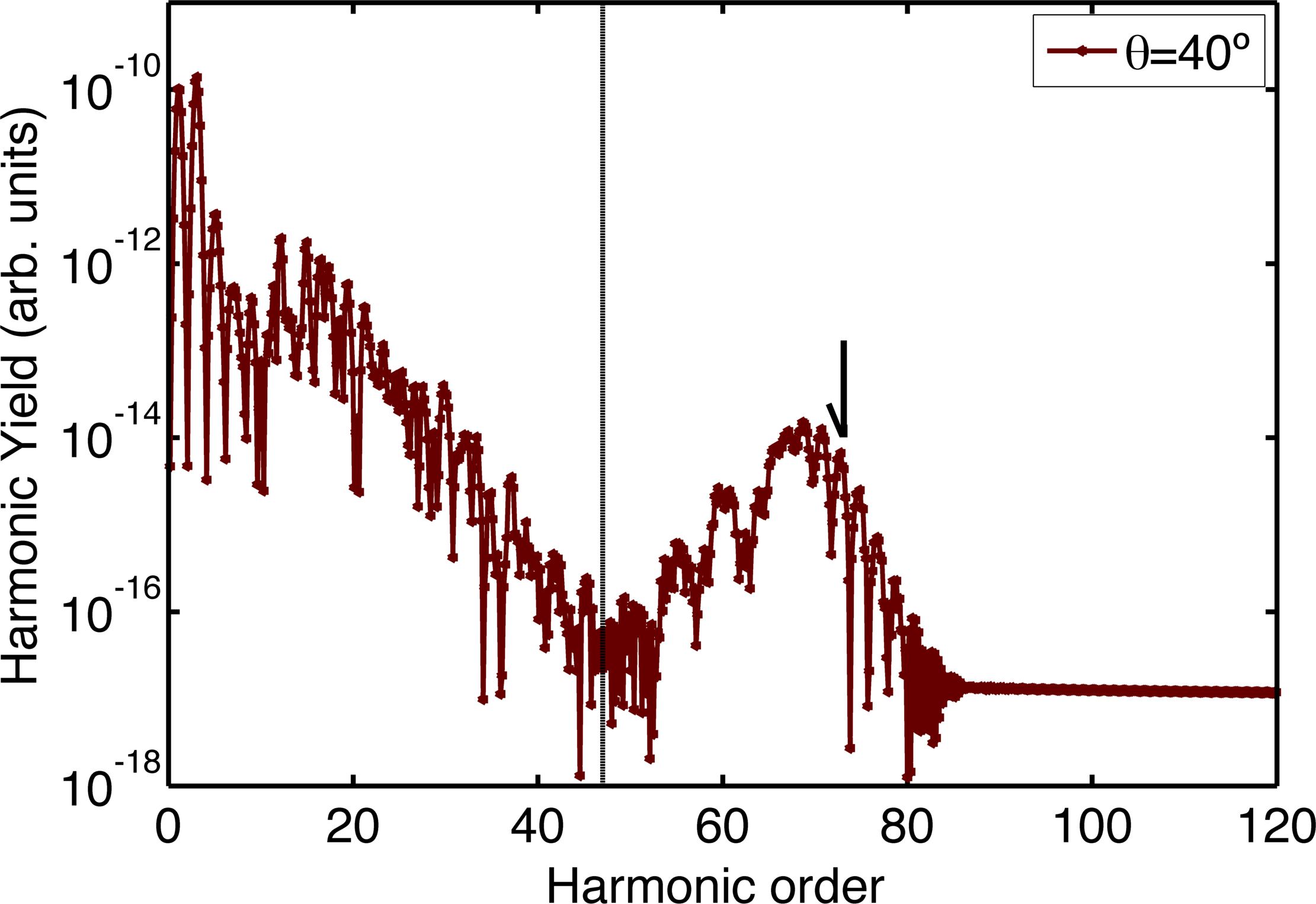}
            \label{fig:subfiga}}
                         \subfigure[]{\includegraphics[width=0.45\textwidth]{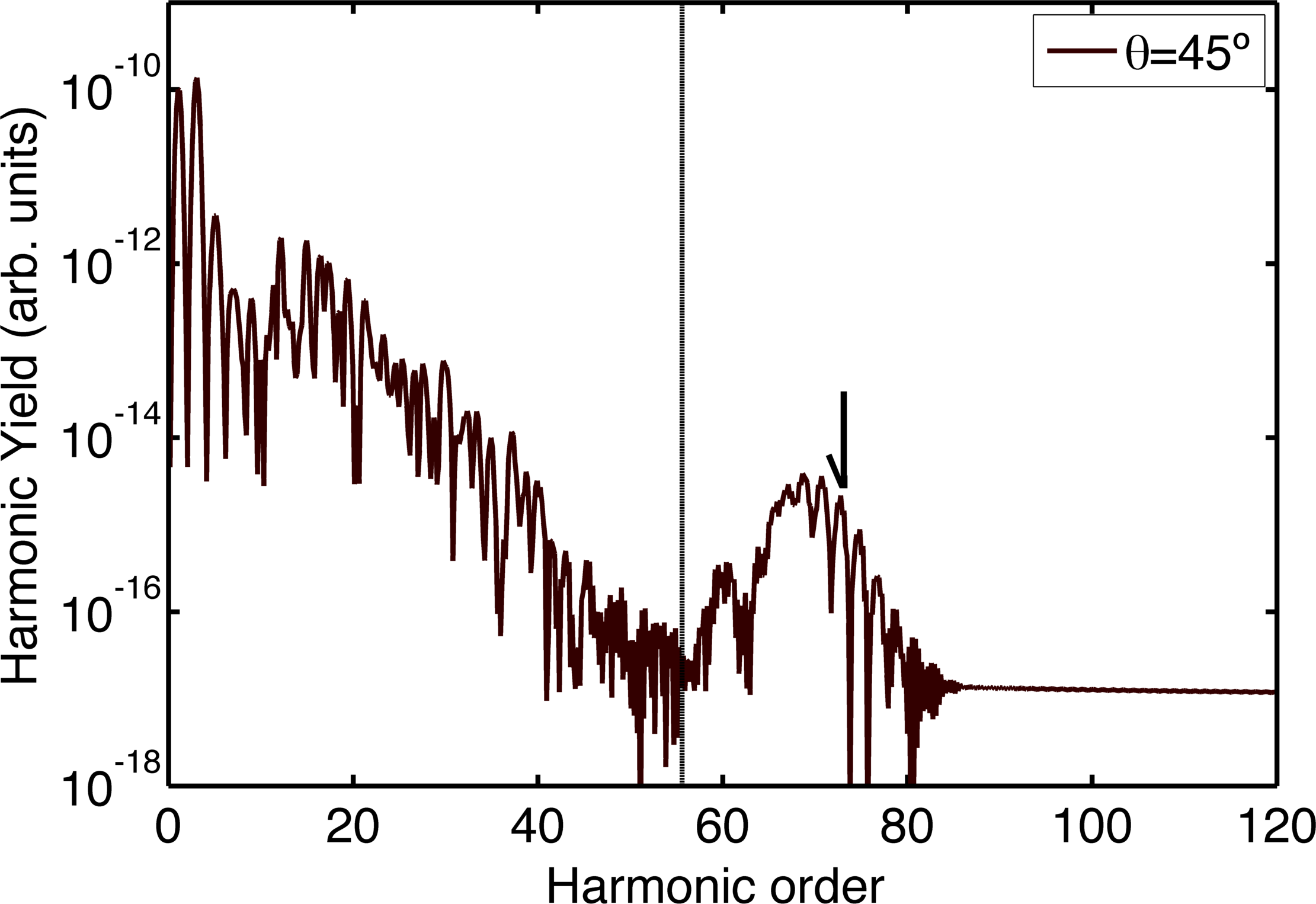}}
		    \caption{(color online) Total harmonic spectra $I_{\mathrm{2N}}(\omega)$ (in logarithmic scale),  Eq.~(\ref{Eq:HHG}), of an H$_2^+$ molecule driven by a strong few-cycle pulse as a function of the harmonic order computed using our quasiclassical SFA. (a) HHG for an H$_2^+$ molecule aligned with the laser pulse polarization axis, i.e.~$\theta=0^{\circ}$; (b) the same as (a) but for $\theta=20^{\circ}$; (c) the same as (a) but for $\theta=40^{\circ}$; (d) the same as (a) but for $\theta=45^{\circ}$. The vertical lines indicate the position of the interference minima of our quasiclassical SFA model and the arrows in all the panels the position of the classical HHG cutoff (see the text for details), respectively.}
		  	\label{Fig:H2pMin}
		\end{figure}

Figure~\ref{Fig:H2pMin} shows the numerically computed HHG spectra for an H$_2^+$ molecule by using the quasiclassical SFA model presented in Sec.~II.B. The H$_2^+$ system is modeled by two identical centers separated by an internuclear distance $R=2.2$~a.u.~(1.16 \AA) and the molecular axis forms a $\theta$ angle with respect to the incident laser electric field polarization, i.e.~$\textbf{R}=(0,0,R\cos\theta)$. The parameters of our non-local potential are set to $\Gamma=1.0$ and $\gamma=0.1$~a.u. in order to reproduce the minimum at the equilibrium internuclear distance, $R_0$ = 2.0 a.u.~(1.06 \AA), in the potential energy surface (PES).  In our short-range potential toy model the total ionization potential extracted from the potential energy surface yields $I_p=0.68$~a.u. (18.50 eV). We compute this electronic ground state energy to fix the asymptotic behavior of the H$_2^+$ potential energy surface (see Ref.~\cite{Molecule2016} for more details). This last value differs from the one obtained with a real Coulomb potential that leads a pure electronic energy of $1.1$~a.u. (30 eV) approximately.

The incident laser field shape is identical to the one used in the atomic case and has a central frequency $\omega_0=0.057$~a.u., corresponding to a wavelength $\lambda=800$~nm and photon energy of $1.55$~eV, respectively. The total number of cycles is $N_c=4$ -this defines a full-width at half-maximum FWHM value of $5.2$~fs- and $\phi_{0}=0$~rad. The time step is set to $\delta t= 0.032$~a.u. and this corresponds to a total of $N_t=20 000$ points for the numerical integration. The time window is $t\in[0,t_{\rm F}]$, where $t_{\rm F}\approx 11$~fs denotes the final time, i.e.~the end of the laser electric field pulse. Finally, the laser peak intensity is set to $I_0=5\times10^{14}$~W$\,\cdot$\,cm$^{-2}$.

In Fig.~\ref{Fig:H2pMin} we display results for a scan of four different molecular orientations, namely Fig.~\ref{Fig:H2pMin}(a) $\theta = 0^{\circ}$ (this value corresponds to the so-called parallel alignment), Fig.~\ref{Fig:H2pMin}(b) $20^{\circ}$, Fig.~\ref{Fig:H2pMin}(c) $40^{\circ}$ and Fig.~\ref{Fig:H2pMin}(d) $45^{\circ}$, respectively. As we can see in all the panels an absolute minimum over the total HHG spectra is clearly visible and the harmonic order where these minima are located increases with the orientation angle. The existence of those minima and their dependency with the alignment angle can be explained by invoking an interference phenomenon as we will see below. In the most simplest picture the minima appears as a consequence of the harmonic emission of two radiant points (see~\cite{PRAMLein2002} for more details).

According to the equation describing the destructive interference of two radiant sources: $R\cos\theta=(2m+1)\lambda_k/2$, where $\lambda_k$ is the de Broglie wavelength of the returning electron and considering the ``fundamental'' instance $m=0$, the minima should be located at the $18^{\mathrm{th}}$, $20^{\mathrm{th}}$, $30^{\mathrm{th}}$ and $36^{\mathrm{th}}$ harmonic order for $\theta=0^{\circ}$, $\theta=20^{\circ}$, $\theta=40^{\circ}$ and $\theta=45^{\circ}$, respectively. The positions of the minima for our SFA calculation are $\approx35^{\mathrm{th}}$, $\approx37^{\mathrm{th}}$, $\approx45^{\mathrm{th}}$ and $\approx54^{\mathrm{th}}$, respectively (see the vertical lines in all the panels of Fig.~\ref{Fig:H2pMin}). We speculate that the shifts in harmonic frequency are related with the kind of potential used in our calculations; the short range potential does not correctly describe the low energy part of the HHG spectra, where the Coulomb potential plays an important role~\cite{marcelo2007}. We note, however, that our SFA calculation for $\theta=40^{\circ}$ is in excellent agreement with the numerical solution of the 2D-TDSE and 3D-TDSE for the H$_2^+$ molecule~\cite{PRAMLein2002, PRAMLein2003}. Lastly, we observe that in all the HHG spectra of Fig.~\ref{Fig:H2pMin} the position of the classical cutoff is in excellent agreement with  Eq.~(\ref{Eq:HHGCutoff}) (see the arrows in all the panels of Fig.~\ref{Fig:H2pMin}). Particularly,  for our H$_2^+$ molecular system and the laser parameters used in our simulations, the cutoff frequency is $\omega_{\textrm{cutoff}}=4.15$~a.u. (112.92 eV), corresponding to the $72^{\mathrm{th}}$ harmonic order. 

\begin{figure}[htb]
            \subfigure[]{ \includegraphics [width=0.45\textwidth] {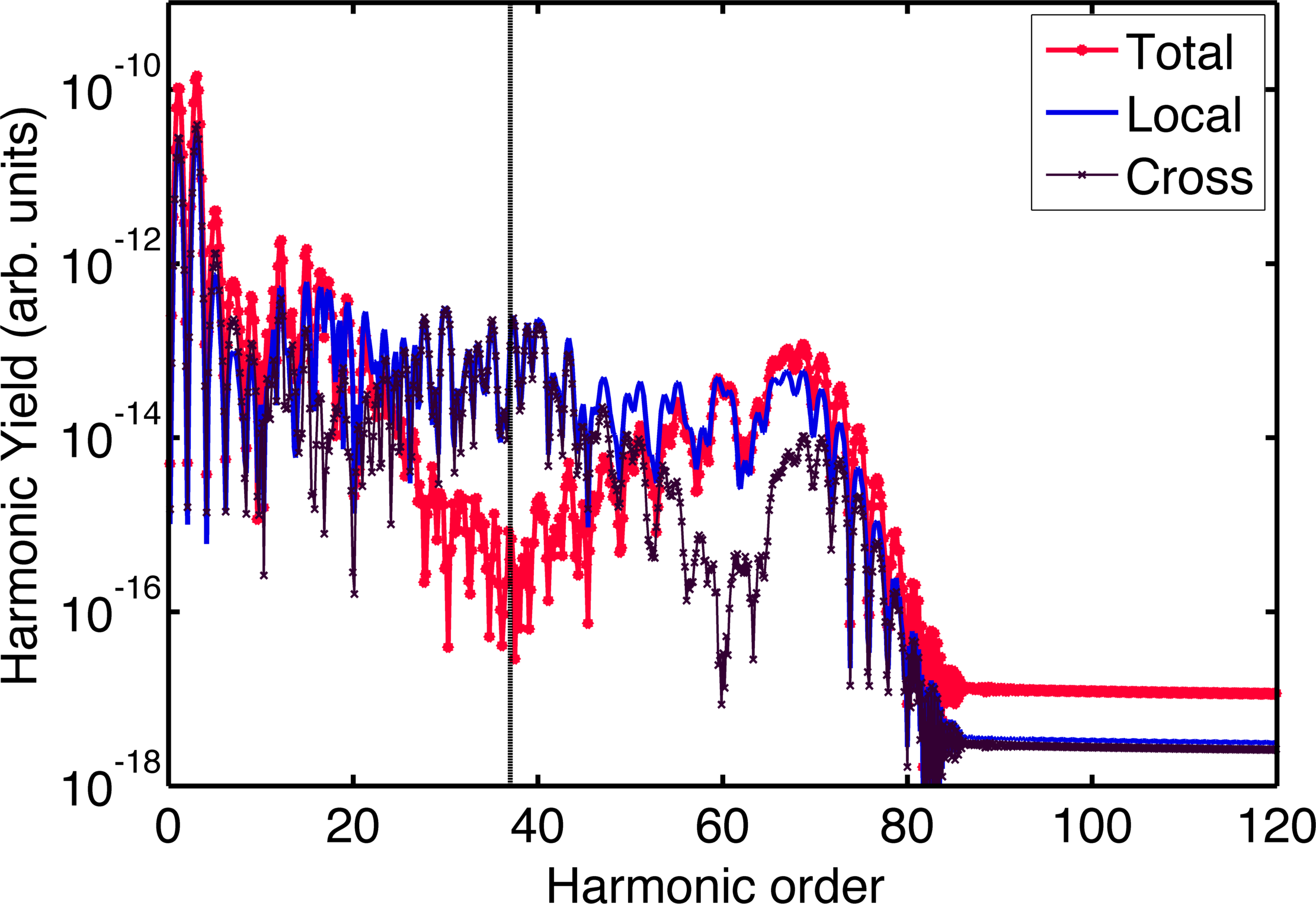}
            \label{fig:subfiga}}
                    \subfigure[]{\includegraphics[width=0.45\textwidth]{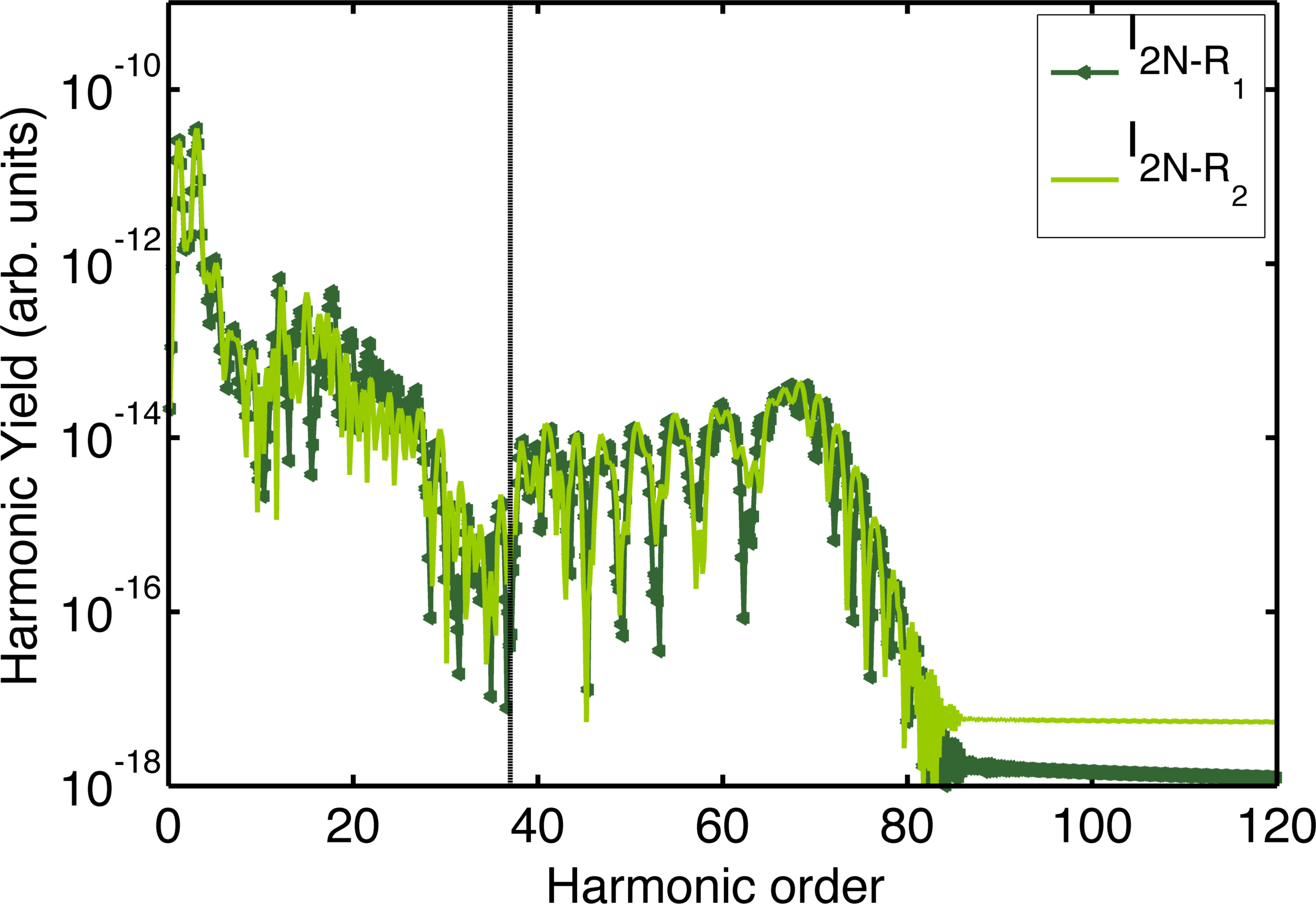}}
		    \caption{(color online) Harmonic spectra $I_{2N}(\omega)$ (in logarithmic scale) of an H$_2^+$ molecule, Eq.~(\ref{Eq:HHG}), as a function of the harmonic order calculated using our quasiclassical SFA and for an orientation angle $\theta= 20^{\circ}$. (a) local, cross and total contributions to the HHG spectrum; (b) contributions depending on the recombination atom. Green circle line: recombination at $\textbf{R}_1$ and light green line: recombination at $\textbf{R}_2$. The vertical lines indicate the position of the interference minima (see the text for details).}
		  	\label{Fig:H2_ContM}
		\end{figure}

Clearly, our quasiclassical molecular SFA model has drawbacks and advantages. The first advantage is from the computational viewpoint; the numerical calculations using the our SFA model are much faster than the numerical solution of the 3D-TDSE. The computation of one single HHG spectrum for a set of fixed parameters takes few seconds. The second, and might be the most important one, is the possibility to disentangle the different components contributing to the final harmonic spectra (see Sec.~II.B).  In order to do so in the Fig.~\ref{Fig:H2_ContM} we plot the different contributions to the HHG spectra for an H$_2^+$ molecule aligned at $\theta=20^{\circ}$ with respect to the laser field polarization. Figure~\ref{Fig:H2_ContM}(a) shows the main contributions of the harmonic spectra, namely the total $I_{2N}(\omega)$ (red circle line) , the local $I_{\rm 2N-Local}(\omega)$ (blue line) and the cross, $I_{\rm2N-Cross}(\omega)$ (dark brown asterisk line) (for details see Sec.~II.B). As we can see from this picture the two-center destructive interference is not present in neither in the local nor in the cross contributions. The latter have a deep minimum but at a different position, about the $60^{\mathrm{th}}$ harmonic order, while the local contribution remains almost constant in yield for all the harmonic frequencies. In order to trace out the origin of the two-center destructive interference present in the total HHG spectra in Fig.~\ref{Fig:H2_ContM}(b) we plot the contributions depending on the recombination atom, calculated as:
\begin{eqnarray}
 I_{\rm 2N-R_1}(\omega)\propto\: \left|\int_{-\infty}^{\infty}{\textit{d}t \: e^{i \omega t}\: [\vec{\mu}_{11}(t) + \vec{\mu}_{21}(t)} ] \right|^2
 \end{eqnarray}
 and 
\begin{eqnarray}
I_{\rm 2N-R_2}(\omega)\propto\: \left|\int_{-\infty}^{\infty}{\textit{d}t \: e^{i \omega t}\: [\vec{\mu}_{22}(t) + \vec{\mu}_{12}(t)}]\right|^2.
\end{eqnarray}

From this figure we can clearly see that there is a deep minimum for both terms and it is located at the same position as for the total HHG spectra. It means that, for the case of the recombination on $\textbf{R}_1$ (dark green circle line), the electron wavepacket ionized at $\textbf{R}_1$ interferes with the one coming from $\textbf{R}_2$ and the other way around. These minima are then generated by the destructive interference of such electron wavepackets.  From the drawbacks side, we have seen that our short range non-local potential is unable to accurately reproduce the interference minima positions for some of the molecular orientation angles. We note, however, that these minima are typically washed out when an average over the molecular orientation is considered, configuration that is commonly used in molecular HHG experiments.

\subsubsection{Time-frequency analysis for H$^+_2$}
  
\begin{figure}[htb]
            \subfigure[]{ \includegraphics [width=0.45\textwidth] {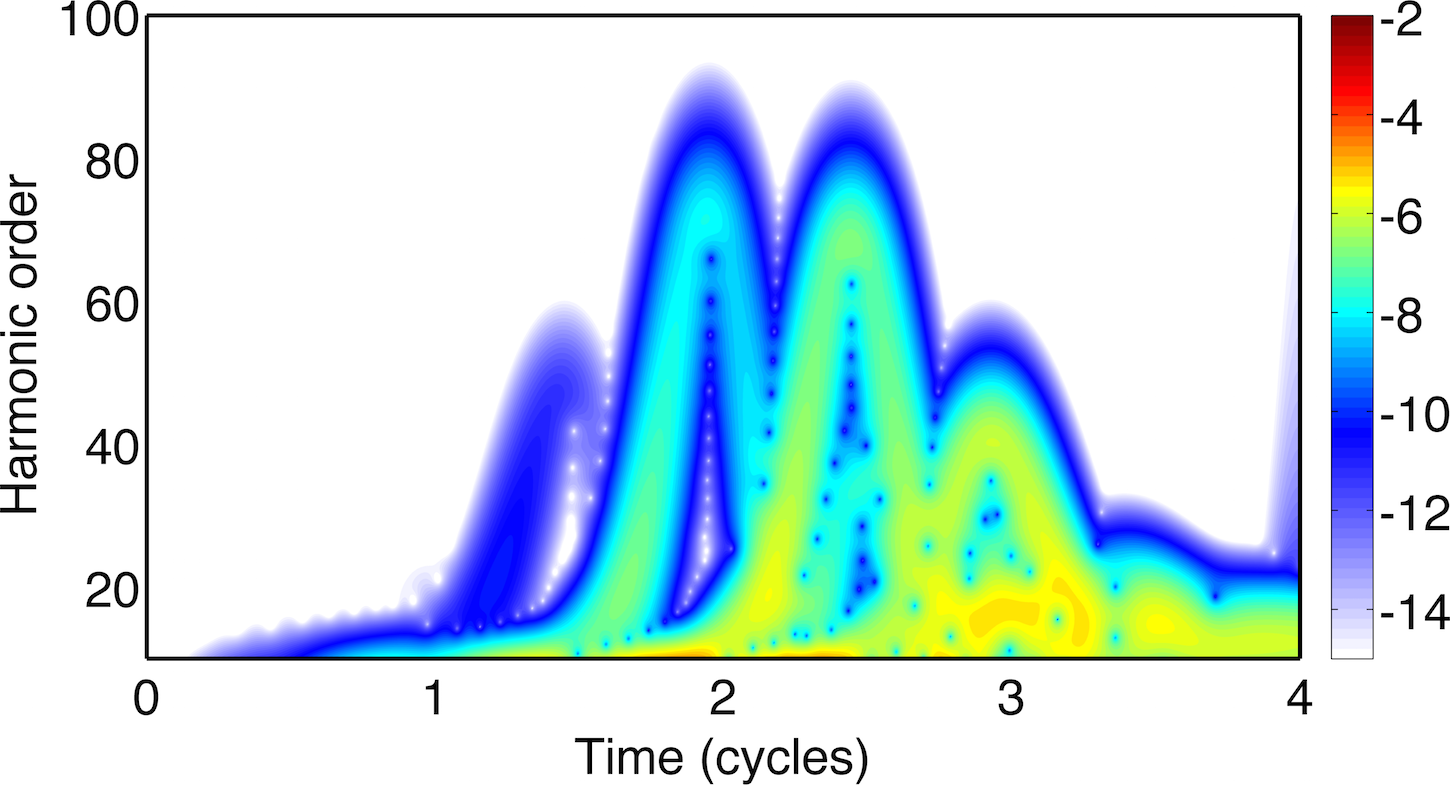}
            \label{fig:subfiga}}
              \subfigure[]{ \includegraphics [width=0.45\textwidth] {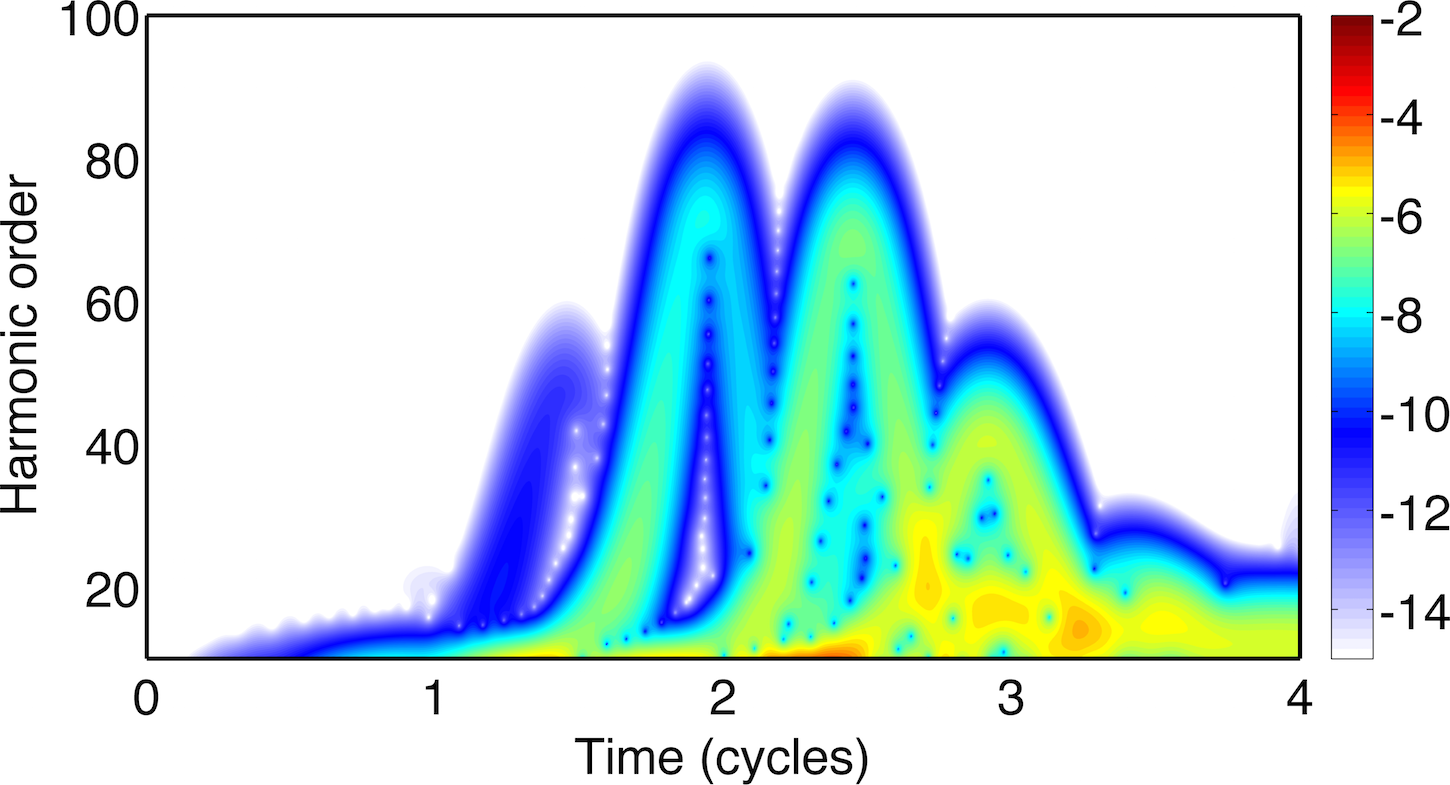}
            \label{fig:subfiga}}
                    \subfigure[]{\includegraphics[width=0.45\textwidth]{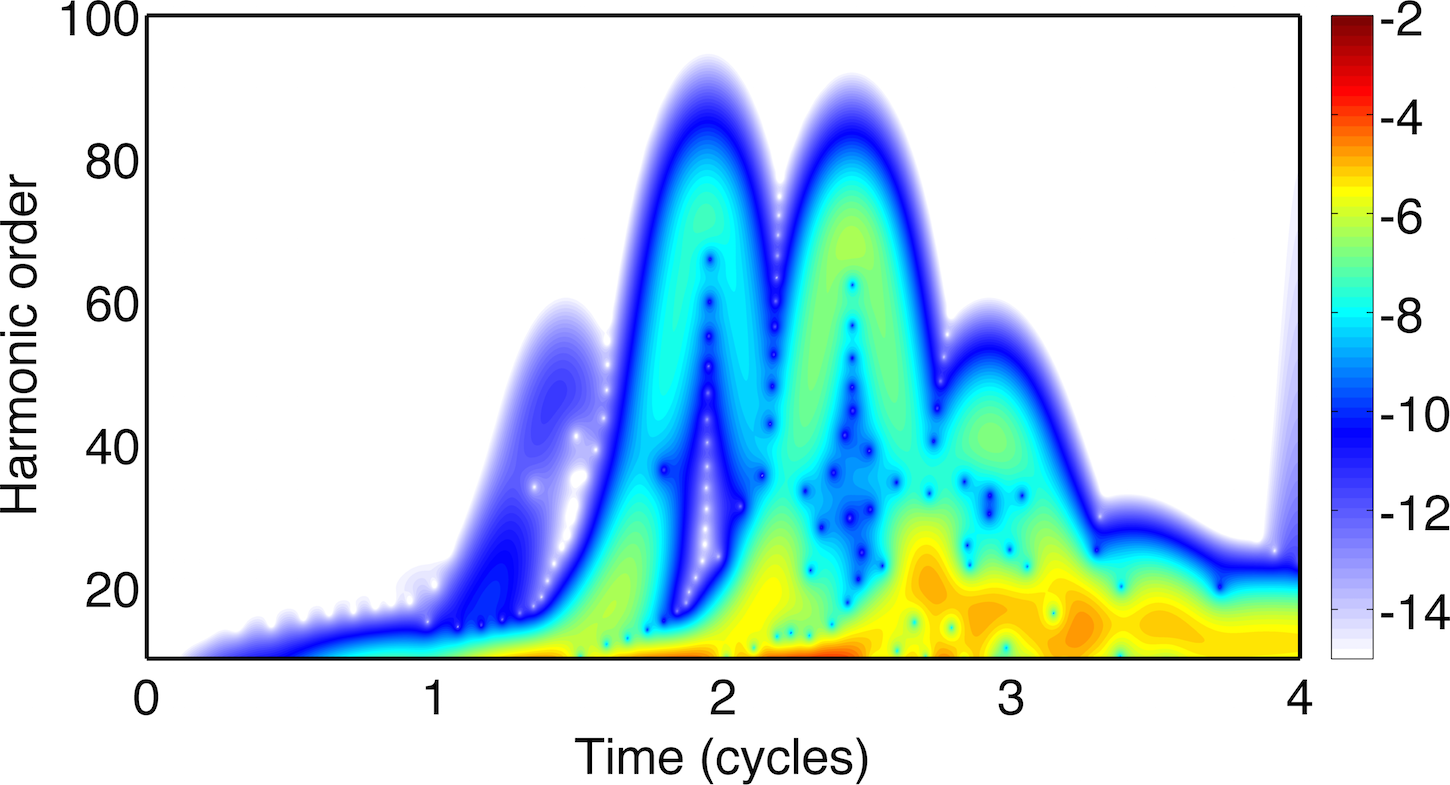}
                     \label{fig:subfiga}}
                      \subfigure[]{\includegraphics[width=0.45\textwidth]{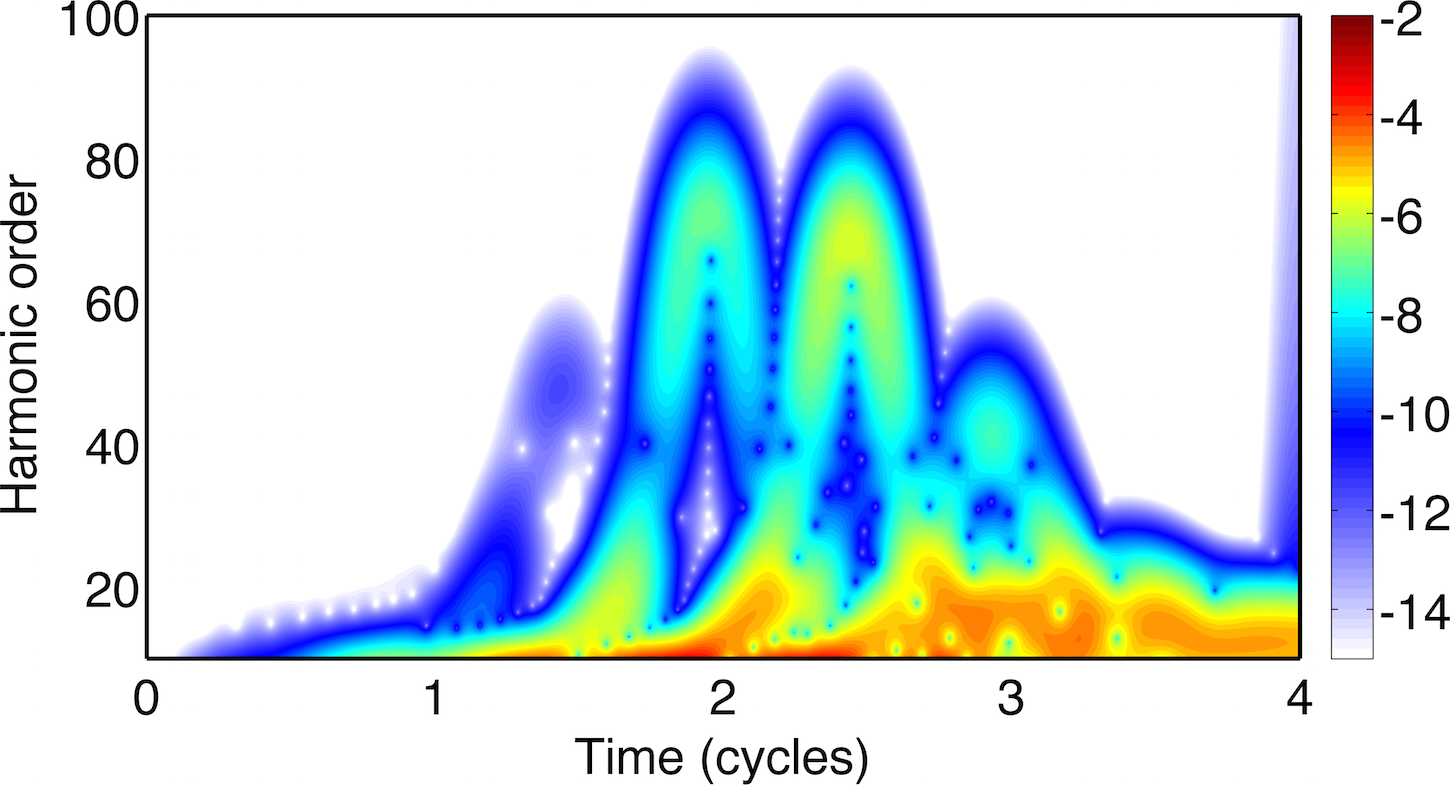}
                    }
                             \caption{(color online) Gabor transformation of the time-dependent dipole moment of an H$_2^+$ diatomic molecule oriented $\theta= 20^{\circ}$ with the laser field. (a) For the case of local analysis for the local process in the core at $\textbf{R}_1$ using the time-dependent dipole moment $\vec{\mu_{11}}(t)$. (b) the same as (a) but for the cross processes using $\vec{\mu_{21}}(t)$. (c) the same as (a) but for the total of recombination processes at $\textbf{R}_1$, here using $\vec{\mu_{11}}(t)+\vec{\mu_{21}}(t)$. (d) Gabor transform for the total time-dependent dipole element, $\vec{\mu}_{2N}(t)$.}
		  	\label{Fig:GaborH2p}
		\end{figure}

 We have seen in Fig.~\ref{Fig:H2_ContM} that the independent processes $\vec{\mu}_{11}(t)/\vec{\mu}_{22}(t)$ and $\vec{\mu}_{21}(t)/\vec{\mu}_{12}(t)$, are the ones interfering and creating the deep minimum in the total HHG spectra. In order to dig deeper about the existence of this distinctive feature a Gabor analysis~\cite{lein2010,gabor} over the different contributions is displayed in Fig.~\ref{Fig:GaborH2p}. The Gabor transformation was performed upon the time-dependent dipole moment calculated using our quasiclassical SFA model. The laser parameters are the same as in Fig.~\ref{Fig:H2_ContM}. 

 This time-frequency analysis allows us to reveal the half-cycle bursts of radiation from which the HHG spectrum is composed and the main trajectories contributing. In Fig.~\ref{Fig:GaborH2p}(a) and Fig.~\ref{Fig:GaborH2p}(b) we show the local and cross process, $\vec{\mu}_{11}(t)$ and $\vec{\mu}_{21}(t)$, respectively. As we can observe from these figures, they both look almost equal and similar to the atomic case. In both cases we have the contribution of the short and long trajectories. For the earlier cycles, the 1$^{\mathrm{st}}$ and 2$^{\mathrm{nd}}$, the short trajectory contributions dominate while for the latest cycles both long and short trajectories have the same weight. The main differences between these two contributions are in the low energy region around the end of the laser pulse, the 3$^{\mathrm{rd}}$ optical cycle, where the contribution of the cross processes $\vec{\mu}_{21}(t)$ is slightly higher than the local ones.

Finally, we plot the mixed, Fig.~\ref{Fig:GaborH2p}(c), and total, Fig.~\ref{Fig:GaborH2p}(d), contributions. From these figures is evident the presence of an interference minimum for the whole temporal window. This means that the, $\vec{\mu}_{11}(t)$ and $\vec{\mu}_{21}(t)$  processes, that describe electrons arriving at the same point $\textbf{R}_1$ from two different atomic sources, $\textbf{R}_1$ and $\textbf{R}_2$ respectively, cancel each other and an interference zone is seen for an harmonic order of around $35^{\mathrm{th}}$. These two contributions are dominated by the short-trajectories, therefore both incoming electron wavepackets arrive at the same time and as a consequence a destructive interference is observed. This feature is inherited to the total time-dependent dipole moment (see Fig.~\ref{Fig:GaborH2p}(d)).

\subsubsection{H$_2$ molecule}

The next simplest diatomic molecule is the H$_2$. In order to investigate the behaviour and versatility of our semiclassical SFA model, we compute HHG spectra using the time-dependent dipole moment presented in Sec.~II.
We consider an H$_2$ molecule in equilibrium where the two H atoms are separated a distance of $R=1.4$~a.u.~(0.74 \AA) The ionization potential of the outer electron predicted by our non-local potential is $I_p=1.5$~a.u.~(40.82 eV) and it was calculated by setting $\Gamma=1.0$ and $\gamma=0.12$~a.u. With these parameters our model reproduce the PES of H$_2$ with a minimum at the equilibrium internuclear distance~\cite{MLein2007}. The driven laser pulse has the same parameters as for the case of H$_2^+$. 

		\begin{figure}[h]
            \subfigure[]{ \includegraphics [width=0.44\textwidth] {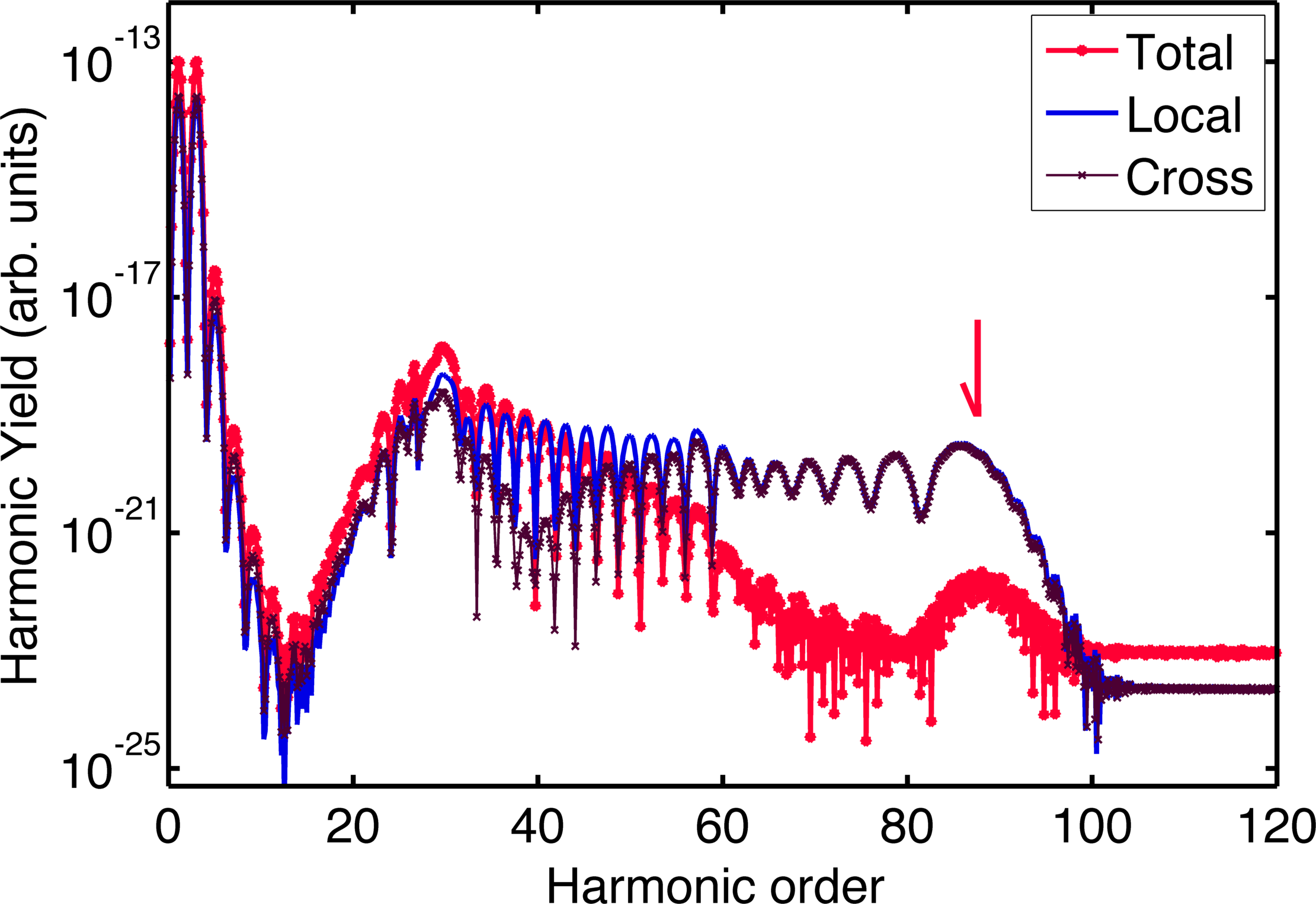}
            \label{fig:subfiga}}
              \subfigure[]{ \includegraphics [width=0.44\textwidth] {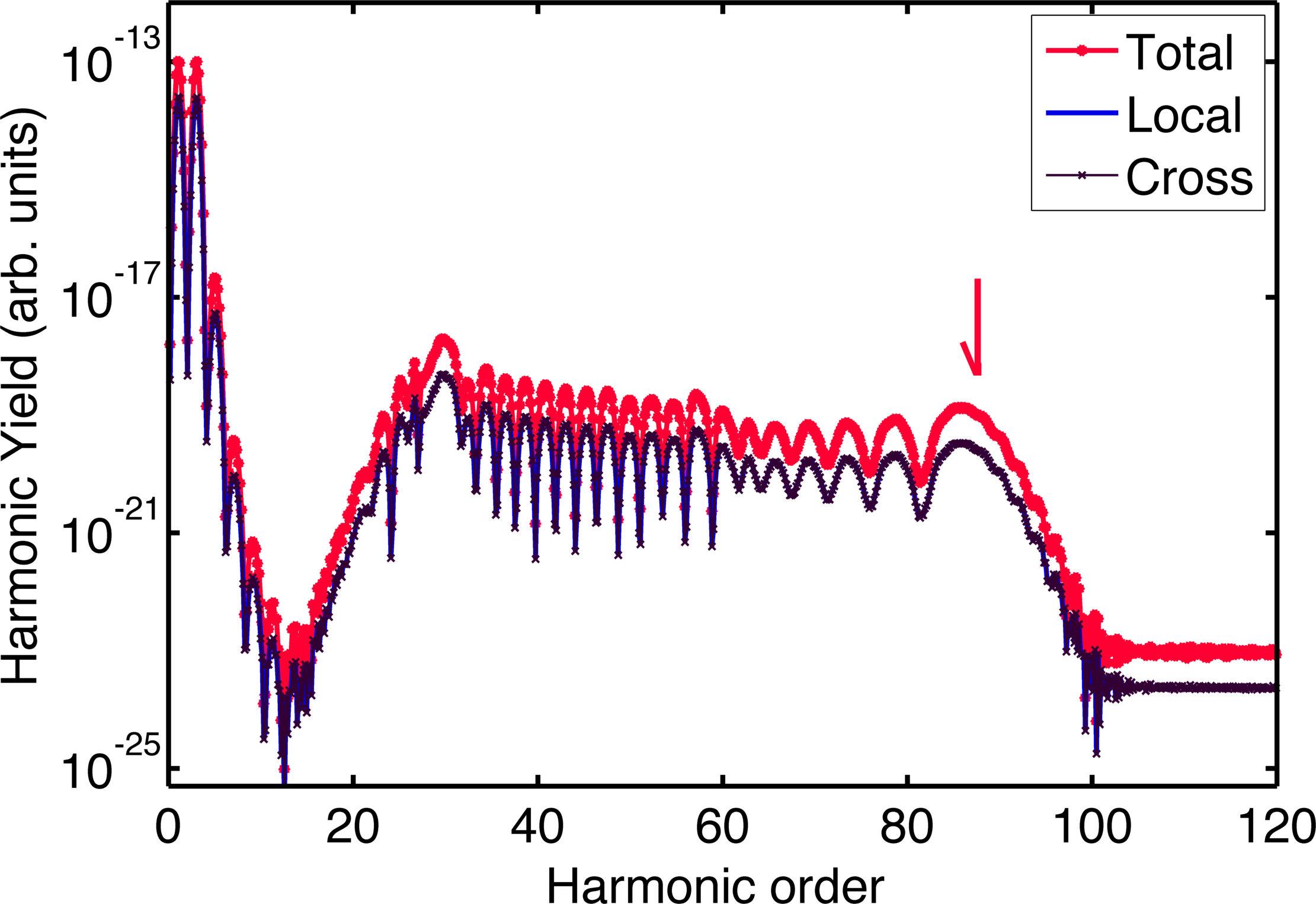}
            \label{fig:subfiga}}
		    \caption{(color online) Different contributions to the molecular HHG spectra (in 
		    logarithmic scale) of an H$_2$ molecule. (a) total, local and cross contributions for 
		    a molecule oriented parallel ($\theta=0^{\circ}$) to the laser field polarization; (b) 
		    the same as in (a) but for $\theta=90^{\circ}$ (perpendicular orientation). }
		  	\label{Fig:H2_Cont}
		\end{figure}

Figure~\ref{Fig:H2_Cont} shows the different contributions to the total HHG spectrum by considering two different molecular orientations: parallel, $\theta=0^{\circ}$, (Fig.~\ref{Fig:H2_Cont}(a)) and perpendicular, $\theta=90^{\circ}$, (Fig.~\ref{Fig:H2_Cont}(b)) with respect to the incident laser pulse polarization. The total HHG spectra (in red) is computed as the sum of all possible processes (see Sec.~II.B for details). In both panels we have grouped two main contributions: (i) the Local and (ii) the Cross ones. The Local contributions (blue line) are processes related with only one atom or position, i.e.~they involve the sum of processes involving only one single atom, meaning ionization from the $\textbf{R}_1/\textbf{R}_2$ and recombination in the same atom. On the other hand, the Cross contributions (in dark brown) include processes involving both of the atoms in the molecule, i.e~ionization from the atom located at $\textbf{R}_1$ and recombination on the atom located at $\textbf{R}_2$ and the other way around.

The first observation regarding Fig.~\ref{Fig:H2_Cont} is that for the case of parallel orientation, Fig.~\ref{Fig:H2_Cont}(a), the total HHG spectrum starts to gradually decrease for harmonic orders higher than the $\approx30^{\mathrm{th}}$. This behaviour is due to a destructive interference of the local and the cross processes. The latter shows a deep minimum around the $\approx40^{\mathrm{th}}$ order. In the case of the molecule perpendicularly oriented, Fig.~\ref{Fig:H2_Cont}(b), an extended plateau with a cutoff around $90^{\mathrm{th}}$ is clearly visible. In both cases, parallel and perpendicular, the molecular HHG spectra shows a deep minimum around the $12^{\mathrm{th}}$ harmonic order. As in the case of H$_2^+$ previously presented, the utilization of a short range potential restricts our results to the higher order harmonics, where the influence of the molecular potential details is less relevant.

	\begin{figure}[h]
	  \centering
                    \includegraphics[width=0.6\textwidth]{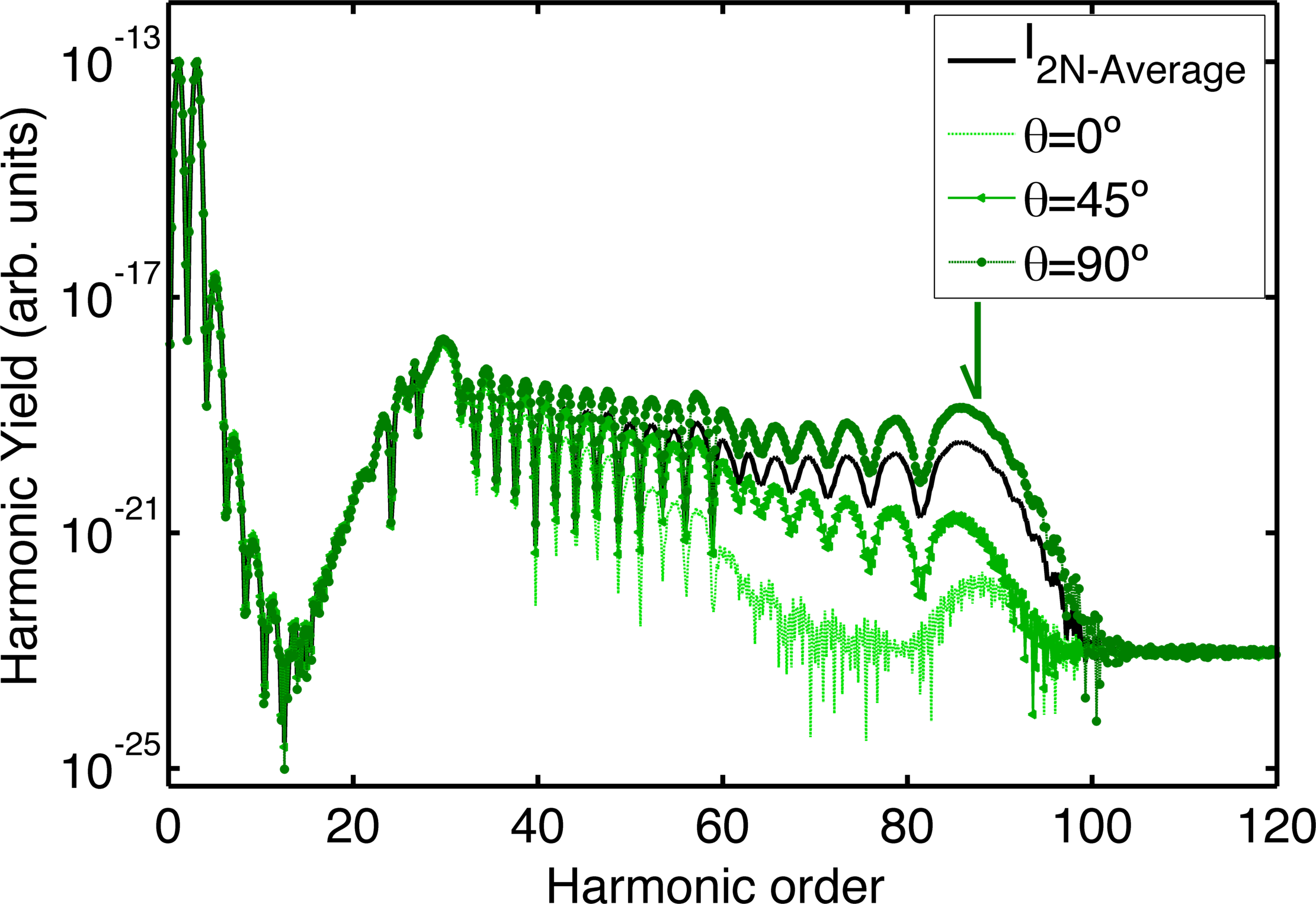}
		    \caption{(color online) Total H$_2$ molecular HHG spectra (in logarithmic scale) for $\theta=0^{\circ}$, $\theta=45^{\circ}$,  $\theta=90^{\circ}$ and averaged over nine different molecular orientations (see the text for more details).}
		  	\label{Fig:H2_Average}
		\end{figure}

It is interesting to note that for the case of perpendicular orientation, $\theta=90^{\circ}$, (Fig.~\ref{Fig:H2_Cont}(b)), both the local and cross terms contribute evenly in the plateau region of the HHG spectra, while for the parallel orientation $\theta=0^{\circ}$, (Fig.~\ref{Fig:H2_Cont}(a)) the local and cross contributions present a different behavior. We can then infer that for the $\theta=90^{\circ}$ case the total harmonic spectrum reaches a maximum yield at the cutoff region. This is due to the fact that, for this favourable orientation, the contribution of each of the processes, local and cross, is comparable.

Finally, in Fig.~\ref{Fig:H2_Average}, we show the total HHG spectra for three different molecular orientations, $\theta=0^{\circ}$, $\theta=45^{\circ}$ and $\theta=90^{\circ}$ and an averaged case over nine values of $\theta$ in the range $[0^{\circ}-360^{\circ}]$. Our diatomic molecule is symmetrical with respect to the origin, i.e.~$\textbf{R}_1=-\textbf{R}/2$ and $\textbf{R}_2=\textbf{R}/2$ and, consequently, the total HHG spectra for $\theta=0^{\circ}$ and $\theta=180^{\circ}$ are identical. The same behaviour is observed for the spectra at $45^{\circ}$, $135^{\circ}$, $225^{\circ}$ and $315^{\circ}$ or for $90^{\circ}$ and $270^{\circ}$.  We can observe in Fig.~\ref{Fig:H2_Average} how different molecular configurations contribute to the total HHG spectra. As we can see the difference in the total spectra for different orientation angles is hardly to notice for lower harmonic orders ($<30^{\mathrm{th}}$). Differences start to be noticeable in the mid-plateau and cutoff regions. In these spectra ranges the larger HHG yield is reached for the perpendicular orientation ($\theta=90^{\circ}$), thus confirming the results presented in Fig.~\ref{Fig:H2_Cont}. Two final remarks are in order, namely (i) the averaged total harmonic spectra is about one order of magnitude lower than the one at perpendicular orientation; (ii) the average procedure washes out any two-slit interference fingerprint.

\subsubsection{Time-frequency analysis for H$_2$}
  
\begin{figure}[htb]
            \subfigure[]{ \includegraphics [width=0.45\textwidth] {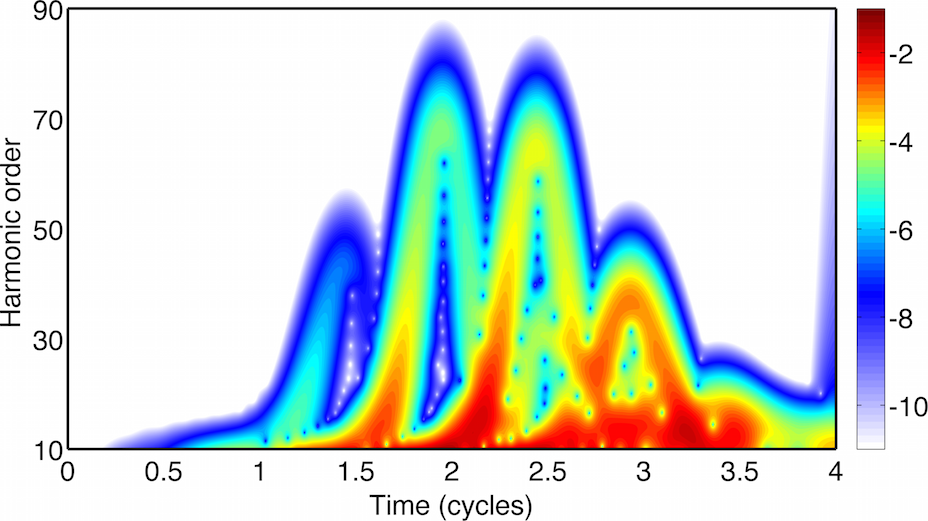}
            \label{fig:subfiga}}
              \subfigure[]{ \includegraphics [width=0.45\textwidth] {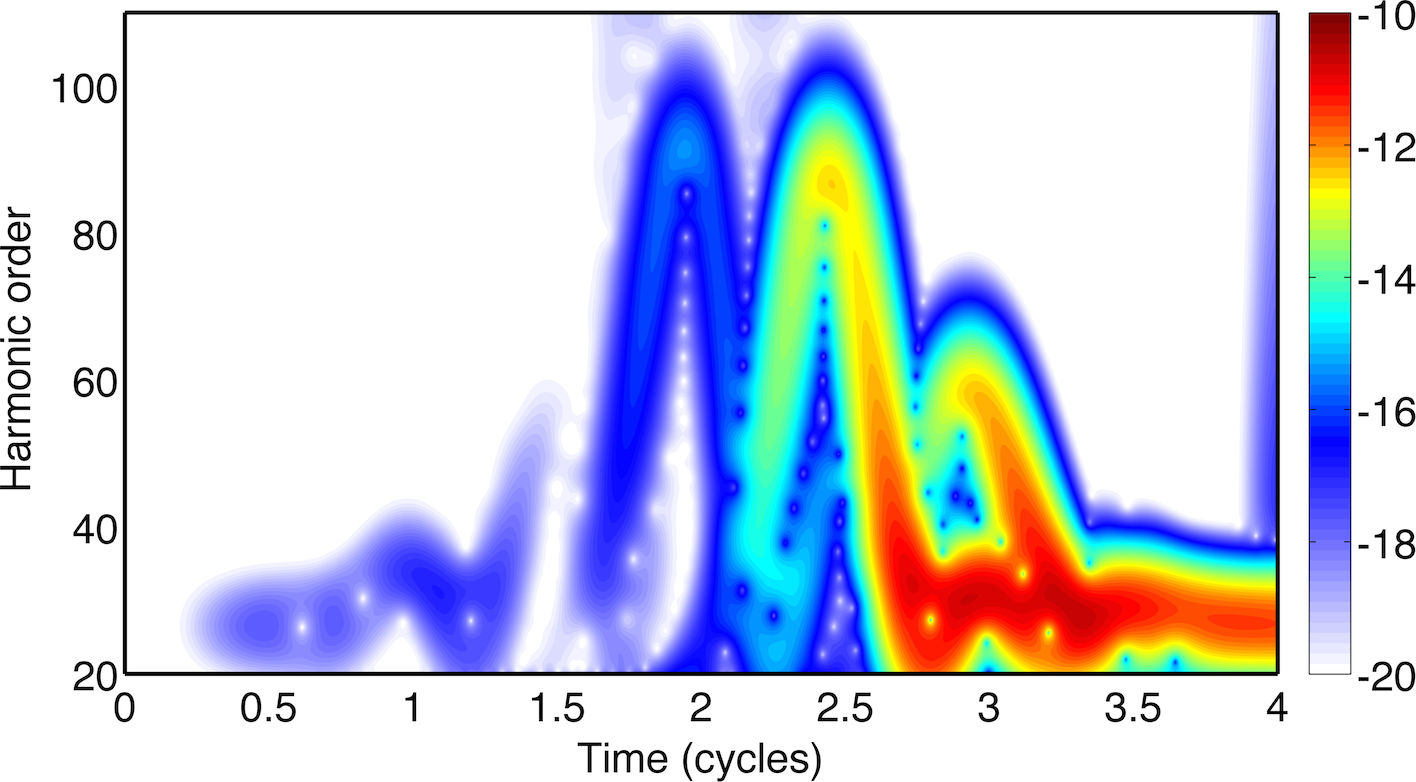}
            \label{fig:subfiga}}
		    \caption{(color online) Gabor transformed time-dependent dipole. (a) H atom driven by a laser pulse with a peak intensity of $I_0=5\times10^{14}$~W$\,\cdot$\,cm$^{-2}$; (b) same as (a) but for an H$_2$ molecule. }
		  	\label{Fig:Gabor}
		\end{figure}

In Fig.~\ref{Fig:Gabor} we perform a Gabor transformation upon the time-dependent dipole moment for both an H atom and our diatomic H$_2$ molecule.  Our aim with this time-energy analysis is to investigate the influence of the short and long trajectories for the molecular system and highlight the differences with the atomic case. In Fig.~\ref{Fig:Gabor}(a) we show the calculation for the H atom, while in Fig.~\ref{Fig:Gabor}(b) we show the same analysis for the molecular  system randomly oriented. In both cases we have considered a laser peak intensity of $I_0=5\times10^{14}$~W$\,\cdot$\,cm$^{-2}$.

In general both figures look quite different. The atomic system, Fig.~\ref{Fig:Gabor}(a), is mostly dominated by the short trajectories while the molecular system, Fig.~\ref{Fig:Gabor}(b), have a prevailing contribution from the long ones. This is so because the orientation average procedure removes any fingerprint of the molecular interferences. 

From a detailed  comparison between the atomic and molecular cases we observe that for the former, even when the short trajectories are dominant at the beginning of the laser pulse (first 2 optical cycles), some contribution of the long ones survives for the later optical cycles where long and short contributions are similar (3 optical cycle). On the contrary, in the molecular system short and long trajectories contribute to different optical cycles. For instance, in the 1$^{\mathrm{st}}$ and 2$^{\mathrm{nd}}$ optical cycles the main contribution is from the short trajectories while for the 3$^{\mathrm{nd}}$ and 4$^{\mathrm{rd}}$ optical cycles a big contribution of the long trajectories appears. In the molecular system the contributions of the long trajectories start to increase, being paramount for the later optical cycles where the contribution of the short one is less significant.


\subsection{Three-center molecular systems}

In order to study systems with more degrees of freedom and describe the different processes contributing to the total HHG spectra, as we have done for diatomics, we apply the model described in Sec.~II.C to both CO$_2$ and H$_2$O molecules.

\subsubsection{The carbon dioxide molecule}
 
The carbon dioxide molecule CO$_2$ is a linear system formed by three atoms, O$=$C$=$O, where the two oxygen atoms are separated by a distance of $R=4.38$~a.u (2.31 \AA) when the system is in equilibrium. At this equilibrium state the parameters of the non-local potential are set to $\Gamma=0.8$ and $\gamma=0.11$~a.u corresponding to an ionization potential of $I_p=0.50$~a.u. (13.6 eV). This value is in excellent agreement with the actual CO$_2$ ionization energy (13.77 eV)~\cite{CO2}. 
The incident laser electric field is defined in Eq.~(\ref{Eq:Efield1}) and we use a laser wavelength and peak intensity of $\lambda=800$~nm and $I_0=1\times10^{14}$~W$\,\cdot$\,cm$^{-2}$, respectively.  The laser pulse has four total cycles (11 fs of total duration) and the CEP is set to $\phi_{0}=0^{\circ}$. 

\begin{figure}[htb]
            \subfigure[]{ \includegraphics [width=0.4\textwidth] {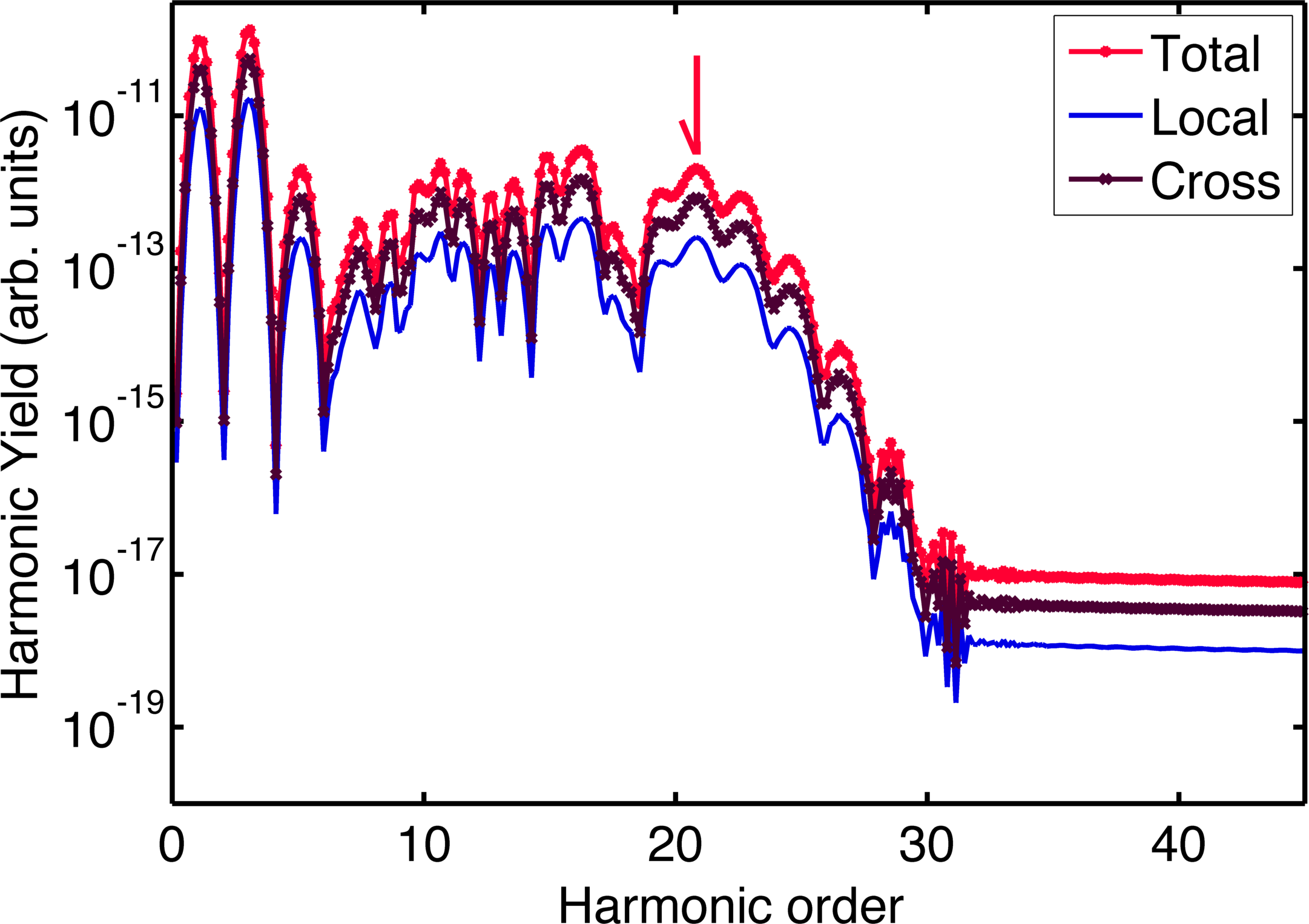}
            \label{fig:subfiga}}
             \subfigure[]{ \includegraphics [width=0.4\textwidth] {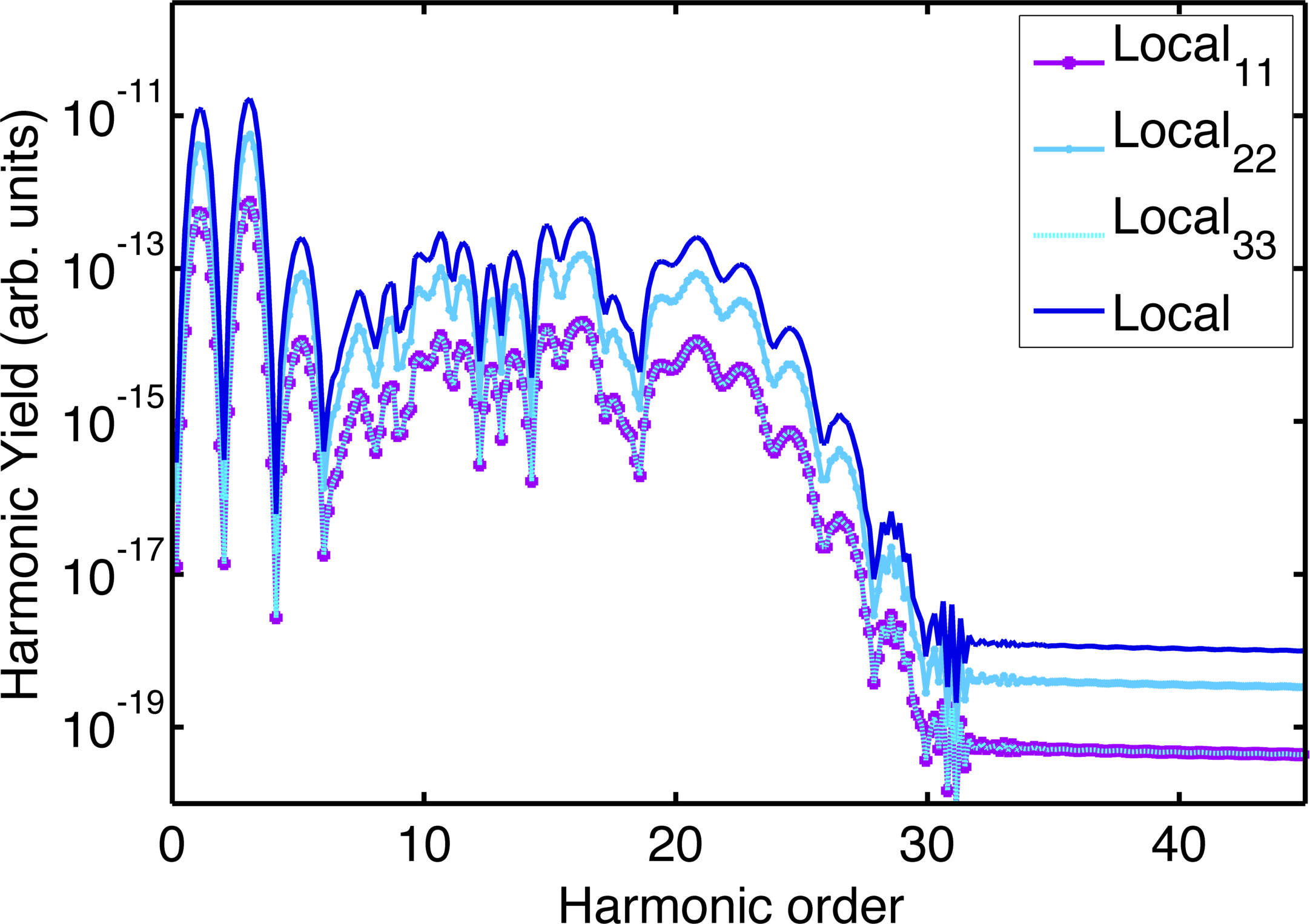}
            \label{fig:subfiga}}
              \subfigure[]{ \includegraphics [width=0.4\textwidth] {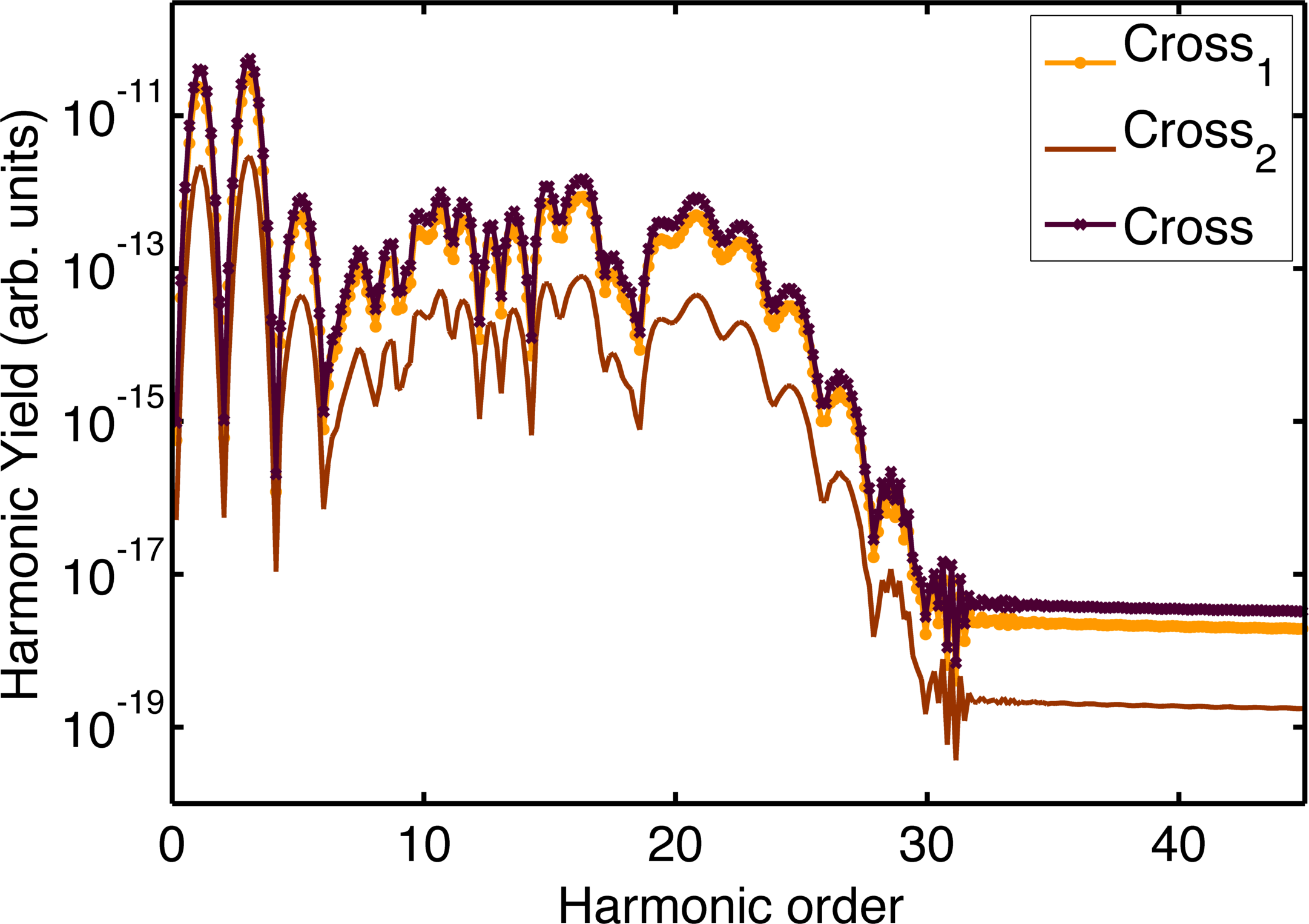}
            \label{fig:subfiga}}
               \subfigure[]{ \includegraphics [width=0.4\textwidth] {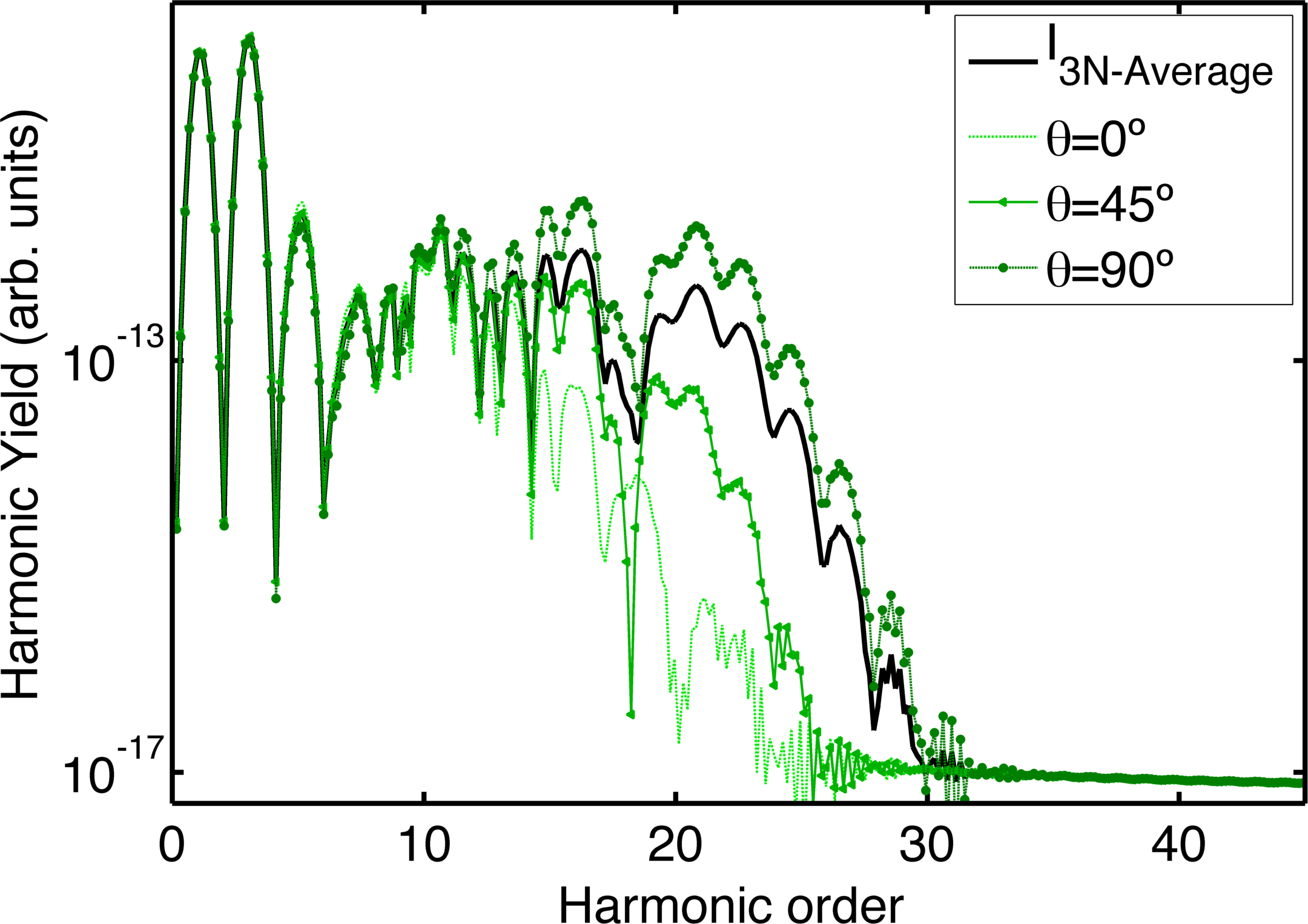}
            \label{fig:subfiga}}
                        
 
		    \caption{(color online) CO$_2$ molecular HHG spectra $I_{3N}(\omega)$ (in logarithmic scale) as a function of the harmonic order calculated by using our quasiclassical SFA  (see text for more details).}
		  	\label{Fig:HHGCO2}
		\end{figure}

The HHG spectra computed by using our quasiclassical SFA model for the CO$_2$ system is presented in Fig.~\ref{Fig:HHGCO2}. In the Fig.~\ref{Fig:HHGCO2}(a) we show the different contributions to the HHG spectra: the total $I_{\rm3N}(\omega)$ (solid line with red circles), calculated from the time-dependent dipole presented in Eq.~(\ref{Eq:mu3N_total}), the local $I_{\rm3N-Local}(\omega)$ (blue solid line), computed with Eq.~(\ref{Eq:mu3N_local}), and the cross $I_{\rm3N-Cross}(\omega)$ (dark brown line with asterisks), extracted from Eq.~(\ref{Eq:mu3N_NLocal}). These calculations show the well known HHG plateau that ends with a cutoff (marked with a red arrow) at around the $21^{\mathrm{th}}$ harmonic order (this last value is in perfect agreement with the one predicted by the semiclassical law Eq.~(\ref{Eq:HHGCutoff})). Both local and cross contributions have almost the same yield over all the frequency range and only minor differences are visible.

In Fig.~\ref{Fig:HHGCO2}(b) we present a split of the local processes, $I_{3N-11}(\omega)$ (solid line with purple circles), $I_{3N-22}(\omega)$ (solid line with light blue squares) and $I_{3N-33}(\omega)$ (dashed line). As we can see the contribution from the O atoms, placed at the end of the molecule, is equal in amplitude and shape and different in yield from the $I_{3N-22}(\omega)$ (corresponding to the C atom placed at the origin). This means that the O atoms contribute a slightly less than the C atom placed at the origin. We notice, however, that the shapes and positions of the minima are the same for the three contributions.

In Fig.~\ref{Fig:HHGCO2}(c) we present each of the contributions that build up the total cross processes. We have separated them depending on how long the ionized electron travels in the continuum before recombination. The Cross$_1$ (solid line with orange circles), Eq.~(\ref{Eq:mu3N_Nlocal1}), and the Cross$_2$ (solid brown line), Eq.~(\ref{Eq:mu3N_Nlocal2}) have similar yields. The main difference between these two HHG spectra is the yield: the cross$_1$ has a bigger contribution than the Cross$_2$. The position of the absolute minima around the $19^{\mathrm{th}}$ harmonic order is present in both contributions as in the local term.  For the calculations in Figs.~\ref{Fig:HHGCO2}(a)-(c) we consider a CO$_2$ molecule aligned perpendicular to the incident laser pulse polarization, i.e.~the internuclear axis vector is forming a angle of $\theta= 90^{\circ}$, being this the most favorable configuration (see Fig.~\ref{Fig:HHGCO2}(d)). 

Finally, in Fig.~\ref{Fig:HHGCO2}(d) we present a set of total HHG spectra for different molecular orientations, namely parallel ($\theta=0^{\circ}$), oblique ($\theta=45^{\circ}$) and perpendicular ($\theta=90^{\circ}$). In addition we include an averaged HHG spectrum, obtained coherently adding \# orientations.  We can observe a similar behavior as for the case of H$_2$ (see Fig.~\ref{Fig:H2_Average}), i.e.~the difference in the total spectra for different orientation angles is hardly to see for lower harmonic orders and starts to be visible in the mid-plateau and cutoff regions. Furthermore, the perpendicular orientation appears to be the dominant one. The comparable behavior between the CO$_2$ and H$_2$ molecules support the fact that the former could be considered as a 'stretched' diatomic O$_2$ molecule for interference minima calculations~\cite{MLein2007}.

\subsubsection{The water molecule}

One of the most important three-center molecules is water (H$_2$O) since it is part of the building blocks of biological life. In this section we theoretically investigate harmonic spectra of the H$_2$O molecule using our semiclassical SFA approach. 

We consider an H$_2$O molecule under the influence of the strong laser field described by Eq.~(\ref{Eq:Efield1}). The H$_2$O molecule is an angular molecule with two H atoms and one O atom. At equilibrium the internuclear distance of the bond H$=$O is about $R=1.8$~a.u. (0.95 \AA) and the angle between the two H atoms $\alpha=104.5^{\circ}$. For these parameters, and considering an ionization potential of $I_p=0.46$~a.u.~(12.52 eV)\cite{H2O}, we set the parameters of our non-local potential to $\Gamma=0.8$ and $\gamma=0.1$~a.u. 

\begin{figure}[htb]
            \subfigure[]{ \includegraphics [width=0.45\textwidth] {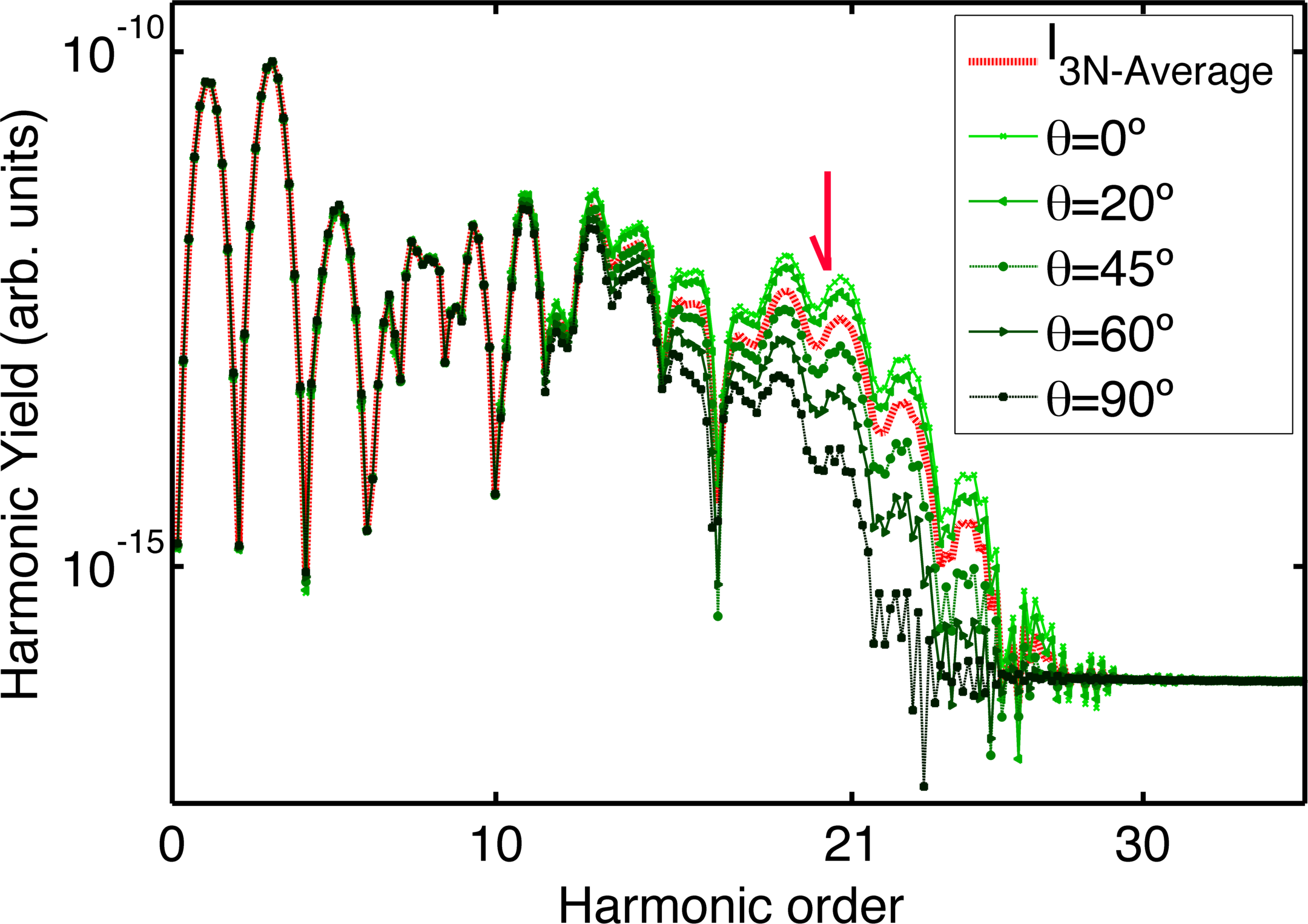}
            \label{fig:subfiga}}
                    \subfigure[]{\includegraphics[width=0.45\textwidth]{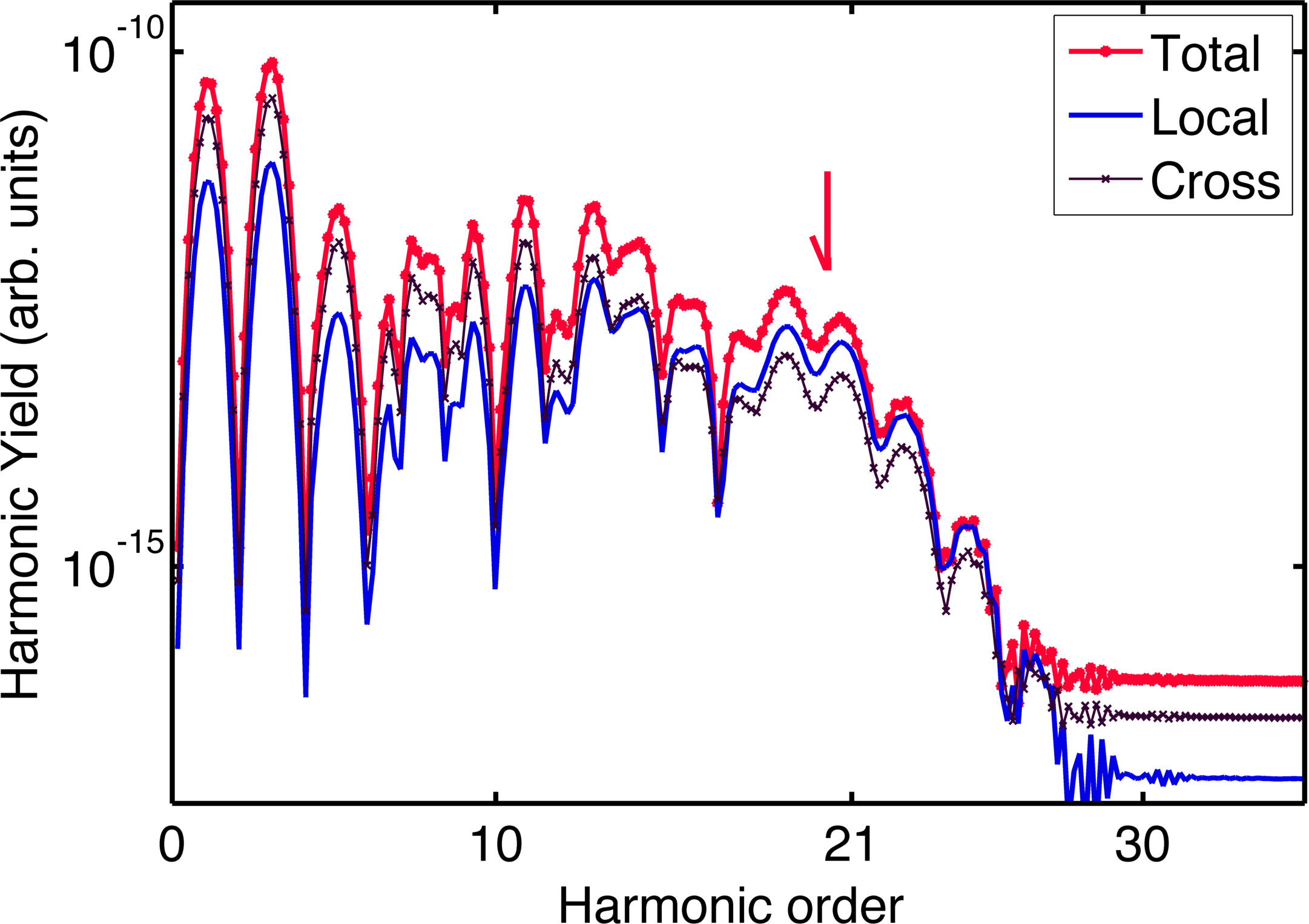}}
		    \caption{(color online) HHG spectra $I_{\rm 3N}(\omega)$ (in logarithmic scale) of an H$_2$O molecule, as a function of the harmonic order, computed using our quasi-classical SFA model. (a) HHG spectra for $\theta=[0^{\circ}, 20^{\circ}, 45^{\circ}, 60^{\circ}, 90^{\circ}]$ and averaged over 8 orientations in the range $\theta=[0^{\circ}-360^{\circ}]$; (b) different contributions to the averaged HHG spectra.}
		  	\label{Fig:HHG_H2O_angle}
		\end{figure}

In Fig.~\ref{Fig:HHG_H2O_angle} we show HHG spectra for a laser wavelength and peak intensity of $\lambda=800$~nm and $I_0=1\times10^{14}$~W$\,\cdot$\,cm$^{-2}$, respectively.  The laser pulse has four total cycles (11 fs of total duration) and the CEP is set to $\phi_{0}=0$~rad.  In Fig.~\ref{Fig:HHG_H2O_angle}(a) we show HHG spectra for both five different molecular orientations,  $\theta=0^{\circ}$, $\theta=20^{\circ}$, $\theta=45^{\circ}$, $\theta=60^{\circ}$ and $\theta=90^{\circ}$ and an averaged case. The molecular axis is fixed in space and forms an angle of $\alpha/2$ with respect to the vector position $\textbf{R}_1$. Furthermore, $\theta$ defines the angle between this molecular axis and the laser electric field polarization (see Fig.~\ref{Fig:1Molecule} and the Appendix A for more details).

The dependency of the HHG spectra with respect to the molecular orientation is quite evident. For lower harmonic orders all the orientations appear to be  equivalent and the main differences start to materialize for the harmonic orders $\gtrsim12^{\mathrm{th}}$. As we can see in  Fig.~\ref{Fig:HHG_H2O_angle}(a) the HHG spectra for $\theta=0^{\circ}$ (solid line with asterisks), $20^{\circ}$ (solid line with left-pointing triangle) and $45^{\circ}$ (dashed line) exhibit a similar structure. The other two orientations, $60^{\circ}$ (right-pointing triangle line) and $\theta= 90^{\circ}$ (square line), present an harmonic yield several orders of magnitude lower in this region. The total HHG spectra for all the molecular orientations show a slight minimum around the $17^{\mathrm{th}}$ harmonic order that could be attributed to interference effects, although it is not an easy task to characterize it using a simple interference formula as in the case of diatomics. 

\begin{figure}[htb]
            \subfigure[]{ \includegraphics [width=0.45\textwidth] {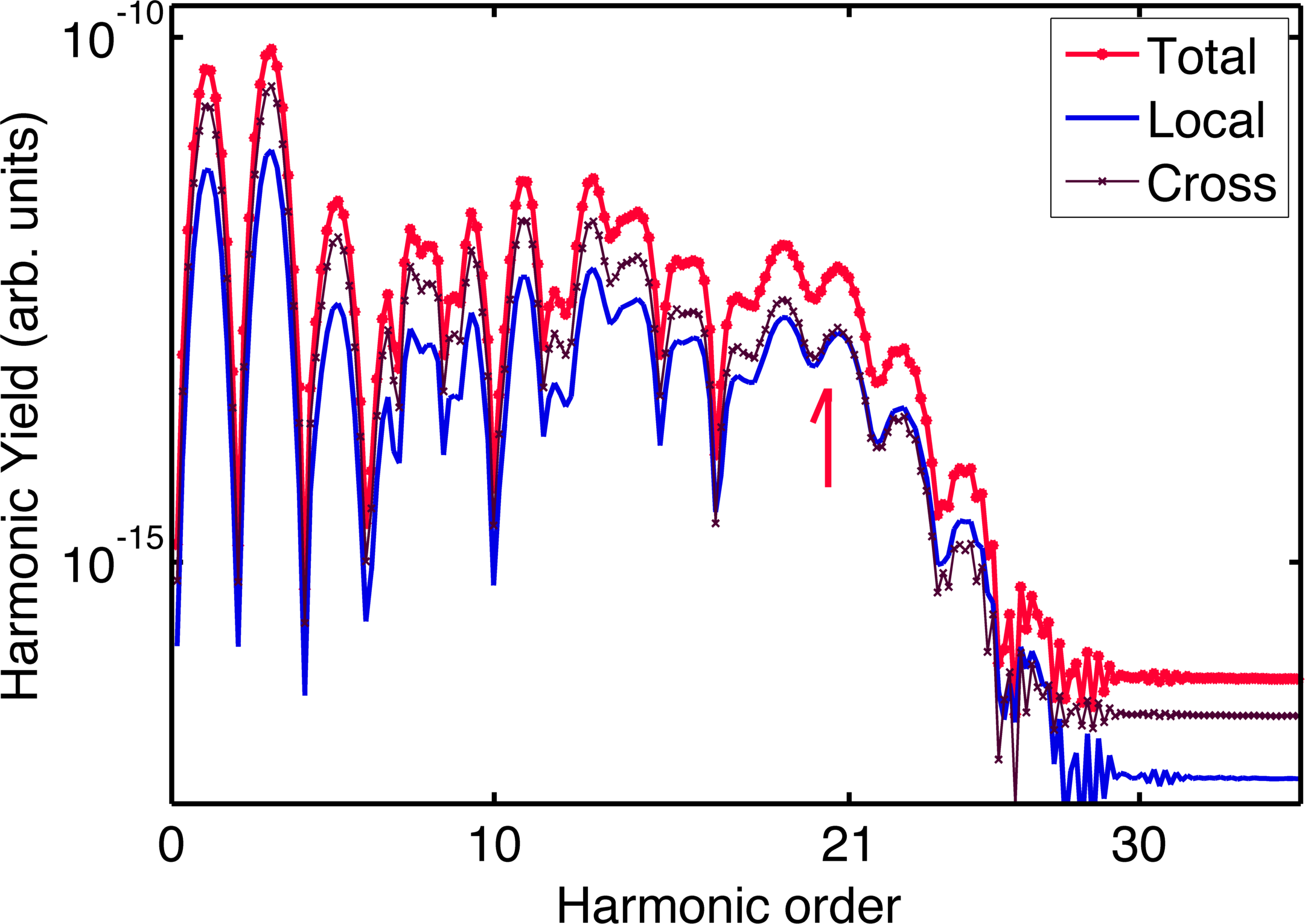}
            \label{fig:subfiga}}
                    \subfigure[]{\includegraphics[width=0.45\textwidth]{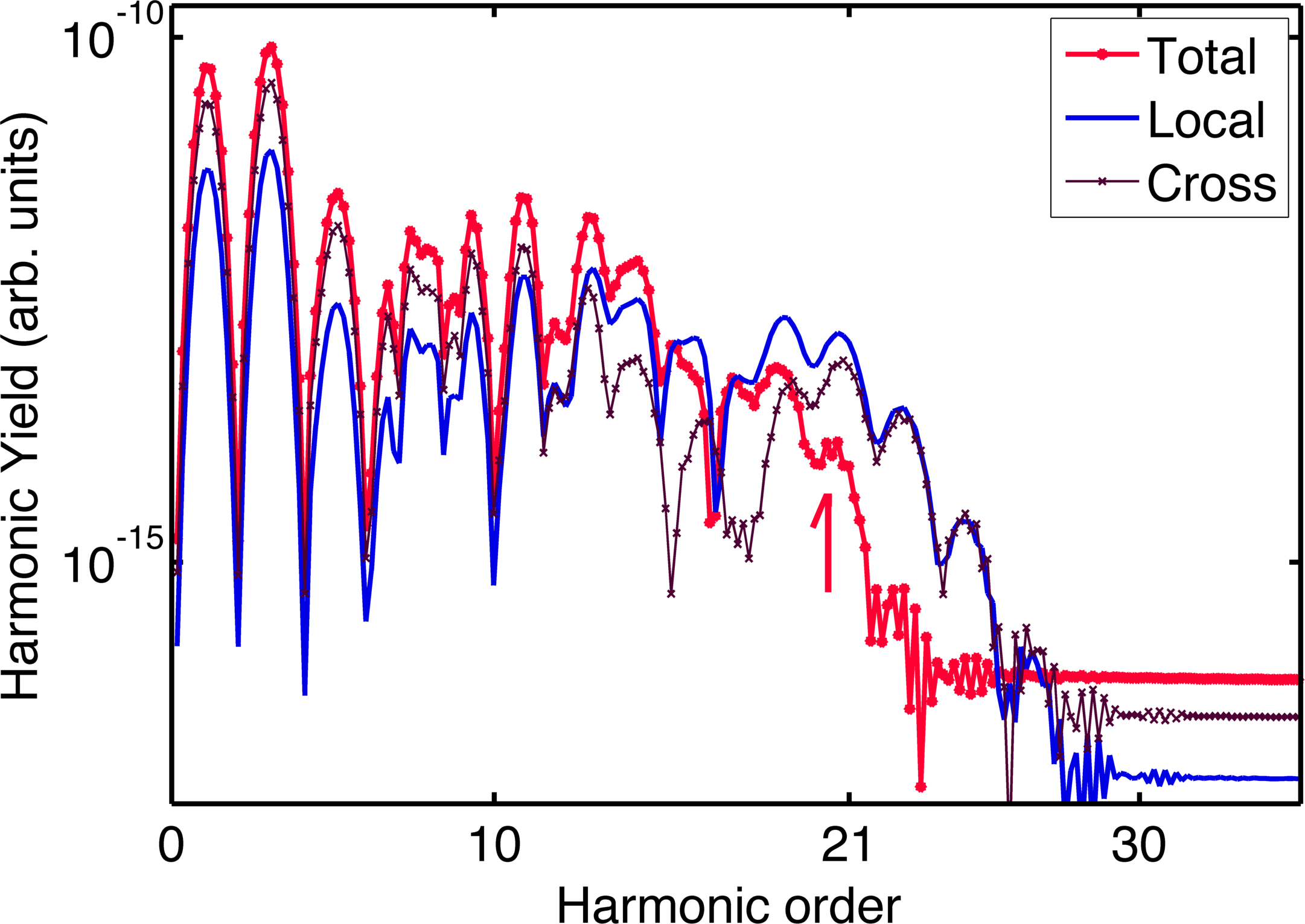}}
		    \caption{(color online) Different contributions to the H$_2$O molecular HHG spectra. (a) $\theta=0^{\circ}$; (b) $\theta=90^{\circ}$.}
		  	\label{Fig:HHG_H2O_Cont}
		\end{figure}

We have also included in Fig.~\ref{Fig:HHG_H2O_angle}(a) an averaged HHG spectra over 8 values of $\theta$ in the range $[0^{\circ}-360^{\circ}]$ (dashed red line). As we can see the minimum survives the orientation average. Furthermore, for $\theta= 90^{\circ}$ (square line) the total HHG spectra rapidly decreases for harmonic orders $>16^{\mathrm{th}}$. This means that the interference between the local and cross processes is destructive and function of the molecular orientation.  This behavior introduces a decrease of the total HHG yield. We note that for H$_2$O, contrarily to the CO$_2$ case, an enhancement of the total HHG spectra is observed when the molecule is oriented parallel,  $\theta=0^{\circ}$, to the laser electric field polarization. As we have done both for diatomics and CO$_2$ in Fig.~\ref{Fig:HHG_H2O_angle} (b) we plot the different terms contributing to the total HHG averaged spectra. Contrarily to the oriented case, here the local and cross processes appear to constructively contribute to the total HHG spectrum. 
 
In order to study more deeply the underlying physics behind the enhancement and decrease of the total HHG spectra for $0^{\circ}$ and $90^{\circ}$ we plot in Fig.~\ref{Fig:HHG_H2O_Cont} the different contributions for these two cases. For the case of $\theta=0^{\circ}$, Fig.~\ref{Fig:HHG_H2O_Cont}(a), the decrease of the HHG yield is evident for harmonic orders higher than the $15^{\mathrm{th}}$. Around this harmonic order both contributions, the local and cross, have a similar yield and the coherent sum develops in a destructive interference decreasing the total HHG spectra in about 3 orders of magnitude. On the other hand,  for $\theta=90^{\circ}$, Fig.~\ref{Fig:HHG_H2O_Cont}(b), we observe a steadily decrease of the cross processes, of about two order of magnitude, in the whole spectral range. Consequently, we can argue that in this case the cross contribution is almost negligible (solid brown line with squares) and the total HHG spectra is dominated by the local processes (solid blue line).


\section{Conclusions and Outlook}

We present a quasiclassical approach that deals with molecular HHG within the SAE. Our model could be considered as a natural extension to the one introduced for above-threshold ionization in atoms~~\cite{PRANoslen2015} and molecules~\cite{Molecule2016}. The focus of our study is on di- and tri-atomic molecular systems, although the extension to more complex systems appears to be straightforward. Firstly, we have validated our formalism comparing the atomic HHG spectra with results extracted from the 3D-TDSE and using a large set of laser intensities and wavelengths. For the molecular systems we have shown our approach is able to capture the interference features, ubiquitously present in every molecular HHG. The main advantages of our model are: (i) the possibility to disentangle the underlying contributions to the HHG spectra. In this way, we could isolate local and cross processes and also treat both fixed and randomly oriented molecules; (ii) the low computational cost. By considering our approach involves only 1D and 2D time integrations, all the other quantities are analytical, it is clear that we compute molecular HHG spectra without too much computational effort; (iii) the concrete feasibility to model complex molecular ground states. For all the studied molecular cases we were able to model reasonable well the initial molecular ground state, varying the parameters of our non-local potential.

\acknowledgments{This work was supported by the project ELI--Extreme Light Infrastructure--phase 2 (CZ.02.1.01/0.0/0.0/15\_008/0000162 ) from European Regional Development Fund, Ministerio de
Econom\'{\i}a y Competitividad through Plan Nacional (FIS2011-30465-C02-01, FrOntiers of QUantum Sciences (FOQUS): Atoms, Molecules, Photons and Quantum Information FIS2013-46768-P, FIS2014-56774-R, and Severo Ochoa Excellence Grant SEV-2015-0522),  the Catalan Agencia de Gestio d'Ajuts Universitaris i de Recerca (AGAUR) with SGR 2014-2016, Fundaci\'o Privada Cellex Barcelona and funding from the European Union’s Horizon 2020 research and innovation Programme under the Marie Sklodowska-Curie grant agreement No. 641272 and Laserlab-Europe (EU-H2020 654148). N.S. was supported by the Erasmus Mundus Doctorate Program Europhotonics (Grant No. 159224-1-2009-1-FR-ERA MUNDUS-EMJD). N. S., A.
C. and M. L. acknowledge ERC AdG OSYRIS and EU FETPRO QUIC. J.B. acknowledges
FIS2014-51478-ERC and the National Science Centre, Poland -- Synfonia grant 2016/20/W/ST4/00314. J.A.P.-H. acknowledges support from Laserlab Europe (Grant No. EU FP7 284464) and the Spanish Ministerio de Econom\'{\i}a y Competitividad (FURIAM Project No. 78 FIS2013-47741-R and PALMA project FIS2016-81056-R)}

\appendix
\section{Strong field approximation for three-center molecular systems}

In this section we develop an analytical model to obtain the direct probability transition amplitude, as well as the bound and scattering states, necessary to calculate the HHG spectra for three-center molecular systems. This approach can be considered an extension of the atomic and diatomic models presented in Refs.~\cite{PRANoslen2015, Molecule2016}. Our quasiclassical formalism takes advantage of both the single active electron and dipole approximations. In addition, we consider the nuclei of the molecule are fixed in space (the so-called frozen core approximation).


\subsection*{Direct transition probability amplitude}

\begin{figure}[htb]
 \centering
            \includegraphics[width=0.6\textwidth] {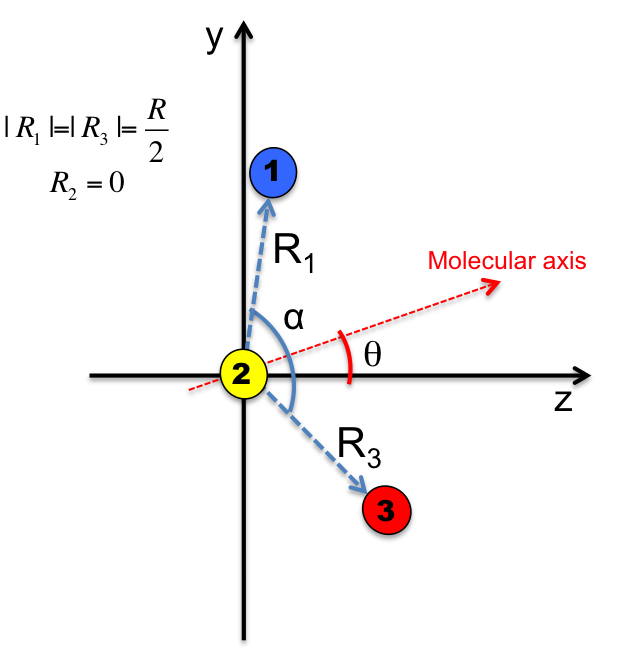}
         \caption{ (color online) Three-center molecular system aligned a $\theta$ angle with the laser field. The red line represents the molecular axis that form an angle of $\alpha/2$ between $\textbf{R}_1=[0, \frac{R}{2}\sin(\frac{\alpha}{2} + \theta), \frac{R}{2}\cos(\frac{\alpha}{2}  + \theta)]$ and $\textbf{R}_3=[0, -\frac{R}{2}\sin(\frac{\alpha}{2} -\theta), \frac{R}{2}\cos(\frac{\alpha}{2} -\theta)]$. }
    \label{Fig:1Molecule}
\end{figure}

We consider a molecular system of three independent atoms, as is shown in Fig.~\ref{Fig:1Molecule}, under the influence of an intense and short laser field. In the limit when the wavelength of the laser, $\lambda_0$, is larger compared with the Bohr radius, $a_0=0.529$ nm, the electric field of the laser beam around the interaction region can be considered as spatially homogeneous. This means that the interacting atoms do not experience any spatial dependence of this driving field. Then, only its time-variation is taken into account (the above asseverations define the so-called dipole approximation). 

Therefore, the laser electric field can be written as:
\begin{equation}
 \textbf{E}(t) = \mathcal{E}_0\:f(t) \sin(\omega_0 \:t + \phi_0)\,{\bf e}_{z}. 
 \label{Eq:Efield}
\end{equation}
The field has a carrier frequency $\omega_0 = \frac{2\pi c}{\lambda_0}$, where $c$ is the speed of light, 
$\mathcal{E}_0$ the laser electric field peak amplitude, and we consider the laser field is linearly polarized 
along the $z$-direction.  In Eq.~(\ref{Eq:Efield}) ${f(t)}$ denotes the envelope of the laser pulse and $\phi_0$ 
defines the CEP (see Sec.~III for more details).

The TDSE that describes the whole laser-molecule interaction can be written as:
\begin{eqnarray}
\nonumber
i\frac{ \partial}{ \partial t} | \Psi(t) \rangle&=&\hat{H} | \Psi(t) \rangle, \\ 
&=&[\hat{H}_0 + \hat{V}_{int}({\bf r},t)]| \Psi(t) \rangle,
\label{Eq:SE}
\end{eqnarray}
where $\hat{H}_0=\frac{{\hat{\bf p}}^2}{2} + \hat{V}(\textbf{r})$ defines the laser-field free Hamiltonian,  with ${\hat{\bf p}}=-i{\nabla}$ the canonical momentum operator and $\hat{V}(\textbf{r})$ the potential operator that describes the interaction of the nuclei with the active electron. $\hat{V}_{int}({\bf r},t)=-q_e\hat{{\bf E}}(t)\cdot\hat{{\bf r}}$ represents the interaction of the molecular system with the laser radiation, written in the dipole approximation and length gauge. $q_e$ denotes the electron charge (in atomic units $q_e=-1.0$~a.u.).

We shall restrict  our model to  the  low ionization  regime, where the  SFA is valid~\cite{ Keldysh1965, Faisal1973,Reiss1980,Lewenstein1986, Lewenstein1994,Lewenstein1995}.  Therefore, we work in the tunneling regime, where the Keldysh parameter  $\gamma=\sqrt{I_p/2U_p}$ ($I_p$ is the ionization potential of the system and $U_p=\frac{\mathcal{E}_0^2}{4\omega_0^2}$ the ponderomotive energy acquired by the electron during its incursion in the laser field) is less than one, i.e.~$\gamma<1$. In addition,  we assume that $V(\textbf{r})$  does not  play an important role in the electron  dynamics  once the electron  appears in the continuum. 

These observations,  and the following three statements, define the standard SFA, namely:
\begin{enumerate}[(i)]
\item Only the ground, $|0 \rangle$, and the continuum  states, $ |\textbf{v}\rangle$, are taken into account in the interaction process.
\item There is no depletion of the ground state $(U_p < U_{sat})$.
\item The continuum  states  are approximated by Volkov states; in the continuum the electron is considered as a free particle solely moving driven by the laser electric field. 
\end{enumerate}
For a more detailed discussion of the validity of the above statements see e.g.~Refs.~\cite{Lewenstein1994,Lewenstein1995, PRANoslen2015}. 

Based on (i), we propose a state, $|\Psi(t) \rangle=\sum_{j=1}^{3}| \Psi_j (t)\rangle$, to describe the time-evolution of the three-center system, i.e.~a superposition of three atomic states. In turn, each independent state, $|\Psi_j (t)\rangle$, is a coherent superposition of ground, $|0\rangle= \sum_{j=1}^{3} |0_j\rangle$, and continuum states, $|\textbf{v}\rangle$ ~\cite{Lewenstein1994,Lewenstein1995}:
\begin{equation}
| \Psi_j (t)\rangle= e^{\textit{i}I_p\textit{t}}\bigg(a(t) |0_j\rangle + \: \int{\textit{d}^3 \textbf{v} \: \textit{b}_j( \textbf{v},t) |\textbf{v}\rangle} \bigg),
\label{Eq:PWavef1}
\end{equation}
where the subscript $j=1,2,3$ refers to the position $\textbf{R}_1$, $\textbf{R}_2$ and $\textbf{R}_3$ of each of the atoms in the three-center molecule, respectively.

The factor, $a(t)$, represents  the  amplitude  of the  ground  state  which is considered constant in time, $a(t) \approx 1$, under the assumption  that  there  is no depletion  of the ground state.  The latter follows directly from statement  (ii). The pre-factor, $e^{\textit{i}I_p\textit{t}}$, represents the phase oscillations which describe the accumulated  electron energy in the ground state ($I_p=-E_0$ is the ionization potential of the molecular target, with $E_0$ the ground state energy of the three-center molecular system). Furthermore, the transition amplitudes to the continuum states are denoted by $\textit{b}_j(\textbf{v},t)$, with $j=1,2,3$ depending on the atomic nuclei. These amplitudes depend both on the kinetic momentum of the outgoing electron and the laser pulse.  Therefore, our main task is to derive a general expression for each transition amplitude ${b}_j({\bf v},t)$. In order to do so, we substitute Eq.~(\ref{Eq:PWavef1}) in Eq.~(\ref{Eq:SE}). We shall consider that $\hat{H_0}|0_{1,2,3}\rangle= -I_p |0_{1, 2, 3}\rangle$ and $\hat{H_0}|\textbf{v}\rangle = \frac{\textbf{v}^2}{2}|\textbf{v}\rangle$ fulfill for the bound and continuum states, respectively. Consequently, the evolution of the transition amplitude ${b}_{j}({\bf v},t)$ becomes:
\begin{eqnarray}
i\int{\textit{d}^3  \textbf{v} \:\dot{b}_j( \textbf{v},t)\: |\textbf{v} \rangle} &=& \,\, \int{\textit{d}^3 \textbf{v} \bigg(\frac{\textbf{v}^2}{2}+ I_p \bigg)\textit{b}_j ( \textbf{v},t) |\textbf{v}}\rangle+\textbf{E}(t) \cdot  \textbf{r} |0_j \rangle \nonumber \\
 && +\: \textbf{E}(t) \cdot \textbf{r}\int{\textit{d}^3 \textbf{v} \: \textit{b}_j( \textbf{v},t)|\textbf{v} \rangle}.
  \label{Eq:TempEq}
\end{eqnarray}

Note that we have assumed that  the electron-nucleus  interaction is neglected once the electron  appears  at the continuum, i.e.~$V({\bf r})|{\bf v} \rangle = 0$, which corresponds to the statement (iii). Therefore, by multiplying Eq.~(\ref{Eq:TempEq}) by $\langle {\bf v}'|$ and after some algebra, the time variation of the transition amplitude $\textit{b}_j(\textbf{v},t)$ reads:
\begin{eqnarray}
\dot{b}_j( \textbf{v},t) &=& \,\,  -i\bigg(\frac{\textbf{v}^2}{2}+ I_p \bigg) \textit{b}_j( \textbf{v},t) - i \: \textbf{E}(t) \cdot\langle \textbf{v} |  \textbf{r}|0_j \rangle -i \: \textbf{E}(t) \cdot \int{\textit{d}^3 \textbf{v}^{\prime} \: \textit{b}_j( \textbf{v}^{\prime},t)\langle \textbf{v} | \textbf{r} |\textbf{v}^{\prime}\rangle}.
\label{Eq:Newbl}
\end{eqnarray}

The first term on the right-hand of Eq.~(\ref{Eq:Newbl}) represents the  phase evolution  of the  electron in the oscillating laser electric field. In the second term  we have defined the bound-free transition dipole matrix  element as:
\begin{equation}
 - \langle \textbf{v} |\textbf{r}|0_j \rangle=\textbf{d}_j( \textbf{v}). 
 \label{Eq:DM1}
\end{equation}
The state $ |\textbf{v} \rangle$ represents a scattering state constructed as a plane wave, $| {\bf v}_p \rangle$ plus corrections on each center position, $| \delta{\bf v}_j \rangle$. 

Based on statement (iii) our formulation only considers the continuum state as a plane wave $| {\bf v}_p \rangle$ for the calculation of the bound-free dipole matrix element. We shall pay special attention to the computation of Eq.~(\ref{Eq:DM1}). Notice that the plane waves are not orthogonal to the bound states due to the fact that the latter are defined depending on the relative position of each of the atoms, $\textbf{R}_j$, with respect to the origin of coordinates (see~\cite{Molecule2016} for more details). So, for the $\textbf{R}_j$ contribution we introduce a correction to the dipole matrix element as:
\begin{equation}
\textbf{d}_j( \textbf{v})= - \langle \textbf{v}_p |\textbf{r}-\textbf{R}_j|0_j \rangle= - \langle \textbf{v}_p |\textbf{r}|0_j \rangle +\textbf{R}_j\langle \textbf{v}_p |0_j \rangle. 
 \label{Eq:d1}
\end{equation}

The third term of Eq.~(\ref{Eq:Newbl}) defines the continuum-continuum transition matrix 
element. In our case we are interested in to describe processes where the electron is ionized 
and goes to the continuum, never returning to the vicinity of the remaining ion core, i.e.~the 
so-called direct processes.  As the direct ionization process should have a larger probability 
compared  with the continuum-continuum one~\cite{Lewenstein1995, PRANoslen2015},  
one might neglect the last term in 
Eq.~(\ref{Eq:Newbl}), $\langle \textbf{v}| \textbf{r} |\textbf{v}^{\prime}\rangle\approx\textbf{0}$. 

This is what we refer as zero$\textsuperscript{th}$ order solution:
\begin{equation}
{\partial }_tb_{0,j}( \textbf{v},t) =-\textit{i}  \bigg(\frac{\textbf{v}^2}{2}+ I_p-\textbf{R}_{j} \cdot \textbf{E}(t) \bigg) \textit{b}_{0,j}( \textbf{v},t)+ \textit{i}\: \textbf{E}(t) \cdot  \textbf{d}_j( \textbf{v}).
\label{Eq:dB1}
\end{equation}
The latter equation is easily solved by conventional integration methods (see e.g.~\cite{Lev}) 
and considering the Keldysh transformation~\cite{Ehlotzky1992,Keldysh1965}. Therefore, the 
solution can be written as:
\begin{eqnarray}
b_{0,j}( \textbf{p},t) &=&\textit{i}   \:\int_0^t{\textit{d} \textit{t}^{\prime}\:\textbf{E}(t^{\prime})}\:\cdot \textbf{d}_{j}\left[ \textbf{p}+\textbf{A}(t^{\prime})\right] \nonumber \\ 
&&\times \exp\left(-\textit{i} \:\int_{t^{\prime}}^t{d{\tilde t} \: \left \{ [ \textbf{p}+\textbf{A}({\tilde t})]^2/2 +I_p -\textbf{R}_{j} \cdot  \textbf{E}({\tilde t}) \right\}}\right). 
\label{Eq:b_10}
\end{eqnarray}

Notice that $j=1,2, 3$ represents either an atom located at $\textbf{R}_1$, 
$\textbf{R}_2$ or $\textbf{R}_3$,  respectively. For instance, to obtain the transition amplitude 
for the atom placed at $\textbf{R}_1$ we need to set $j={1}$ in Eq.~(\ref{Eq:b_10}).

Note that the above equation is written in terms of the canonical 
momentum~${\bf p}= {\bf v} - {\bf A}(t)$ ~\cite{Lewenstein1994}. 
Here, we have considered that the electron appears in the continuum 
with kinetic momentum ${\bf v}(t')={\bf v}-{\bf A}(t)+{\bf A}(t')$ at the 
time $t'$, where {\bf v}~is the final kinetic momentum (note that in 
atomic units ${\bf p}= {\bf v}$), and
$\textbf{A}(t) =-\int^{t}{ \textbf{E}(t^{\prime})dt^{\prime}}$ 
is the associated vector potential.

Equation (\ref{Eq:b_10}) has a direct physical interpretation: it can be understood as the sum of all the ionization events that occur from the time $t'$ to $t$. Then, the instantaneous transition probability amplitude of an electron at a time $t'$, at which it appears into the continuum with momentum ${\bf v}(t')= \textbf{p}+\textbf{A}(t^{\prime})$, is defined by the argument of the $[0,t]$ integral in Eq.~(\ref{Eq:b_10}). Furthermore, the exponent phase factor denotes the ``semi-classical action", ${S}_{j}({\bf p},t,t^{\prime})$, that defines a possible electron trajectory  from the birth time $t'$, at position $\textbf{R}_{j}$, until the ``recombination" one $t$ as:
 \begin{equation}
{S}_{j}({\bf p},t,t^{\prime}) = \int_{t^{\prime}}^{t}{\:d{\tilde t}\left\{[{\bf p}+\textbf{A}({\tilde t})]^2/2 +I_p -\textbf{R}_{j} \cdot  \textbf{E}({\tilde t})  \right \}}.
\label{Eq:S_R}
\end{equation}
Note that the transition amplitude equations obtained so far depend on the position from which the electron is tunnel ionized to the continuum. The semi-classical action ${S}_{j}({\bf p},t,t^{\prime})$ contains this dependency as well.

Considering we are interested in to obtain the transition amplitude $b_{0,j}({\bf p},t)$ at the end of the laser pulse, the time $t$ is set at $t=t_{\rm F}$. Consequently, we shall define the integration time window as $t\in[0,t_{\rm F}]$. Furthermore we set ${\bf E}(0) = {\bf E}(t_{\rm F}) = {\bf 0}$, in such a way to make sure that the laser electric field is a time  oscillating wave and  does not  contain static components (the  same arguments apply to the vector  potential  ${\bf A}(t)$). Finally, the total transition amplitude for the direct process taking place on our three-center molecular system reads as:
 \begin{eqnarray}
 b_0( \textbf{p},t)= \sum_{j=1}^{3} b_{0,j}( \textbf{p},t).
 \label{Eq:b_0}
\end{eqnarray}

\subsection*{Bound states calculation}

In this section, we are going to develop analytical expressions to obtain the bound states for our three-center molecular system. As in the atomic and diatomic cases we have chosen a non-local potential to describe the interaction of the electrons with the nuclei. We consider a molecular system with three fixed nuclei under the single active electron approximation. Our purpose is to find the analytical dependency of the bound state wavefunctions, that allow us to compute the bound-continuum matrix transition dipole and the direct transition amplitudes, Eq.~(\ref{Eq:b_0}).

The hamiltonian for the molecular system in the momentum representation can be written in a similar way as for the diatomic case, i.e.~:
 \begin{equation}
\hat{H}_{\textbf{M}}(\textbf{p},\textbf{p}^{\prime}) = \frac{\hat{p}^2}{2}\delta(\textbf{p}-\textbf{p}^{\prime}) + \hat{V}_{\textbf{M}}(\textbf{p},\textbf{p}^{\prime})
\label{Hm}
\end{equation}
where the first term on the right-hand side is the kinetic energy operator and the second one describes the interacting non-local potential between the active electron and each molecular nuclei:
\begin{equation}
\hat{V}_{\textbf{M}}(\textbf{p},\textbf{p}^{\prime})= -\gamma' \:\phi (\textbf{p}) \: \phi (\textbf{p}^{\prime}) \: \sum_{j=1}^{3} \:e^{-\textit{i}\textbf{R}_{j} \cdot  (\textbf{p}-\textbf{p}^{\prime}) },
\end{equation}
where $\phi({\bf p})=\frac{1}{\sqrt{{\bf p}^2 +\Gamma^2}}$ is an auxiliary function and $\gamma'= \frac{\gamma}{3}$ is a parameter related with the shape of the ground state. 
By using $\hat{H}_{\textbf{M}}(\textbf{p},\textbf{p}^{\prime})$ from Eq.~(\ref{Hm}), we write the stationary  Schr\"odinger equation as follows:
\begin{equation}
\hat{H}_{\textbf{M}}(\textbf{p},\textbf{p}^{\prime})\Psi(\textbf{p}) = \int{\textit{d}^3 \textbf{p}^{\prime}\hat{H}_{\textbf{M}}(\textbf{p},\textbf{p}^{\prime})\Psi(\textbf{p}^{\prime})} = E_0 \: \Psi(\textbf{p}),
\end{equation}
where $E_0$ denotes the energy of the bound state. Thus, for our three-center system the Schr\"odinger equation reads:
\begin{eqnarray} 
\bigg(\frac{p^2}{2}  + I_p  \bigg) \Psi_{0\textbf{M}}(\textbf{p}) &= &\: \gamma' \: \phi (\textbf{p})\:\sum_{j=1}^{3}\:e^{- \textit{i} \textbf{R}_j\cdot \textbf{p}} \int{ \textit{d}^3 \textbf{p}^{\prime} \Psi_{0\textbf{M}}(\textbf{p}^{\prime}) \phi (\textbf{p}^{\prime}) e^{\textit{i}\textbf{R}_j\cdot \textbf{p}^{\prime} } }.
\label{Eq:Sch1}
\end{eqnarray} 
Defining new variables $\check{\varphi} _{j}$ as: 
 \begin{equation}
\check{\varphi} _{j} = \int{ \textit{d}^3 \textbf{p}^{\prime} \Psi_{0\textbf{M}}(\textbf{p}^{\prime}) \phi (\textbf{p}^{\prime}) e^{\textit{i}\textbf{R}_j \cdot \textbf{p}^{\prime}}} = \int{ \frac{ \textit{d}^3 \textbf{p}^{\prime}  \Psi_{0\textbf{M}}(\textbf{p}^{\prime}) e^{\textit{i}\textbf{R}_j \cdot \textbf{p}^{\prime} }}{\sqrt{  {p^{\prime}}^2 + \Gamma^2}}},
\label{Eq:gamma1}
\end{equation}
we could analytically obtain the bound states by solving Eq.~(\ref{Eq:Sch1}) in the momentum representation. Explicitly we can write:
\begin{equation}
\bigg(\frac{p^2}{2}  + I_p  \bigg) \Psi_{0\textbf{M}}(\textbf{p})= \gamma' \: \phi (\textbf{p}) \:\sum_{j=1}^{3}\: e^{- \textit{i}\textbf{R}_j\cdot \textbf{p}} \: \check{\varphi}_j,
\label{Eq:Sch12}
\end{equation}
where $I_p$ denotes the ionization potential that is related to the ground state potential energy by $E_{0} = -I_p$. The wavefunction $\Psi_{0\textbf{M}}(\textbf{p})$ for the bound state in momentum space is defined as:
 \begin{equation}
\Psi_{0\textbf{M}}(\textbf{p}) =  \frac{ \gamma' }{\sqrt{(p^2 + \Gamma^2})(\frac{p^2}{2} + I_p)}\:\sum_{j=1}^{3}\: \check{\varphi}_j e^{- \textit{i}\textbf{R}_j \cdot \textbf{p} } .
 \label{Eq:MWF1}
\end{equation}
In order to find the value of the constants we multiply and divide Eq.~(\ref{Eq:MWF1}) by $e^{- \textit{i}  \textbf{R}_{j} \cdot \textbf{p} }$ and $\sqrt{p^2 + \Gamma ^2}$, respectively.  After some algebra we find that:
\begin{eqnarray}
\check{\varphi}_1I_1 &=& \check{\varphi}_2 I_2 + \check{\varphi}_3 I_3,\nonumber \\
\check{\varphi}_2I_1&=&\check{\varphi}_1 I_2 + \check{\varphi}_3 I_2,\nonumber \\
\check{\varphi}_3I_1&=&\check{\varphi}_2 I_2 + \check{\varphi}_1 I_3,
\label{Eq:SystmEq}
\end{eqnarray}
where $I_1$, $I_2$ and $I_3$ read as:
\begin{eqnarray}
I_1=1-\gamma'    \int{ \frac{\textit{d}^3 \textbf{p} \:  \phi^2 (\textbf{p}) } {( \frac{p^2}{2} + I_p )}},
\end{eqnarray}
\begin{eqnarray}
I_2 &=&\gamma'  \int{ \frac{\textit{d}^3 \textbf{p} \: \phi^2 (\textbf{p}) } {( \frac{p^2}{2} + I_p )} \: e^{\textit{i}(\textbf{R}_1 -\textbf{R}_2 ) \cdot \textbf{p}}}=\gamma'  \int{ \frac{\textit{d}^3 \textbf{p}  \phi^2 (\textbf{p}) } {( \frac{p^2}{2} + I_p )} e^{\textit{i}(\textbf{R}_3 -\textbf{R}_2 ) \cdot \textbf{p}}}
\end{eqnarray}
and
\begin{eqnarray}
I_3=\gamma'  \int{ \frac{\textit{d}^3 \textbf{p} \phi^2 (\textbf{p}) } {( \frac{p^2}{2} + I_p )} e^{\textit{i}(\textbf{R}_1 -\textbf{R}_3 ) \cdot \textbf{p}}}=\gamma'  \int{ \frac{\textit{d}^3 \textbf{p} \phi^2 (\textbf{p}) } {( \frac{p^2}{2} + I_p )} e^{-\textit{i}(\textbf{R}_1 -\textbf{R}_3 ) \cdot \textbf{p}}},
\end{eqnarray}
respectively.

Solving the system of equations Eq.~(\ref{Eq:SystmEq}) we find the relations between the $\check{\varphi} _{j}$ defined by Eqs.~(\ref{Eq:gamma1}) and $I_1$, $I_2$ and $I_3$. The system Eq.~(\ref{Eq:SystmEq}) is solved with the restriction:
\begin{eqnarray}
I_3 = \frac{I^2_1 -2\:I^2_2}{I_1}\:;\:\:\:\: I_1\neq0,
\label{Eq:Int1}
\end{eqnarray}
and 
\begin{eqnarray}
\check{\varphi}_1=\check{\varphi}_3= \frac{ I_2}{I_1-I_3}\: \check{\varphi}_2.
\label{Eq:Int2}
\end{eqnarray}
$I_1$, $I_2$ and $I_3$ are written in spherical coordinates as:
\begin{equation}
\textit{I}_1 =1-\frac{4 \pi^2 \gamma'}{\Gamma + \sqrt{2I_p}},
\end{equation}
\begin{equation}
\textit{I}_2  =  \frac{8 \pi^2 \gamma'}{R} \Bigg\{ \frac{ e^{-\frac{R\Gamma}{2}} -e^{-\frac{R \sqrt{2I_p}}{2}} }{2I_p- \Gamma^2} \Bigg\},
\end{equation}
and
\begin{eqnarray}
I_3=\frac{4 \pi^2 \gamma'}{R \sin(\alpha/2)} \Bigg\{ \frac{  e^{-R \sin(\alpha/2)\Gamma} -e^{-R\sin(\alpha/2) \sqrt{2I_p}}  }{2I_p- \Gamma^2} \Bigg\}.
\end{eqnarray}
Finally, Eq.~(\ref{Eq:MWF1}), can be written as:
  \begin{equation}
\Psi_{0\textbf{M}}(\textbf{p}) =  \frac{\mathcal{M}_{3N}}{\sqrt{(p^2 + \Gamma^2})(\frac{p^2}{2} + I_p)}  \bigg[ \Big(\frac{ I_2}{I_1-I_3} \Big) \:e^{- \textit{i}\textbf{R}_1 \cdot \textbf{p} }   +   e^{- \textit{i} \textbf{R}_2 \cdot \textbf{p} } +  \Big( \frac{ I_2}{I_1-I_3}\Big) \:e^{- \textit{i} \textbf{R}_3 \cdot \textbf{p} } \bigg],
 \label{Eq:MWF2}
 \end{equation} 
where $\mathcal{M}_{3N}= \gamma' \: \check{\varphi}_1= \frac{\gamma}{3} \: \check{\varphi}_1$ is a normalization constant. It can be calculated using the usual normalization condition:
 \begin{equation}
  \int {\textit{d}^3 \textbf{p} \: \Psi_{0\textbf{M}}(\textbf{p})^* \: \Psi_{0\textbf{M}}(\textbf{p})} = 1.
\end{equation}
From the above equation ${\mathcal{M}_{3N}}$ can be written as: 
\begin{equation} 
1=  {{\mathcal{M}_{3N}}}^2 \: I_4,
\end{equation}
with $I_4$ defined as:
\begin{eqnarray}
I_4 &&= \Big(\frac{2\:I_2}{I_1-I_3}\Big)^2 \: \Bigg\{ \frac{4\pi^2} {R(2I_p -\Gamma^2)^2} \Bigg[ e^{-R \Gamma} - e^{-R \sqrt{2I_p}} \Bigg( \frac{2 \sqrt{2I_p} + R(2I_p -\Gamma^2)}{2\sqrt{2I_p}} \Bigg) \Bigg]+ \\
&& \frac{4\pi^2} {R \cos(\alpha/2)(2I_p -\Gamma^2)^2} \Bigg[ e^{-R \cos(\alpha/2) \Gamma} - e^{-R \cos(\alpha/2) \sqrt{2I_p}} \Bigg( \frac{2 \sqrt{2I_p} + R \cos(\alpha/2)(2I_p -\Gamma^2)}{2\sqrt{2I_p}} \Bigg) \Bigg] \Bigg\}\nonumber\\
&& +\frac{8\:I_2^2}{I_1-I_3}  \times \frac{4\pi^2} {\frac{R}{2}(2I_p -\Gamma^2)^2} \Bigg[ e^{-\frac{R}{2} \Gamma} - e^{-\frac{R}{2} \sqrt{2I_p}} \Bigg( \frac{2 \sqrt{2I_p} + \frac{R}{2}(2I_p -\Gamma^2)}{2\sqrt{2I_p}} \Bigg) \Bigg] 
+  \frac{4\pi^2 (\sqrt{2I_p} - \Gamma)^2}{\sqrt{2I_p}(2I_p -\Gamma^2)^2}.\nonumber 
\label{Eq:M}
\end{eqnarray}
With the exact knowledge of $\mathcal{M}_{3N}$ we have now defined the bound state in our three-center molecular system from 
Eq.~(\ref{Eq:MWF2}).  Notice that the dependency of the system energy with the internuclear distance appears from the solution of the system of equations, Eq.~(\ref{Eq:SystmEq}). In order to find potential energy surface describing the whole molecule we need to use Eq.~(\ref{Eq:Int1}) to obtain the relation: 
\begin{eqnarray} I_3 I_1 = I_1^2- 2 I_2^2.
\label{Eq:Int12}
\end{eqnarray}

\subsection*{Bound-continuum transition matrix element}
The total dipole matrix element for the three-center molecular system is defined as a sum: 
\begin{equation}
\textbf{d}_{3N}( \textbf{v})=-\sum_{j=1}^{3} \Big( \langle \textbf{v}_p |\textbf{r}|0_j \rangle +\textbf{R}_j \langle \textbf{v}_p |0_j\rangle \Big).
 \label{Eq:d3m1}
\end{equation}
Equation (\ref{Eq:d3m1}) can be explicitly written as:
\begin{eqnarray}
\textbf{d}_{3N}( \textbf{p}_0)&=&  -2i\:\mathcal{M}_{3N} \mathcal{ A} ( \textbf{p}_0)
\Bigg[ \frac{I_2}{I_1-I_3} \: \Big(e^{-i\textbf{R}_1 \cdot  \textbf{p}_0} + e^{-i\textbf{R}_3 \cdot  \textbf{p}_0}\Big) + 1 \Bigg],
 \label{Eq:d3mfull}
\end{eqnarray} 
where:
\begin{equation}
 \mathcal{ A} ( \textbf{p}_0)=\nabla_{\textbf{p}} \Bigg[\frac{1}{(p^2+\Gamma^2)^{\frac{1}{2}}(p^2 + 2 I_p)}\Bigg]\Bigg\rvert_{{\bf p}_0}=- \textbf{p}_0\: \frac{(3p_0^2 + 2I_p +2\Gamma^2)}{(p_0^2 + \Gamma^2)^{\frac{3}{2}}(p_0^2 + 2I_p)^2}.
\end{equation}
Finally notice that we could extract the contributions of each center from Eq.~(\ref{Eq:d3mfull}), i.e.~
\begin{equation}
\textbf{d}_{1}( \textbf{p}_0)  = - 2\textit{i}\: \mathcal{M}_{3N} \mathcal{A}( \textbf{p}_0)\: \Big( \frac{I_2}{I_1-I_3} \Big) e^{-i\textbf{R}_1 \cdot \textbf{p}_0},
\end{equation}
\begin{equation}
\textbf{d}_{2}( \textbf{p}_0)  = - 2\textit{i}\: \mathcal{M}_{3N} \mathcal{A}( \textbf{p}_0),
\label{Eq:d+1}
\end{equation}
and 
\begin{equation}
\textbf{d}_{3}( \textbf{p}_0)  = - 2\textit{i}\: \mathcal{M}_{3N} \mathcal{A}( \textbf{p}_0)\:\Big( \frac{I_2}{I_1-I_3} \Big)  e^{-i\textbf{R}_3 \cdot \textbf{p}_0 } .
\label{Eq:d+3}
\end{equation}

%


\end{document}